%% file: main.tex
\documentclass[preprint,11pt,3p,review]{elsarticle}

    \journal{%
        Journal of the Mechanical Behavior of Biomedical Materials
    }

    \usepackage{local-colors}
    \usepackage{paper-style}
    \usepackage{tensor-operators}

    \usepackage{tikz}
    \usetikzlibrary{arrows.meta}

    \tikzset{
        pics/picStatsTest/.style args={#1/#2}{%
            code={%
                \draw[%
                    densely dotted,
                    {Circle[open,length=2pt]}-{Circle[open,length=2pt]},
                ]   (-0.5cm,0cm)
                    -- node[%
                            midway,
                            fill=white,
                            inner sep=0pt,
                            label={[fill=white,inner sep=2pt]#1:{\small #2}}
                        ]   {$p$}
                    ++(1cm,0cm);
            }
        }
    }

    \loadglsentries{glossary.tex}
    \loadglsentries{acronyms.tex}
    \loadglsentries{local-glossary.tex}
    \makeglossaries

    \usepackage{chngcntr}

    \usepackage{nth}

    \usepackage[capitalise]{cleveref}

\begin{document}

    \import{tex/}{elsevier-header.tex}


    \sisetup{%
        detect-all,
        output-decimal-marker={.},
        multi-part-units=single,
        separate-uncertainty=true,
        per-mode=symbol,
    }%

    \glsresetall

        \import{tex/}{introduction.tex}
        \import{tex/}{methodology.tex}
        \import{tex/}{results.tex}
        \import{tex/}{discussion.tex}

    \section{Acknowledgements}

        \noindent
        This research was supported by grants 2017/18514-1 and 2019/19098-7 of
        the \gls{fapesp} (the funder had no involvment in any step of the
        production of this manuscript). It was also supported by resources
        supplied by the Center for Scientific Computing (NCC/GridUNESP) of the
        \gls{unesp} (\url{www2.unesp.br/portal#!/gridunesp}), by ACENET
        (\url{www.ace-net.ca}) through Dalhousie University and Compute Canada
        (\url{www.computecanada.ca}).

    \bibliographystyle{elsarticle-num-names}
    \bibliography{main}

    \newpage
    \appendix
    \import{tex/}{supplementaryMaterial.tex}

\end{document}

%% file: glossary.tex
%
%

\newglossaryentry{radial}{
    type=hidden,
    name={\ensuremath{r}},
    description={Radial coordinate of the cylindric coordinate system}
}
\newglossaryentry{azimuthal}{
    type=hidden,
    name={\ensuremath{\theta}},
    description={Circunferential coordinate of the cylindric coordinate system}
}
\newglossaryentry{z}{
    type=hidden,
    name={\ensuremath{z}},
    description={Axial coordinate of the cylindric coordinate system}
}

\newglossaryentry{xAxis}{
    type=hidden,
    name={\ensuremath{x}},
    description={%
        First coordinate of the Cartesian system (normally, $x_1$)
    }
}
\newglossaryentry{yAxis}{
    type=hidden,
    name={\ensuremath{y}},
    description={%
        Second coordinate of the Cartesian system (normally, $x_2$)
    }
}
\newglossaryentry{zAxis}{
    type=hidden,
    name={\ensuremath{z}},
    description={%
        Third coordinate of the Cartesian system (normally, $x_3$)
    }
}

\newglossaryentry{average}{
    type=subscript,
    name={\ensuremath{avg}},
    description={Indicates a time and surface averaged variable}
}
\newglossaryentry{maximum}{
    type=subscript,
    name={\ensuremath{max}},
    description={Indicates the maximum value of a sample}
}
\newglossaryentry{minimum}{
    type=subscript,
    name={\ensuremath{min}},
    description={Indicates the minimum value of a sample}
}
\newglossaryentry{percentil95}{
    type=subscript,
    name={\ensuremath{95}},
    description={Indicates the 95th percentile of a distribution}
}
\newglossaryentry{percentil99}{
    type=subscript,
    name={\ensuremath{99}},
    description={Indicates the 99th percentile of a distribution}
}
\newglossaryentry{solid}{
    type=superscript,
    name={\ensuremath{\mathrm{s}}},
    description={Used for solid-related variables}
}
\newglossaryentry{fluid}{
    type=superscript,
    name={\ensuremath{\mathrm{f}}},
    description={Used for fluid-related variables}
}
\newglossaryentry{fs}{
    type=superscript,
    name={\ensuremath{\mathrm{fs}}},
    description={Used for variables related to the fluid-solid interface}
}
\newglossaryentry{zeroTime}{
    type=superscript,
    name={\ensuremath{\mathrm{0}}},
    description={%
        Variables at the initial instant or the reference configuration
    }
}
\newglossaryentry{timeIndex}{
    type=superscript,
    name={\ensuremath{\mathrm{n}}},
    description={Time index}
}
\newglossaryentry{fsiIteration}{
    type=subscript,
    name={\ensuremath{k}},
    description={%
        Index for a coupling iteration of the outer loop of the IQN-ILS
        algorithm
    }
}
\newglossaryentry{iteration}{
    type=hidden,
    name={\ensuremath{m}},
    description={%
        Index for iterations when solving a linear system  $\matrix{A}
        \cvec{x} = \cvec{b}$ with an iterative method
    }
}
\newglossaryentry{normal}{
    type=subscript,
    name={\ensuremath{n}},
    description={%
        Indicate the normal component of a vector to a surface, i.e.
        $\vec{u}_{n} = \vec{n}(\vec{n} \dprod \vec{u})$
    }
}
\newglossaryentry{tangential}{
    type=subscript,
    name={\ensuremath{t}},
    description={%
        Indicate the tangential component of a vector to a surface, i.e.
        $\vec{u}_{t} = (\tensor{I} - \vec{n}\vec{n}) \dprod \vec{u}$
    }
}
\newglossaryentry{boundary}{
    type=subscript,
    name={\ensuremath{b}},
    description={%
        Indicate a property on a boundary of a discretized domain in the
        context of the Finite Volume Method.
    }
}

\newglossaryentry{time}{
    type=latin,
    name={\ensuremath{t}},
    description={Time},
    unit={\si{\second}}
}
\newglossaryentry{temperature}{
    type=latin,
    name={\ensuremath{T}},
    description={Temperature},
    unit={\si{\kelvin}}
}
\newglossaryentry{deltat}{
    type=greek,
    name={\ensuremath{\Delta \gls{time}}},
    description={Time-step},
    unit={\si{\second}}
}
\newglossaryentry{mass}{
    type=latin,
    name={\ensuremath{m}},
    description={Mass},
    unit={\si{\kilo\gram}}
}
\newglossaryentry{length}{
    type=latin,
    name={\ensuremath{L}},
    description={Length},
    unit={\si{\meter}}
}
\newglossaryentry{perimeter}{
    type=latin,
    name={\ensuremath{P}},
    description={Perimeter of a contour},
    unit={\si{\meter}}
}
\newglossaryentry{surfaceArea}{
    type=latin,
    name={\ensuremath{A}},
    description={Area of a surface or region},
    unit={\si{\square\meter}}
}
\newglossaryentry{volume}{
    type=latin,
    name={\ensuremath{V}},
    description={Volume of a region in spatial coordinates},
    unit={\si{\cubic\meter}}
}
\newglossaryentry{surface}{
    type=latin,
    name={\ensuremath{S}},
    description={%
        Generic representation of surface (or patch of another surface)
    },
    unit={\si{\square\meter}}
}
\newglossaryentry{density}{
    type=greek,
    name={\ensuremath{\rho}},
    description={Density},
    unit={\si{\kilo\gram\per\cubic\meter}}
}
\newglossaryentry{force}{
    type=latin,
    name={\ensuremath{\vec{f}}},
    description={Force vector},
    unit={\si{\newton}}
}
\newglossaryentry{linearMomentum}{
    type=latin,
    name={\ensuremath{\vec{p}}},
    description={Linear momentum vector},
    unit={\si{\kilo\gram\meter\per\second}}
}
\newglossaryentry{forceComp}{
    type=hidden,
    name={\ensuremath{f}},
    description={Component of the force vector},
    unit={\si{\newton}}
}
\newglossaryentry{traction}{
    type=latin,
    name={\ensuremath{\vec{t}}},
    description={Traction vector},
    unit={\si{\newton\per\square\meter}}
}
\newglossaryentry{CauchyStress}{
    type=greek,
    name={\ensuremath{\tensor{\sigma}}},
    description={Cauchy stress tensor},
    unit={\si{\pascal}}
}
\newglossaryentry{CauchyStressComp}{
    type=hidden,
    name={\ensuremath{\sigma}},
    description={Cauchy stress tensor component},
    unit={\si{\pascal}}
}
\newglossaryentry{bodyForcesPerVolume}{
    type=latin,
    name={\ensuremath{\vec{f}_b}},
    description={Body forces per unit volume},
    unit={\si{\newton\per\cubic\meter}}
}
\newglossaryentry{bodyForcesPerMass}{
    type=latin,
    name={\ensuremath{\vec{b}}},
    description={Body forces per unit mass},
    unit={\si{\newton\per\kilo\gram}}
}
\newglossaryentry{bodyForcesPerMassComp}{
    type=hidden,
    name={\ensuremath{{b}}},
    description={Body forces per unit mass component},
    unit={\si{\newton\per\kilo\gram}}
}
\newglossaryentry{gravity}{
    type=latin,
    name={\ensuremath{\vec{g}}},
    description={Gravitational acceleration vector},
    unit={\si{\meter\per\square\second}}
}
\newglossaryentry{gravityComp}{
    type=hidden,
    name={\ensuremath{g}},
    description={Gravitational acceleration vector},
    unit={\si{\meter\per\square\second}}
}

%
%

\newglossaryentry{particle}{
    type=latin,
    name={\ensuremath{\xi}},
    description={Representation of a material particle},
    unit={-}
}
\newglossaryentry{EulerCoord}{
    type=latin,
    name={\ensuremath{\vec{x}}},
    description={Eulerian or spatial coordinates system (Cartesian coordinates)},
    unit={\si{\meter}}
}
\newglossaryentry{meshCoord}{
    type=greek,
    name={\ensuremath{\gvec{\chi}}},
    description={%
        Referential coordinates system for the arbitrary Lagrangian-Eulerian
        description
    },
    unit={\si{\meter}}
}
\newglossaryentry{LagrangeCoord}{
    type=greek,
    name={\ensuremath{\gvec{\xi}}},
    description={Lagrangian or material coordinates system},
    unit={\si{\meter}}
}
\newglossaryentry{meshCoordComp}{
    type=hidden,
    name={\ensuremath{\chi}},
    description={Referential, ALE or mesh coordinates system components},
    unit={\si{\meter}}
}
\newglossaryentry{EulerCoordComp}{
    type=hidden,
    name={\ensuremath{x}},
    description={Eulerian or spatial coordinates components},
    unit={\si{\meter}}
}
\newglossaryentry{LagrangeCoordComp}{
    type=hidden,
    name={\ensuremath{\xi}},
    description={Lagrangian or material coordinates components},
    unit={\si{\meter}}
}

\newglossaryentry{domain}{
    type=greek,
    name={\ensuremath{\Omega}},
    description={Spatial solution domain},
    unit={\si{\cubic\meter}}
}
\newglossaryentry{domainBoundary}{
    type=greek,
    name={\ensuremath{\Gamma}},
    description={Boundary of the domain \gls{domain}},
    unit={\si{\square\meter}}
}
\newglossaryentry{contour}{
    type=greek,
    name={\ensuremath{\Gamma}},
    description={%
        Generic three-dimensional contour; boundary of an open surface
        \gls{surface}, $\partial \gls{surface}$, in spatial coordinates
    },
    unit={\si{\meter}}
}
\newglossaryentry{nablaSpatial}{
    type=other,
    name={\ensuremath{\nabla}},
    description={Nabla operator in Eulerian coordinates},
    unit={\si{\per\meter}}
}
\newglossaryentry{nablaMaterial}{
    type=other,
    name={\ensuremath{\nabla_0}},
    description={Nabla operator in the initial material coordinates},
    unit={\si{\per\meter}}
}
\newglossaryentry{materialDomain}{
    type=greek,
    name={\ensuremath{\Omega_{\xi}}},
    description={Solution domain as seen by the material coordinates},
    unit={\si{\cubic\meter}}
}
\newglossaryentry{meshDomain}{
    type=greek,
    name={\ensuremath{\Omega_{\chi}}},
    description={Solution domain as seen by the reference coordinates},
    unit={\si{\cubic\meter}}
}
\newglossaryentry{materialVolume}{
    type=latin,
    name={\ensuremath{V_{\xi}}},
    description={Volume of an amount of particles in material coordinates},
    unit={\si{\cubic\meter}}
}
\newglossaryentry{meshVolume}{
    type=latin,
    name={\ensuremath{V_{\chi}}},
    description={Control volume in ALE coordinates},
    unit={\si{\cubic\meter}}
}
\newglossaryentry{sNormalVector}{
    type=latin,
    name={\ensuremath{\vec{n}}},
    description={%
        Unit normal vector to the surface \gls{surface} pointing outwards
    },
    unit={-}
}

\newglossaryentry{funcMaterial}{
    type=latin,
    name={\ensuremath{f_{\xi}}},
    description={Function $f$ representing any scalar variable of the material},
    unit={-}
}
\newglossaryentry{funcSpatial}{
    type=latin,
    name={\ensuremath{f}},
    description={%
        Spatial function $f$ representing any scalar variable of the material
    },
    unit={-}
}
\newglossaryentry{funcMesh}{
    type=latin,
    name={\ensuremath{f_{\chi}}},
    description={%
        ALE description of a function $f$ representing any scalar variable of
        the material
    },
    unit={-}
}
\newglossaryentry{materialSpatialMap}{
    type=greek,
    name={\ensuremath{\gvec{\Phi}}},
    description={Mapping of material coordinates onto spatial coordinates},
    unit={-}
}
\newglossaryentry{meshSpatialMap}{
    type=greek,
    name={\ensuremath{\gvec{\Psi}}},
    description={Mapping of mesh coordinates onto spatial coordinates},
    unit={-}
}
\newglossaryentry{materialMeshMap}{
    type=greek,
    name={\ensuremath{\gvec{\Xi}}},
    description={Mapping of material coordinates onto mesh coordinates},
    unit={-}
}
\newglossaryentry{identityTensor}{
    type=latin,
    name={\ensuremath{\tensor{I}}},
    description={Second-order identity tensor},
    unit={-}
}
\newglossaryentry{kroneckerDelta}{
    type=hidden,
    name={\ensuremath{\delta_{ij}}},
    description={Kronecker's delta},
    unit={-}
}
\newglossaryentry{fourthIdentityTensor}{
    type=latin,
    name={\ensuremath{\fourthTensor{I}}},
    description={Fourth-order identity tensor},
    unit={-}
}

\newglossaryentry{Jacobian}{
    type=latin,
    name={\ensuremath{J}},
    description={Jacobian (determinant of the deformation gradient)},
    unit={-}
}
\newglossaryentry{materialJacobianMatrix}{
    type=latin,
    name={\ensuremath{\mathbf{J}_{\gls{materialSpatialMap}}}},
    description={Jacobian matrix of the material-spatial mapping},
    unit={-}
}
\newglossaryentry{meshJacobianMatrix}{
    type=latin,
    name={\ensuremath{\mathbf{J}_{\gls{meshSpatialMap}}}},
    description={Jacobian matrix of the mesh-spatial mapping},
    unit={-}
}
\newglossaryentry{materialJacobian}{
    type=latin,
    name={\ensuremath{\mathrm{J}_{\gls{materialSpatialMap}}}},
    description={Jacobian of the material-spatial mapping},
    unit={-}
}
\newglossaryentry{meshJacobian}{
    type=latin,
    name={\ensuremath{\mathrm{J}_{\gls{meshSpatialMap}}}},
    description={Jacobian of the mesh-spatial mapping},
    unit={-}
}

\newglossaryentry{velocity}{
    type=latin,
    name={\ensuremath{\vec{v}}},
    description={Material velocity},
    unit={\si{\meter\per\second}}
}
\newglossaryentry{angularVelocity}{
    type=latin,
    name={\ensuremath{\omega}},
    description={Angular velocity},
    unit={\si{\radian\per\second}}
}
\newglossaryentry{velocityComp}{
    type=hidden,
    name={\ensuremath{v}},
    description={Material velocity},
    unit={\si{\meter\per\second}}
}
\newglossaryentry{xVelocity}{
    type=latin,
    name={\ensuremath{v_x}},
    description={%
        Velocity component on the x direction of the spatial coordinates
    },
    unit={\si{\meter\per\second}}
}
\newglossaryentry{yVelocity}{
    type=latin,
    name={\ensuremath{v_y}},
    description={%
        Velocity component on the y direction of the spatial coordinates
    },
    unit={\si{\meter\per\second}}
}
\newglossaryentry{zVelocity}{
    type=latin,
    name={\ensuremath{v_z}},
    description={%
        Velocity component on the z direction of the spatial coordinates
    },
    unit={\si{\meter\per\second}}
}
\newglossaryentry{meshVelocity}{
    type=greek,
    name={\ensuremath{\gvec{\omega}}},
    description={Mesh velocity of a cell centroid},
    unit={\si{\meter\per\second}}
}
\newglossaryentry{xMeshVelocity}{
    type=hidden,
    name={\ensuremath{\omega_x}},
    description={%
        Mesh velocity component on the x direction of the spatial coordinates
    },
    unit={\si{\meter\per\second}}
}
\newglossaryentry{yMeshVelocity}{
    type=hidden,
    name={\ensuremath{\omega_y}},
    description={%
        Mesh velocity component on the y direction of the spatial coordinates
    },
    unit={\si{\meter\per\second}}
}
\newglossaryentry{zMeshVelocity}{
    type=hidden,
    name={\ensuremath{\omega_z}},
    description={%
        Mesh velocity component on the z direction of the spatial coordinates
    },
    unit={\si{\meter\per\second}}
}
\newglossaryentry{meshVelocityNode}{
    type=greek,
    name={\ensuremath{\gvec{\omega}_{node}}},
    description={%
        Mesh velocity at the geometric nodes of the Finite Volume mesh
    },
    unit={\si{\meter\per\second}}
}
\newglossaryentry{materialMeshVelocity}{
    type=latin,
    name={\ensuremath{\hat{\vec{v}}}},
    description={%
        Material velocity measured from the ALE coordinates point of view
    },
    unit={\si{\meter\per\second}}
}
\newglossaryentry{convectiveVelocity}{
    type=latin,
    name={\ensuremath{\vec{c}}},
    description={Convective velocity},
    unit={\si{\meter\per\second}}
}

%
%


\newglossaryentry{massFlowRate}{
    type=latin,
    name={\ensuremath{\dot{m}}},
    description={Mass flow rate},
    unit={\si{\kilo\gram\per\second}}
}

\newglossaryentry{numberMoles}{
    type=latin,
    name={\ensuremath{N}},
    description={Number of moles of a species $i$ in a system},
    unit={\si{\mol}}
}
\newglossaryentry{numberSpecies}{
    type=latin,
    name={\ensuremath{n}},
    description={Number of moles of a species $i$ in a system},
    unit={}
}

\newglossaryentry{specificVolume}{
    type=latin,
    name={\ensuremath{v}},
    description={Specific volume -- volume per mass},
    unit={\si{\cubic\meter\per\kilogram}}
}

\newglossaryentry{internalEnergy}{
    type=latin,
    name={\ensuremath{U}},
    description={Internal energy of a system},
    unit={\si{\joule}}
}
\newglossaryentry{intInternalEnergy}{
    type=latin,
    name={\ensuremath{u}},
    description={Intensive internal energy of a system},
    unit={\si{\joule\per\kilo\gram}}
}

\newglossaryentry{enthalpy}{
    type=latin,
    name={\ensuremath{H}},
    description={Enthalpy of a system},
    unit={\si{\joule}}
}
\newglossaryentry{intEnthalpy}{
    type=latin,
    name={\ensuremath{h}},
    description={Intensive enthalpy of a system},
    unit={\si{\joule\per\kilo\gram}}
}

\newglossaryentry{entropy}{
    type=latin,
    name={\ensuremath{S}},
    description={Entropy of a system},
    unit={\si{\joule\per\kelvin}}
}
\newglossaryentry{intEntropy}{
    type=latin,
    name={\ensuremath{s}},
    description={Intensive entropy of a system},
    unit={\si{\joule\per\kilo\gram\per\kelvin}}
}

\newglossaryentry{HelmholtzFreeEnergy}{
    type=latin,
    name={\ensuremath{F}},
    description={Helmholtz free-energy of a system},
    unit={\si{\joule}}
}
\newglossaryentry{intHelmholtzFreeEnergy}{
    type=latin,
    name={\ensuremath{f}},
    description={Intensive Helmholtz free-energy of a system},
    unit={\si{\joule\per\kilo\gram}}
}

\newglossaryentry{GibbsFreeEnergy}{
    type=latin,
    name={\ensuremath{G}},
    description={Gibbs free-energy of a system},
    unit={\si{\joule}}
}
\newglossaryentry{intGibbsFreeEnergy}{
    type=latin,
    name={\ensuremath{g}},
    description={Intensive Gibbs free-energy of a system},
    unit={\si{\joule\per\kilo\gram}}
}

\newglossaryentry{entropyCreated}{
    type=latin,
    name={\ensuremath{S_{ger}}},
    description={%
        Entropy created or generated in a process undergone by a system
    },
    unit={\si{\joule\per\kelvin}}
}
\newglossaryentry{intEntropyCreated}{
    type=latin,
    name={\ensuremath{s_{ger}}},
    description={%
        Intensive entropy created or generated in a process undergone by a
        system
    },
    unit={\si{\joule\per\kelvin}}
}
\newglossaryentry{entropyCreatedRate}{
    type=latin,
    name={\ensuremath{\dot{S}_{ger}}},
    description={%
        Entropy created or generated in a process undergone by a system
    },
    unit={\si{\watt\per\kelvin}}
}
\newglossaryentry{intEntropyCreatedRate}{
    type=latin,
    name={\ensuremath{\dot{s}_{ger}}},
    description={%
        Intensive entropy created or generated in a process undergone by a
        system
    },
    unit={\si{\watt\per\kelvin}}
}

\newglossaryentry{chemicalPotential}{
    type=greek,
    name={\ensuremath{\mu}},
    description={%
        Chemical potential of a species crossing the boundary of a system
    },
    unit={\si{\joule\per\kelvin}}
}
\newglossaryentry{intIrreversibility}{
    type=latin,
    name={\ensuremath{I}},
    description={Intensive irreversibility irreversibility},
    unit={\si{\joule}}
}
\newglossaryentry{thermodynamicSubstate}{
    type=greek,
    name={\ensuremath{\upsilon}},
    description={%
        Generic thermodynamic sub-state of a system, identified as the
        intensive counterpart of the generalized displacement producing work
        transfers
    },
    unit={-}
}
\newglossaryentry{generalizedForce}{
    type=greek,
    name={\ensuremath{\tau}},
    description={Generalized force applied to a system},
    unit={-}
}
\newglossaryentry{generalizedDisplacement}{
    type=greek,
    name={\ensuremath{\Upsilon}},
    description={Generalized displacement applied to a system},
    unit={-}
}

\newglossaryentry{heatTransfer}{
    type=latin,
    name={\ensuremath{Q}},
    description={Heat transfer interaction on the boundary of a system},
    unit={\si{\joule}}
}
\newglossaryentry{heatFlowVector}{
    type=latin,
    name={\ensuremath{\vec{q}}},
    description={%
        Heat flux or heat flow vector, defined as the heat transfer per unit
        surface area
    },
    unit={\si{\watt\per\square\meter}}
}

\newglossaryentry{workTransfer}{
    type=latin,
    name={\ensuremath{W}},
    description={Generic work transfer interaction on the boundary of a system},
    unit={\si{\joule}}
}
\newglossaryentry{internalMechanicalPower}{
    type=latin,
    name={\ensuremath{\dot{w}_{int}}},
    description={%
        Internal mechanical power in a system, caused by the stress and
        deformation
    },
    unit={\si{\watt}}
}

\newglossaryentry{internalMechanicalWork}{
    type=latin,
    name={\ensuremath{w_{int}}},
    description={%
        Internal mechanical work in a system, caused by the stress and
        deformation
    },
    unit={\si{\joule}}
}
\newglossaryentry{numberWorkSources}{
    type=latin,
    name={\ensuremath{m}},
    description={Number of work sources to a system},
    unit={}
}

\newglossaryentry{isothermalCompressibility}{
    type=greek,
    name={\ensuremath{\psi}},
    description={Isothermal compressibility of a fluid},
    unit={\si{\square\second\per\square\meter}}
}

\newglossaryentry{bulkModulus}{
    type=greek,
    name={\ensuremath{\kappa}},
    description={Bulk modulus},
    unit={\si{\pascal}}
}

%
%

\newglossaryentry{cvNumber}{
    type=latin,
    name={\ensuremath{N_{CV}}},
    description={Number of control volumes in a finite volume mesh},
    unit={-}
}

\newglossaryentry{centroid}{
    type=subscript,
    name={\ensuremath{C}},
    description={%
        Indicates variables evaluated at the centroid of a control volume
    }
}
\newglossaryentry{nbCentroid}{
    type=subscript,
    name={\ensuremath{N}},
    description={%
        Indicates variables evaluated at the centroid of a neighbor control
        volume to cell $V_C$
    }
}
\newglossaryentry{owner}{
    type=subscript,
    name={\ensuremath{\gls{centroid}}},
    description={%
        Indicates variables evaluated at the centroid of a the face owner
        control volume
    }
}
\newglossaryentry{neighbor}{
    type=subscript,
    name={\ensuremath{\gls{nbCentroid}}},
    description={%
        Indicates variables evaluated at the centroid of a face neighbor
        control volume
    }
}
\newglossaryentry{face}{
    type=subscript,
    name={\ensuremath{f}},
    description={%
        Indicate a face of a polyhedral control volume and a property evaluated
        at it
    }
}
\newglossaryentry{skewFace}{
    type=subscript,
    name={\ensuremath{\gls{face}^{\prime}}},
    description={%
        Indicates the point where the vector joining two face adjacent cell's
        centroids intercepts the face
    }
}

\newglossaryentry{fvCellVolume}{
    type=latin,
    name={\ensuremath{\gls{volume}_{\gls{centroid}}}},
    description={Volume of a cell with centroid $C$ in a finite volume mesh},
    unit={\si{\cubic\meter}}
}
\newglossaryentry{fvCellSurface}{
    type=latin,
    name={\ensuremath{\gls{surface}_{\gls{centroid}}}},
    description={Surface of a cell with centroid $C$ in a finite volume mesh},
    unit={\si{\square\meter}}
}
\newglossaryentry{fvFaceNormal}{
    type=latin,
    name={\ensuremath{\vec{S}_{\gls{face}}}},
    description={Normal vector to the face $f$ of a polyhedral cell},
    unit={\si{\square\meter}}
}
\newglossaryentry{joinCentroidVector}{
    type=latin,
    name={\ensuremath{\vec{d}_{\gls{centroid}\gls{nbCentroid}}}},
    description={%
        Vector joining the centroids of adjacent cells in a finite volume mesh
    },
    unit={\si{\meter}}
}
\newglossaryentry{skewnessVector}{
    type=latin,
    name={\ensuremath{\vec{m}}},
    description={%
        Vector joining the face centroid with the interception point between
        the face and the line between the owner and neighbour cells.
    },
    unit={\si{\meter}}
}
\newglossaryentry{facePyramidHeigth}{
    type=latin,
    name={\ensuremath{\vec{h}_{\gls{face}\gls{centroid}}}},
    description={%
        Heigth of a cell face pyramid -- is the opposite of the vector joining
        the cell centroid and the face centroid
    },
    unit={\si{\meter}}
}
\newglossaryentry{centroidFaceVector}{
    type=latin,
    name={\ensuremath{\vec{d}_{\gls{centroid}\gls{face}}}},
    description={%
        Vector joining the centroids of a cell \gls{fvCellVolume} and the
        centroid of a face \gls{face}
    },
    unit={\si{\meter}}
}
\newglossaryentry{ownerFaceVector}{
    type=latin,
    name={\ensuremath{\vec{d}_{\gls{owner}\gls{face}}}},
    description={%
        Vector joining the centroid of a face owner cell and the centroid of
        the face \gls{face}
    },
    unit={\si{\meter}}
}
\newglossaryentry{neighFaceVector}{
    type=latin,
    name={\ensuremath{\vec{d}_{\gls{neighbor}\gls{face}}}},
    description={%
        Vector joining the centroid of a face neighbor cell and the centroid of
        the face \gls{face}
    },
    unit={\si{\meter}}
}

\newglossaryentry{projCentroidNormalVec}{
    type=hidden,
    name={\ensuremath{\vec{d}_{fn}}},
    description={%
        Vector formed by the projection of \gls{joinCentroidVector}
        on \gls{sNormalVector}
    },
    unit={\si{\meter}}
}
\newglossaryentry{projCentroidNormalNorm}{
    type=latin,
    name={\ensuremath{d_{fn}}},
    description={%
        Projection of the vector \gls{joinCentroidVector} on the face normal
        \gls{sNormalVector}
    },
    unit={\si{\meter}}
}
\newglossaryentry{advWeightFactor}{
    type=hidden,
    name={\ensuremath{w}},
    description={%
        Weight of cell \gls{centroid} on the advective interpolation profile
    },
    unit={-}
}
\newglossaryentry{faceVolumeSwept}{
    type=latin,
    name={\ensuremath{\dot{V}_f}},
    description={%
        Volume swept by face \gls{face} of control volume \gls{fvCellVolume}
    },
    unit={\si{\cubic\meter\per\second}}
}
\newglossaryentry{normalDecomposition1}{
    type=latin,
    name={\ensuremath{\vec{\beta}}},
    description={%
        Vector joining the centroid of face \gls{face} and the neighbor cell
        centroid \gls{nbCentroid}
    },
    unit={-}
}
\newglossaryentry{normalDecomposition1Norm}{
    type=hidden,
    name={\ensuremath{\beta}},
    description={%
        Modulus of the vector joining the centroid of face \gls{face} and the
        neighbor cell centroid \gls{nbCentroid}
    },
    unit={-}
}
\newglossaryentry{normalDecomposition2}{
    type=greek,
    name={\ensuremath{\vec{\kappa}}},
        description={Complement vector of the decomposition of
        \gls{fvFaceNormal} in \gls{normalDecomposition1}},
        unit={-}
}

\newglossaryentry{transportVariable}{
    type=greek,
    name={\ensuremath{\phi}},
    description={%
        Generic intensive (per unit mass) variable transported by a fluid flow
    },
    unit={$[\phi]/$\si{\kilo\gram}}
}
\newglossaryentry{integralTransportVariable}{
    type=greek,
    name={\ensuremath{\Phi}},
    description={%
        Generic intensive variable transported by a fluid flow integrated in a
        finite control volume
    },
    unit={$[\Phi]$}
}
\newglossaryentry{transportVariablePerVolume}{
    type=greek,
    name={\ensuremath{\varphi}},
    description={%
        Generic intensive (per unit volume) variable transported by a fluid
        flow
    },
    unit={$[\Phi]/$\si{\cubic\meter}}
}
\newglossaryentry{transportDiffusivity}{
    type=greek,
    name={\ensuremath{\Gamma^{\phi}}},
    description={Diffusivity of \gls{transportVariable}},
    unit={\si{\square\meter\per\second}}
}
\newglossaryentry{transportSource}{
    type=latin,
    name={\ensuremath{S^{\phi}}},
    description={%
        Source and/or sink terms of the differential transport equation of
        \gls{transportVariable}
    },
    unit={$[\phi]/$\si{\cubic\meter\per\second}}
}
\newglossaryentry{linearTransportSource}{
    type=latin,
    name={\ensuremath{\bar{S}^{\phi}}},
    description={%
        Mean source and/or sink terms of the integral transport equation of
        \gls{transportVariable}
    },
    unit={$[\phi]/$\si{\cubic\meter\per\second}}
}
\newglossaryentry{fvCoeffs}{
    type=latin,
    name={\ensuremath{a}},
    description={%
        Generic coefficients of the linearized transport equation for the
        volume $V_C$
    },
    unit={\si{\kilo\gram\per\second}}
}
\newglossaryentry{fvDiagCoeffs}{
    type=latin,
    name={\ensuremath{\gls{fvCoeffs}_{\gls{centroid}}}},
    description={%
        Diagonal coefficients of the linearized transport equation for the
        volume $V_C$
    },
    unit={\si{\kilo\gram\per\second}}
}
\newglossaryentry{fvOffDiagCoeffs}{
    type=latin,
    name={\ensuremath{\gls{fvCoeffs}_{\gls{nbCentroid}}}},
    description={%
        Off-diagonal coefficients of the linearized transport equation for the
        volume $V_C$
    },
    unit={\si{\kilo\gram\per\second}}
}
\newglossaryentry{fvMatrix}{
    type=latin,
    name={\ensuremath{\matrix{A}}},
    description={%
        Coefficients matrix of a system of equations assembled for a finite
        volume mesh
    },
    unit={$[\phi]$}
}
\newglossaryentry{sourceScalar}{
    type=latin,
    name={\ensuremath{b}},
    description={%
        Source term of the linearized transport equation for a control volume
        $V_C$
    },
    unit={$[\phi]$\si{\kilo\gram\per\second}}
}
\newglossaryentry{fvSource}{
    type=hidden,
    name={\ensuremath{\cvec{b}}},
    description={%
        Source term of the linearized system equation for a finite volume mesh
    },
    unit={-}
}
\newglossaryentry{sourceVector}{
    type=latin,
    name={\ensuremath{\vec{b}}},
    description={%
        Source vector of the linearized momentum equations for a control volume
    },
    unit={$[\phi]$\si{\kilo\gram\per\second}}
}
\newglossaryentry{pressureCorrection}{
    type=latin,
    name={\ensuremath{p'}},
    description={%
        Pressure correction field in the pressure-velocity algorithms to solve
        the coupled Navier-Stokes equations
    },
    unit={\si{\pascal}}
}
\newglossaryentry{transportOperator}{
    type=hidden,
    name={\ensuremath{\matrix{H}}},
    description={Transport operator},
    unit={-}
}
\newglossaryentry{sourceOperator}{
    type=hidden,
    name={\ensuremath{\matrix{B}}},
    description={Source operator},
    unit={-}
}
\newglossaryentry{pressureDiffOperator}{
    type=hidden,
    name={\ensuremath{\matrix{D}}},
    description={Pressure diffusive operator},
    unit={-}
}

\newglossaryentry{meshNonOrthog}{
    type=latin,
    name={\ensuremath{mesh_{nonOrtho}}},
    description={Mesh non-orthogonality}
}

\newglossaryentry{meshSkewness}{
    type=latin,
    name={\ensuremath{mesh_{skew}}},
    description={Mesh skewness}
}

\newglossaryentry{faceTwist}{
    type=latin,
    name={\ensuremath{mesh_{faceTwist}}},
    description={Mesh face twist}
}

\newglossaryentry{faceWeight}{
    type=latin,
    name={\ensuremath{mesh_{faceWeight}}},
    description={Mesh face weight}
}

\newglossaryentry{faceVolumeRatio}{
    type=latin,
    name={\ensuremath{mesh_{faceVolRatio}}},
    description={Mesh face volume ratio}
}

\newglossaryentry{faceTriangTwist}{
    type=latin,
    name={\ensuremath{mesh_{faceTriTwist}}},
    description={Mesh face twist}
}

\newglossaryentry{cellConvexity}{
    type=latin,
    name={\ensuremath{mesh_{cellConcave}}},
    description={Mesh cells convexity: it measures if a cell is concave.}
}

\newglossaryentry{pyramidVolume}{
    type=latin,
    name={\ensuremath{mesh_{cellPyrVol}}},
    description={Mesh cells pyramids volume}
}

\newglossaryentry{cellTetQuality}{
    type=latin,
    name={\ensuremath{mesh_{tetQuality}}},
    description={Mesh cells tetrahedral decomposition quality}
}
\newglossaryentry{polyCellTensor}{
    type=latin,
    name={\ensuremath{\tensor{T}}},
    description={%
        Second-order tensor formed by the outer product of the face vector of a
        generic polyhedral cell
    }
}
\newglossaryentry{cellDeterminant}{
    type=latin,
    name={\ensuremath{mesh_{cellDeterminant}}},
    description={Mesh face twist}
}

%
%
\newglossaryentry{dynamicViscosity}{
    type=greek,
    name={\ensuremath{\mu^{\gls{fluid}}}},
    description={Dynamic viscosity of a fluid},
    unit={\si{\pascal\second}}
}
\newglossaryentry{secondViscosity}{
    type=greek,
    name={\ensuremath{\lambda^{\gls{fluid}}}},
    description={%
        Second viscosity of a fluid (sometimes referred to as
        \emph{bulk viscosity})
    },
    unit={\si{\pascal\second}}
}
\newglossaryentry{kinematicViscosity}{
    type=greek,
    name={\ensuremath{\nu^{\gls{fluid}}}},
    description={Kinematic viscosity of a fluid},
    unit={\si{\square\meter\per\second}}
}
\newglossaryentry{pressure}{
    type=latin,
    name={\ensuremath{p}},
    description={Pressure field},
    unit={\si{\pascal}}
}
\newglossaryentry{meshDiffusivity}{
    type=greek,
    name={\ensuremath{\gamma}},
    description={%
        Mesh diffusivity for the Laplace equation governing the mesh motion
    },
    unit={-}
}
\newglossaryentry{centroidToBoundaryDistance}{
    type=latin,
    name={\ensuremath{l}},
    description={Cell centroid distance to a selected boundary},
    unit={\si{\meter}}
}
\newglossaryentry{viscousCauchyStress}{
    type=greek,
    name={\ensuremath{\gvec{\tau}}},
    description={Viscous part of the Cauchy stress tensor of a fluid flow},
    unit={\si{\pascal}}
}
\newglossaryentry{viscousCauchyStressComp}{
    type=hidden,
    name={\ensuremath{\tau}},
    description={Viscous part of the Cauchy stress tensor of a fluid flow},
    unit={\si{\pascal}}
}
\newglossaryentry{ReynoldsNumber}{
    type=latin,
    name={\ensuremath{Re}},
    description={Reynolds number},
    unit={-}
}
\newglossaryentry{CourantNumber}{
    type=latin,
    name={\ensuremath{Co}},
    description={Courant number},
    unit={-}
}
\newglossaryentry{shearRate}{
    type=latin,
    name={\ensuremath{\dot{\gamma}}},
    description={%
        Shear rate component in the flow between two coaxial cylinders
    },
    unit={\si{\per\second}}
}
\newglossaryentry{yieldStress}{
    type=latin,
    name={\ensuremath{\tau_C}},
    description={Yield stress of a Herschel-Bulkley fluid},
    unit={\si{\pascal}}
}
\newglossaryentry{consistency}{
    type=latin,
    name={\ensuremath{k_n}},
    description={Consistency of a Herschel-Bulkley fluid},
    unit={\si{\pascal\second^n}}
}
\newglossaryentry{indexFlow}{
    type=latin,
    name={\ensuremath{n}},
    description={Index flow of a Herschel-Bulkley fluid},
    unit={-}
}
\newglossaryentry{apparentViscosity}{
    type=greek,
    name={\ensuremath{\eta^{\gls{fluid}}}},
    description={Apparent viscosity of a non-Newtonian fluid},
    unit={\si{\pascal\second}}
}
\newglossaryentry{CassonCoefficient}{
    type=latin,
    name={\ensuremath{k}},
    description={Consistency coefficient of Casson's model},
    unit={\si{\pascal\second}}
}

\newglossaryentry{firstInvariant}{
    type=latin,
    name={\ensuremath{I_1}},
    description={First principal invariant of a second order tensor},
    unit={-}
}
\newglossaryentry{secondInvariant}{
    type=latin,
    name={\ensuremath{I_2}},
    description={Second principal invariant of a second-order tensor},
    unit={-}
}
\newglossaryentry{thirdInvariant}{
    type=latin,
    name={\ensuremath{I_3}},
    description={Third principal invariant of a second order tensor},
    unit={-}
}
\newglossaryentry{modifiedSecondInvariant}{
    type=latin,
    name={\ensuremath{J_2}},
    description={Modified second invariant of the rate of deformation tensor},
    unit={-}
}

%
%
\newglossaryentry{elasticModuliTensor}{
    type=latin,
    name={\ensuremath{\fourthTensor{C}}},
    description={Elastic moduli tensor},
    unit={\si{\pascal}}
}
\newglossaryentry{elasticModuliComp}{
    type=latin,
    name={\ensuremath{C}},
    description={Elastic moduli tensor component},
    unit={\si{\pascal}}
}
\newglossaryentry{YoungModulus}{
    type=latin,
    name={\ensuremath{E}},
    description={Young's modulus of a solid},
    unit={\si{\pascal}}
}
\newglossaryentry{shearModulus}{
    type=latin,
    name={\ensuremath{G}},
    description={Shear modulus of a solid},
    unit={\si{\pascal}}
}
\newglossaryentry{linearShearModulus}{
    type=latin,
    name={\ensuremath{\mu}},
    description={Shear modulus of the linearized theory of Solid Mechanics},
    unit={\si{\pascal}}
}
\newglossaryentry{PoissonRatio}{
    type=greek,
    name={\ensuremath{\nu^{\gls{solid}}}},
    description={Poisson's ratio of a solid},
    unit={-}
}
\newglossaryentry{strainEnergyFunction}{
    type=greek,
    name={\ensuremath{\Psi}},
    description={%
        Strain energy function of a hyperelastic law
    },
    unit={\si{\joule\per\cubic\meter}}
}
\newglossaryentry{deformationWorkFunction}{
    type=latin,
    name={\ensuremath{\Psi}},
    description={%
        Deformation work function or strain energy function -- function of
        \gls{rightCauchyGreenDeformation}. Energy from deformation work per
        unit volume
    },
    unit={\si{\joule\per\cubic\meter}}
}
\newglossaryentry{lameConst1}{
    type=greek,
    name={\ensuremath{\mu^{\gls{solid}}}},
    description={First Lamé's constant of a solid},
    unit={\si{\pascal}}
}
\newglossaryentry{lameConst2}{
    type=greek,
    name={\ensuremath{\lambda^{\gls{solid}}}},
    description={Second Lamé's constant of a solid},
    unit={\si{\pascal}}
}
\newglossaryentry{nominalStress}{
    type=latin,
    name={\ensuremath{\tensor{P}}},
    description={Nominal stress tensor},
    unit={\si{\pascal}}
}
\newglossaryentry{nominalStressComp}{
    type=hidden,
    name={\ensuremath{P}},
    description={Nominal stress tensor component},
    unit={\si{\pascal}}
}
\newglossaryentry{rateDeformationTensor}{
    type=latin,
    name={\ensuremath{\tensor{D}}},
    description={Rate of deformation tensor},
    unit={\si{\per\second   }}
}
\newglossaryentry{rateDeformationTensorComp}{
    type=hidden,
    name={\ensuremath{D}},
    description={Rate of deformation tensor component},
    unit={\si{\per\second}}
}
\newglossaryentry{deformationGradient}{
    type=latin,
    name={\ensuremath{\tensor{F}}},
    description={Deformation gradient tensor},
    unit={-}
}
\newglossaryentry{rotationTensor}{
    type=latin,
    name={\ensuremath{\tensor{R}}},
    description={Rotation tensor},
    unit={-}
}

\newglossaryentry{stretch}{
    type=greek,
    name={\ensuremath{\lambda}},
    description={%
        Stretch of a solid
    },
    unit={-}
}
\newglossaryentry{deformationGradComp}{
    type=hidden,
    name={\ensuremath{F}},
    description={Deformation gradient tensor component},
    unit={-}
}
\newglossaryentry{solidDisplacement}{
    type=latin,
    name={\ensuremath{\vec{u}}},
    description={Solid displacement vector},
    unit={m}
}
\newglossaryentry{solidDisplComp}{
    type=hidden,
    name={\ensuremath{u}},
    description={Solid displacement vector component},
    unit={-}
}
\newglossaryentry{xSolidDisplacement}{
    type=hidden,
    name={\ensuremath{u_{\gls{xAxis}}}},
    description={%
        Solid displacement vector component in the x direction of Cartesian
        coordinates
    },
    unit={-}
}
\newglossaryentry{ySolidDisplacement}{
    type=hidden,
    name={\ensuremath{u_{\gls{yAxis}}}},
    description={%
        Solid displacement vector component in the y direction of Cartesian
        coordinates
    },
    unit={-}
}
\newglossaryentry{zSolidDisplacement}{
    type=hidden,
    name={\ensuremath{u_{\gls{zAxis}}}},
    description={%
        Solid displacement vector component in the z direction of Cartesian
        coordinates
    },
    unit={-}
}
\newglossaryentry{2PKstress}{
    type=latin,
    name={\ensuremath{\tensor{S}}},
    description={Second Piola-Kirchhoff stress tensor},
    unit={\si{\pascal}}
}
\newglossaryentry{KirchhoffStress}{
    type=greek,
    name={\ensuremath{\tensor{\sigma}^{\mathrm{K}}}},
    description={Kirchhoff stress tensor},
    unit={\si{\pascal}}
}
\newglossaryentry{KirchhoffStressComp}{
    type=hidden,
    name={\ensuremath{\sigma^{\mathrm{K}}}},
    description={Kirchhoff stress tensor component},
    unit={\si{\pascal}}
}
\newglossaryentry{GLstrain}{
    type=latin,
    name={\ensuremath{\tensor{E}}},
    description={Green-Lagrange strain tensor},
    unit={-}
}
\newglossaryentry{leftCauchyGreenDeformation}{
    type=latin,
    name={\ensuremath{\tensor{B}}},
    description={Left Green-Cauchy deformation tensor},
    unit={-}
}
\newglossaryentry{rightCauchyGreenDeformation}{
    type=latin,
    name={\ensuremath{\tensor{C}}},
    description={Right Green-Cauchy deformation tensor},
    unit={-}
}
\newglossaryentry{2PKstressComp}{
    type=hidden,
    name={\ensuremath{S}},
    description={Second Piola-Kirchhoff stress tensor component},
    unit={\si{\pascal}}
}
\newglossaryentry{GLstrainComp}{
    type=hidden,
    name={\ensuremath{E}},
    description={Green-Lagrange strain tensor component},
    unit={-}
}
\newglossaryentry{leftCauchyGreenDeformationComp}{
    type=hidden,
    name={\ensuremath{B}},
    description={Left Green-Cauchy deformation tensor component},
    unit={-}
}
\newglossaryentry{rightCauchyGreenDeformationComp}{
    type=hidden,
    name={\ensuremath{C}},
    description={Right Green-Cauchy deformation tensor component},
    unit={-}
}
\newglossaryentry{linearStrain}{
    type=greek,
    name={\ensuremath{\tensor{\varepsilon}}},
    description={Linear or engineering strain tensor},
    unit={-}
}
\newglossaryentry{linearStrainComp}{
    type=hidden,
    name={\ensuremath{\varepsilon}},
    description={Linear or engineering strain tensor},
    unit={-}
}

\newglossaryentry{trueLinearStrain}{
    type=greek,
    name={\ensuremath{\tensor{e}}},
    description={True or logarithmic strain tensor},
    unit={-}
}


\newglossaryentry{tensorInvariant}{
    type=hidden,
    name={\ensuremath{I}},
    description={},
    unit={-}
}
\newglossaryentry{fourthInvariant}{
    type=latin,
    name={\ensuremath{I_4}},
    description={%
        Fourth invariant of the right Cauchy-Green deformation tensor with
        direction information
    },
    unit={-}
}
\newglossaryentry{fifthInvariant}{
    type=latin,
    name={\ensuremath{I_5}},
    description={%
        Third invariant of the right Cauchy-Green deformation tensor with
        directional information
    },
    unit={-}
}
\newglossaryentry{sixthInvariant}{
    type=latin,
    name={\ensuremath{I_6}},
    description={%
        Sixth invariant of the right Cauchy-Green deformation tensor with
        directional information
    },
    unit={-}
}
\newglossaryentry{seventhInvariant}{
    type=latin,
    name={\ensuremath{I_7}},
    description={%
        Seventh invariant of the right Cauchy-Green deformation tensor with
        directional information
    },
    unit={-}
}
\newglossaryentry{eighthInvariant}{
    type=latin,
    name={\ensuremath{I_8}},
    description={%
        Eighth invariant of the right Cauchy-Green deformation tensor with
        directional information
    },
    unit={-}
}

\newglossaryentry{MooneyCoeffs}{
    type=hidden,
    name={\ensuremath{c}},
    description={%
        Coefficients of Rivlin series expansion of the strain-energy function
    },
    unit={\si{\kilo\joule\per\cubic\meter}}
}
\newglossaryentry{OgdenCoeffs}{
    type=hidden,
    name={\ensuremath{\alpha}},
    description={Exponents of the Ogden strain-energy function},
    unit={-}
}
\newglossaryentry{FungArgumentFunction}{
    type=latin,
    name={\ensuremath{Q}},
    description={%
        Function of the strain tensor components used in the generic Fung-type
        models
    },
    unit={-}
}
\newglossaryentry{fiberDirection}{
    type=latin,
    name={\ensuremath{\vec{m}}},
    description={%
        Vector of the preferred direction of a fiber reinforced material
    },
    unit={\si{-}}
}
\newglossaryentry{orientationTensor}{
    type=latin,
    name={\ensuremath{\tensor{M}}},
    description={Orientation tensor},
    unit={\si{-}}
}
\newglossaryentry{orientationTensorComp}{
    type=hidden,
    name={\ensuremath{M}},
    description={Orientation tensor component},
    unit={\si{-}}
}
\newglossaryentry{fiberDirectionComp}{
    type=hidden,
    name={\ensuremath{m}},
    description={%
        Vector component of the preferred direction of a fiber reinforced
        material
    },
    unit={\si{-}}
}
\newglossaryentry{groundSubstance}{
    type=subscript,
    name={\ensuremath{g}},
    description={Ground substance}
}
\newglossaryentry{collagen}{
    type=subscript,
    name={\ensuremath{c}},
    description={Collagen fibers}
}
\newglossaryentry{wallLumenRatio}{
    type=latin,
    name={\ensuremath{\mathit{WLR}}},
    description={Wall-to-lumen ratio},
    unit={\si{-}}
}

%
%
\newglossaryentry{fsiInterface}{
    type=greek,
    name={\ensuremath{\gls{domainBoundary}^{\gls{fs}}}},
    description={Fluid-solid interface  = $\Gamma^f \cap \Gamma^s$},
    unit={-}
}
\newglossaryentry{fsiInterfaceFluid}{
    type=greek,
    name={\ensuremath{\gls{domainBoundary}_{fluid}^{\gls{fs}}}},
    description={%
        Fluid mesh boundary corresponding to the fluid-solid interface
    },
    unit={-}
}
\newglossaryentry{fsiInterfaceSolid}{
    type=greek,
    name={\ensuremath{\gls{domainBoundary}_{solid}^{\gls{fs}}}},
    description={%
        Solid mesh boundary corresponding to the fluid-solid interface
    },
    unit={-}
}
\newglossaryentry{fsiTolerance}{
    type=greek,
    name={\ensuremath{{\epsilon}_0}},
    description={Tolerance of the FSI coupling iterations},
    unit={-}
}
\newglossaryentry{fluidSolver}{
    type=latin,
    name={\ensuremath{\bm{\mathcal{F}}}},
    description={Symbolic representation of the fluid solver},
    unit={-}
}
\newglossaryentry{solidSolver}{
    type=latin,
    name={\ensuremath{\bm{\mathcal{S}}}},
    description={Symbolic representation of the solid solver},
    unit={-}
}
\newglossaryentry{addedMassOperator}{
    type=latin,
    name={\ensuremath{\bm{\mathcal{M}_{\mathrm{a}}}}},
    description={Symbolic representation of the added-mass operator},
    unit={-}
}
\newglossaryentry{addedMassEigenvalue}{
    type=latin,
    name={\ensuremath{\mu_\mathrm{a}}},
    description={Eigenvalue of the added-mass operator},
    unit={-}
}
\newglossaryentry{fsiTraction}{
    type=latin,
    name={\ensuremath{\cvec{y}}},
    description={%
        Solid solver input: traction exerted by the flow on the FSI interface
    },
    unit={\si{\pascal}}
}
\newglossaryentry{fsiInterfaceDispl}{
    type=latin,
    name={\ensuremath{\cvec{d}}},
    description={Position of the FSI interface},
    unit={\si{\meter}}
}
\newglossaryentry{fsDisplOutput}{
    type=latin,
    name={\ensuremath{\cvec{\tilde{d}}}},
    description={FSI interface position after a coupling iteration},
    unit={\si{\meter}}
}
\newglossaryentry{fsiResidualOperator}{
    type=latin,
    name={\ensuremath{\bm{\mathcal{R}}}},
    description={%
        Residual operator resulted from the fixed-point formulation of the FSI
        problem
    },
    unit={\si{\meter}}
}
\newglossaryentry{fsiInterfaceDoF}{
    type=latin,
    name={\ensuremath{N^{\gls{fs}}}},
    description={Number of degrees of freedom on the FSI interface},
    unit={-}
}
\newglossaryentry{fsiResidual}{
    type=latin,
    name={\ensuremath{\cvec{r}}},
    description={Discrete residual vector of the coupling iterations},
    unit={\si{\meter}}
}
\newglossaryentry{fsiReuse}{
    type=latin,
    name={\ensuremath{q}},
    description={Number of time-steps reused in the IQN-ILS algorithm},
    unit={-}
}

%
%
\newglossaryentry{aneurysm}{
    type=subscript,
    name={\ensuremath{ia}},
    description={%
        Variables or properties of an aneurysm surface
    }
}
\newglossaryentry{convexHull}{
    type=subscript,
    name={\ensuremath{ch}},
    description={%
        Indicates variables defined or integrated on the aneurysm surface
        convex hull
    }
}
\newglossaryentry{ostium}{
    type=subscript,
    name={\ensuremath{o}},
    description={%
        Indicates variables defined or integrated on the aneurysm ostium
        surface
    }
}

\newglossaryentry{domeHeight}{
    type=latin,
    name={\ensuremath{h_d}},
    description={%
        Maximum height of an aneurysm's dome perpendicular to ostium
        surface
    },
    unit={\si{\meter}}
}
\newglossaryentry{maximumDomeHeight}{
    type=latin,
    name={\ensuremath{h_{max}}},
    description={%
        Maximum height of an aneurysm's dome, measured from the ostium center
    },
    unit={\si{\meter}}
}
\newglossaryentry{bulgeHeight}{
    type=latin,
    name={\ensuremath{h_b}},
    description={%
        Bulge height of an aneurysm dome, i.e. the distance from the
        neck to the section with largest diameter
    },
    unit={\si{\meter}}
}
\newglossaryentry{neckDiameter}{
    type=latin,
    name={\ensuremath{d_n}},
    description={Diameter of an aneurysm's neck},
    unit={\si{\meter}}
}
\newglossaryentry{domeDiameter}{
    type=latin,
    name={\ensuremath{d_d}},
    description={Diameter of an aneurysm dome},
    unit={\si{\meter}}
}
\newglossaryentry{aspectRatio}{
    type=latin,
    name={\ensuremath{\mathit{AR}}},
    description={%
        Aspect ratio of an aneurysm, defined as the ratio between dome height
        and neck diameter
    },
    unit={-}
}
\newglossaryentry{meanBloodFlowRate}{
    type=latin,
    name={\ensuremath{\bar{q}_a}},
    description={Mean blood flow rate through an artery's section},
    unit={\si{\milli\litre\per\second}}
}
\newglossaryentry{timeBloodFlowRate}{
    name={\ensuremath{q_p}},
    description={%
        Blood flow rate profile through an artery section of a specific patient
    },
    unit={\si{\milli\litre\per\second}}
}
\newglossaryentry{normBloodFlowRate}{
    name={\ensuremath{q^{*}}},
    description={%
        Normalized blood flow rate profile through the ICA section measured
        experimentally for a population.
    },
    unit={-}
}
\newglossaryentry{arteryDiameter}{
    type=latin,
    name={\ensuremath{d_a}},
    description={Diameter of an aneurysm's parent artery section},
    unit={\si{\meter}}
}

\newglossaryentry{bottleneckFactor}{
    type=latin,
    name={\ensuremath{\mathit{BF}}},
    description={Bottleneck factor},
    unit={-}
}
\newglossaryentry{bulgeLocation}{
    type=latin,
    name={\ensuremath{\mathit{BL}}},
    description={Bulge location},
    unit={-}
}
\newglossaryentry{sizeRatio}{
    type=latin,
    name={\ensuremath{\mathit{SR}}},
    description={Size ratio},
    unit={-}
}
\newglossaryentry{convexityRatio}{
    type=latin,
    name={\ensuremath{\mathit{CR}}},
    description={Convexity ratio},
    unit={-}
}
\newglossaryentry{isoperimetricRatio}{
    type=latin,
    name={\ensuremath{\mathit{IPR}}},
    description={Isoperimetric ratio},
    unit={-}
}
\newglossaryentry{undulationIndex}{
    type=latin,
    name={\ensuremath{\mathit{UI}}},
    description={Undulation index},
    unit={-}
}
\newglossaryentry{ellipticityIndex}{
    type=latin,
    name={\ensuremath{\mathit{EI}}},
    description={Ellipticity index},
    unit={-}
}
\newglossaryentry{nonsphericityIndex}{
    type=latin,
    name={\ensuremath{\mathit{NSI}}},
    description={Nonsphericity index},
    unit={-}
}
\newglossaryentry{conicityParameter}{
    type=latin,
    name={\ensuremath{\mathit{CP}}},
    description={Conicity Parameter},
    unit={-}
}
\newglossaryentry{areaAvgGaussianCurvature}{
    type=latin,
    name={\ensuremath{\mathit{GAA}}},
    description={Area-averaged Gaussian curvature of a surface},
    unit={\si{\per\square\milli\meter}}
}
\newglossaryentry{areaAvgMeanCurvature}{
    type=latin,
    name={\ensuremath{\mathit{MAA}}},
    description={Area-averaged mean curvature of a surface},
    unit={\si{\per\milli\meter}}
}
\newglossaryentry{l2NormGaussianCurvature}{
    type=latin,
    name={\ensuremath{\mathit{GLN}}},
    description={L2-norm of the Gaussian curvature field of a surface},
    unit={-}
}
\newglossaryentry{l2NormMeanCurvature}{
    type=latin,
    name={\ensuremath{\mathit{MLN}}},
    description={L2-norm of the mean curvature of a surface},
    unit={-}
}

\newglossaryentry{wallShearStressVector}{
    type=greek,
    name={\ensuremath{ \vec{\tau}_w }},
    description={%
        Wall shear stress vector defined on a vessel and aneurysm surface
    },
    unit={\si{\pascal}}
}
\newglossaryentry{wallShearStressMag}{
    type=greek,
    name={\ensuremath{{\tau}_w}},
    description={%
        Wall shear stress magnitude defined on a vessel and aneurysm surface
    },
    unit={\si{\pascal}}
}
\newglossaryentry{timeAvgWallShearStress}{
    type=latin,
    name={\ensuremath{\mathit{TAWSS}}},
    description={%
        Time-average WSS field
    },
    unit={\si{\pascal\per\meter}}
}

\newglossaryentry{wallShearStressGradient}{
    type=latin,
    name={\ensuremath{\vec{G}}},
    description={%
        Wall shear stress gradient vector, defined on the surface
        $pq$-coordinate system
    },
    unit={\si{\pascal\per\meter}}
}
\newglossaryentry{timeAvgWallShearStressGradient}{
    type=latin,
    name={\ensuremath{\mathit{TAWSSG}}},
    description={%
        Surface gradient of the time-averaged wall shear stress gradient
        vector, defined on the surface $p$-coordinate direction (the flow
        direction)
    },
    unit={\si{\pascal\per\meter}}
}
\newglossaryentry{avgShearDirectionVector}{
    type=latin,
    name={\ensuremath{\vec{p}}},
    description={$p$-coordinate direction (the flow direction)},
    unit={-}
}
\newglossaryentry{transShearDirectionVector}{
    type=latin,
    name={\ensuremath{\vec{q}}},
    description={%
        $q$-coordinate direction -- transversal to average flow direction
    },
    unit={-}
}
\newglossaryentry{oscillatoryShearIndex}{
    type=latin,
    name={\ensuremath{\mathit{OSI}}},
    description={The oscillatory shear index},
    unit={-}
}
\newglossaryentry{relativeResidenceTime}{
    type=latin,
    name={\ensuremath{\mathit{RRT}}},
    description={The relative residence time indicator},
    unit={-}
}
\newglossaryentry{gradientOscillatoryNumber}{
    type=latin,
    name={\ensuremath{\mathit{GON}}},
    description={The gradient oscillatory number},
    unit={-}
}
\newglossaryentry{aneurysmFormationIndicator}{
    type=latin,
    name={\ensuremath{\mathit{AFI}}},
    description={The \textit{potential} aneurysm formation indicator},
    unit={-}
}
\newglossaryentry{cardiacPeriod}{
    type=latin,
    name={\ensuremath{T}},
    description={Cardiac period},
    unit={\si{\second}}
}
\newglossaryentry{lowShearArea}{
    type=latin,
    name={\ensuremath{\mathit{lsa}}},
    description={Normalized area of low WSS defined on the aneurysm surface},
    unit={-}
}
\newglossaryentry{avgParentArteryWSS}{
    type=latin,
    name={\ensuremath{\avg{\tau}_{w,pa}}},
    description={%
        Averaged parent artery WSS, defined in the anterior portion of the
        parent artery
    },
    unit={\si{\pascal}}
}

\newglossaryentry{lowWSSReference}{
    type=latin,
    name={\ensuremath{(\gls{wallShearStressMag})_{low}}},
    description={Low WSS reference},
    unit={-}
}

%
%
\newglossaryentry{speedup}{
    type=latin,
    name={\ensuremath{s}},
    description={Speedup},
    unit={-}
}
\newglossaryentry{residualVector}{
    type=latin,
    name={\ensuremath{\cvec{r}}},
    description={%
        Residual vector of a linear system $\matrix{A} \cvec{x} = \cvec{b}$
    },
    unit={-}
}
\newglossaryentry{residualNorm}{
    type=hidden,
    name={\ensuremath{R_s}},
    description={%
        Scaled normalized residual of a linear system in an iterative solver
    },
    unit={-}
}
\newglossaryentry{scaleFactor}{
    type=latin,
    name={\ensuremath{F_s}},
    description={Scale factor used for scaling the residual norm},
    unit={-}
}
\newglossaryentry{identityMatrix}{
    type=hidden,
    name={\ensuremath{\matrix{I}}},
    description={Identity matrix of $\realnumbers^{N \times N}$},
    unit={-}
}

%% file: acronyms.tex

\newacronym[category=standard,type=hidden]{1d}{1D}{one-dimensional}
\newacronym[category=standard,type=hidden]{2d}{2D}{two-dimensional}
\newacronym[category=standard,type=hidden]{3d}{3D}{three-dimensional}

\newacronym{cae}{CAE}{Computer-Aided Engineering}
\newacronym{fea}{FEA}{Finite Element Analysis}
\newacronym{cem}{CEM}{Computational Electromagnetic}
\newacronym{mbd}{MBD}{Multi-body Dynamics}

\newacronym{pde}{PDE}{partial differential equations}
\newacronym{fdm}{FDM}{Finite Difference Method}
\newacronym{fem}{FEM}{Finite Element Method}
\newacronym[type=hidden]{ale}{ALE}{arbitrary Lagrangian-Eulerian}
\newacronym{iqn-ils}{IQN-ILS}{Interface Quasi-Newton-Implicit Jacobian Least-Squares}
\newacronym{simple}{SIMPLE}{Semi-Implicit Method for Pressure Linked Equations}
\newacronym[category=only-short,type=hidden]{piso}{PISO}{Pressure Implicit with Splitting Operators}

\newacronym{bc}{BC}{boundary condition}
\newacronym{am}{AM}{added-mass}
\newacronym{dnf}{DNF}{Dirichlet-Neumann formulation}
\newacronym{dof}{DOF}{degrees of freedom}

\newacronym{ia}{IA}{intracranial aneurysm}
\newacronym{aaa}{AAA}{abdominal aortic aneurysm}
\newacronym[category=standard,type=hidden]{vmtk}{VMTK}{Vascular Modeling Toolkit}
\newacronym{dsa}{DSA}{digital subtraction angiography}
\newacronym{cta}{CTA}{computational tomography angiography}
\newacronym{wlr}{WLR}{wall-to-lumen ratio}
\newacronym{pae}{PAE}{peri-aneurysmal environment}
\newacronym{sah}{SAH}{subarachnoid hemorrhage}
\newacronym[category=standard,type=hidden]{vtk}{VTK}{Visualization Toolkit}

\newacronym{wss}{WSS}{wall shear stress}
\newacronym{pswss}{PSWSS}{peak systole wall shear stress}
\newacronym{tawss}{TAWSS}{time-averaged wall shear stress}
\newacronym{tawssg}{TAWSSG}{time-averaged WSS gradient}
\newacronym{wsstg}{WSSTG}{maximum wall shear stress temporal gradient}
\newacronym{wsssg}{WSSSG}{wall shear stress surface gradient}
\newacronym{wsspi}{WSSPI}{wall shear stress pulsatility index}
\newacronym{transwss}{transWSS}{transversal wall shear stress}
\newacronym{osi}{OSI}{oscillatory shear index}
\newacronym{rrt}{RRT}{relative residence time}
\newacronym{afi}{AFI}{aneurysm formation indicator}
\newacronym{gon}{GON}{gradient oscillatory number}
\newacronym{lsa}{LSA}{low shear area}

\newacronym{mca}{MCA}{middle cerebral artery}
\newacronym{ica}{ICA}{internal carotid artery}
\newacronym{aca}{ACA}{anterior cerebral artery}
\newacronym{acoa}{ACoA}{anterior communicating artery}
\newacronym{pica}{PICA}{posterior inferior cerebellar artery}
\newacronym{pca}{PCA}{posterior cerebral artery}
\newacronym{ba}{BA}{basilar artery}
\newacronym{va}{VA}{vertebral arteries}
\newacronym{oa}{OA}{ophthalmic artery}

\newacronym[type=hidden]{fapesp}{FAPESP}{São Paulo Research Foundation}
\newacronym[type=hidden]{unesp}{UNESP}{São Paulo State University}

%% file: local-glossary.tex

\newacronym{1wfsi}{1WFSI}{one-way fluid-solid interaction}
\newacronym[category=never-short,type=hidden]{2wfsi}{2WFSI}{two-way fluid-solid interaction}

\newacronym{mr}{MR}{Mooney-Rivlin}
\newacronym{ul}{UL}{updated Lagrangian}
\newacronym{tl}{TL}{total Lagrangian}
\newacronym[category=never-short,type=hidden]{svk}{SVK}{St.\ Venant-Kirchhoff}

\newacronym[category=never-short,type=hidden]{mri}{MRI}{Magnetic Resonance Imaging}
\newacronym[category=never-short,type=hidden]{uwm}{UWM}{uniform-wall model}
\newacronym[category=never-short,type=hidden]{awm}{AWM}{abnormal-wall model}
\newacronym[category=never-short,type=hidden]{voi}{VOI}{volume of interest}
\newacronym[category=never-short,type=hidden]{sd}{SD}{standard deviation}

\newacronym[category=standard,type=hidden]{cfd}{CFD}{Computational Fluid Dynamics}
\newacronym[category=standard,type=hidden]{csd}{CSD}{Computational Solid Dynamics}
\newacronym[category=standard,type=hidden]{fvm}{FVM}{Finite Volume Method}
\newacronym[category=standard,type=hidden]{fsi}{FSI}{Fluid-Solid Interaction}

\newacronym[type=hidden]{case50}{rMCA1}{}
\newacronym[type=hidden]{case23}{rMCA2}{}
\newacronym[type=hidden]{case37}{rICA1}{}
\newacronym[type=hidden]{case1}{rMCA3}{}
\newacronym[type=hidden]{case4ruptured}{rICA2}{}
\newacronym[type=hidden]{case2}{rICA3}{}

\newacronym[type=hidden]{case51}{urMCA1}{}
\newacronym[type=hidden]{case74}{urICA1}{}
\newacronym[type=hidden]{case4unruptured}{urMCA2}{}
\newacronym[type=hidden]{case13}{urICA2}{}
\newacronym[type=hidden]{case8}{urMCA3}{}
\newacronym[type=hidden]{case64}{urMCA4}{}
\newacronym[type=hidden]{case11}{urMCA5}{}

\newacronym[type=hidden]{m1}{\ensuremath{M_1}}{}
\newacronym[type=hidden]{m2}{\ensuremath{M_2}}{}
\newacronym[type=hidden]{m3}{\ensuremath{M_3}}{}

\newacronym[type=hidden]{dt1}{\ensuremath{\gls{deltat}_1}}{}
\newacronym[type=hidden]{dt2}{\ensuremath{\gls{deltat}_2}}{}
\newacronym[type=hidden]{dt3}{\ensuremath{\gls{deltat}_3}}{}

\newcommand{\case}[1]{\acrshort{#1}}

\makeatletter
\newcommand*{\Xbar}{}%
\DeclareRobustCommand*{\Xbar}{%
  \mathpalette\@Xbar{}%
}
\newcommand*{\@Xbar}[2]{%
  \sbox0{$#1\mathrm{X}\m@th$}%
  \sbox2{$#1X\m@th$}%
  \rlap{%
    \hbox to\wd2{%
      \hfill
      $\overline{%
        \vrule width 0pt height\ht0 %
        \kern\wd0 %
      }$%
    }%
  }%
  \copy2 %
}
\makeatother



\usepackage{xifthen}
\newcommand{\isoinvariant}[2]{%
    \ifthenelse{\isempty{#2}}%
        {\ensuremath{I^{*}_{#1}}}
        {\ensuremath{I^{*}_{#1,\gls{#2}}}}
}

\newcommand{\materialcoeff}[3]{%
    \ifthenelse{\isempty{#3}}%
        {\ensuremath{\gls{#1}_{#2}}}
        {\ensuremath{\gls{#1}_{#2#3}}}
}

\newglossaryentry{VonMisesStress}{
    type=latin,
    name={\ensuremath{\gls{CauchyStressComp}_{0}}},
    description={Von Mises or equivalent stress},
    unit={\si{\pascal}}
}
\newglossaryentry{maxPrincipalCauchyStress}{
    type=latin,
    name={\ensuremath{\gls{CauchyStressComp}_{1}}},
    description={Largest principal eigenvalue of the Cauchy stress tensor},
    unit={\si{\pascal}}
}
\newglossaryentry{maxStretch}{
    type=latin,
    name={\ensuremath{\gls{stretch}_{1}}},
    description={Largest eigenvalue of the stretch tensor},
    unit={\si{\pascal}}
}

\newglossaryentry{penaltyParameter}{
    type=greek,
    name={\ensuremath{\kappa}},
    description={Penalty parameter},
    unit={\si{\pascal}}
}

\newglossaryentry{pressureDiffusivity}{
    type=greek,
    name={\ensuremath{\gamma}},
    description={%
        Diffusivity of the Rhie-Chow correction in the pressure
        equation
    },
    unit={\si{\pascal}}
}

\newglossaryentry{modifiedCompressibility}{
    type=greek,
    name={\ensuremath{\gls{isothermalCompressibility}^{\gls{fluid}^{*}}}},
    description={%
        Modified isothermal compressibility of a weakly compressible fluid
    },
    unit={\si{\square\second\per\square\meter}}
}

\newglossaryentry{MurnaghanCoeff}{
    type=latin,
    name={\ensuremath{n^{\gls{fluid}}}},
    description={Material constant defined in the Murnaghan equation of state},
    unit={-}
}

\newglossaryentry{referenceDensity}{
    type=greek,
    name={\ensuremath{ \gls{density}^{\gls{fluid}}_{\mathrm{ref}}}},
    description={Reference density of the weakly compressible model},
    unit={\si{\kilo\gram\per\cubic\meter}}
}

\newglossaryentry{vasculature}{
    type=subscript,
    name={\ensuremath{v}},
    description={%
        Indicates variables defined or integrated on the vasculature surface
    }
}
\newglossaryentry{parentArtery}{
    type=subscript,
    name={\ensuremath{pa}},
    description={%
        Indicates variables defined or integrated on the parent artery surface
        of an aneurysm
    }
}
\newglossaryentry{branch}{
    type=subscript,
    name={\ensuremath{b}},
    description={%
        Variables or properties of the vascular branches
    }
}

\newglossaryentry{peakSystole}{
    type=subscript,
    name={\ensuremath{ps}},
    description={%
        Indicates variables defined at the peak systole instant
    }
}
\newglossaryentry{lowDiastole}{
    type=subscript,
    name={\ensuremath{ld}},
    description={%
        Indicates variables defined at the low diastole instant
    }
}

\newglossaryentry{uniformWallThickness}{
    type=latin,
    name={\ensuremath{{e_{uw}}}},
    description={Uniform thickness of the artery walls},
    unit={\si{\milli\meter}}
}

\newglossaryentry{wallThickness}{
    type=latin,
    name={\ensuremath{{e_w}}},
    description={%
        Thickness field of a vascular wall (i.e. aneurysms and branches)
    },
    unit={\si{\milli\meter}}
}
\newglossaryentry{branchesThickness}{
    type=latin,
    name={\ensuremath{{e_{\gls{branch}}}}},
    description={Thickness field of the branches walls},
    unit={\si{\milli\meter}}
}
\newglossaryentry{aneurysmThickness}{
    type=latin,
    name={\ensuremath{e_{\gls{aneurysm}}}},
    description={Thickness field of an aneurysm's wall},
    unit={\si{\milli\meter}}
}

\newglossaryentry{aneurysmThicknessFactor}{
    type=latin,
    name={\ensuremath{f_a}},
    description={%
        Scale factor used on the calculation of the aneurysm thickness
    },
    unit={\si{-}}
}
\newglossaryentry{abnormalHemodynamics}{
    type=subscript,
    name={\ensuremath{ab}},
    description={%
        Properties of an aneurysm wall due to abnormal hemodynamics
    },
}

\newglossaryentry{abnormalThicknessFactor}{
    type=latin,
    name={\ensuremath{f_{\gls{abnormalHemodynamics}}}},
    description={%
        Scale factor used on the calculation of the aneurysm thickness on
        abnormal hemodynamics patches
    },
    unit={\si{-}}
}

\newglossaryentry{aneurysmPulsatility}{
    type=greek,
    name={\ensuremath{\delta_{v}}},
    description={Pulsatility of an intracranial aneurysm},
    unit={-}
}

\newglossaryentry{distanceToNeckLine}{
    type=latin,
    name={\ensuremath{g_n}},
    description={%
        Geodesic distance of the vasculature to the aneurysm neck line
    },
    unit={\si{\milli\meter}}
}

\newglossaryentry{lumenDiameter}{
    type=latin,
    name={\ensuremath{d_l}},
    description={%
        Vasculature lumen diameter at a specified section
    },
    unit={\si{\milli\meter}}
}

\newglossaryentry{intracranialPressure}{
    type=latin,
    name={\ensuremath{\gls{pressure}_{o}}},
    description={Intracranial pressure},
    unit={\si{\pascal}}
}

\newglossaryentry{SpearmanCoeff}{
    type=greek,
    name={\ensuremath{\rho_{\mathrm{S}}}},
    description={Spearman's correlation coefficient},
    unit={-}
}
\newglossaryentry{PearsonCoeff}{
    type=greek,
    name={\ensuremath{\rho_{\mathrm{P}}}},
    description={Pearson's correlation coefficient},
    unit={-}
}

\newglossaryentry{HookeanCauchyStress}{
    type=greek,
    name={\ensuremath{{\gls{CauchyStress}}_{H}}},
    description={Cauchy stress tensor of the Hookean materal model},
    unit={\si{\pascal}}
}

\newglossaryentry{HookeanCauchyStressComp}{
    type=hidden,
    name={\ensuremath{{\gls{CauchyStressComp}}_{H}}},
    description={Cauchy stress tensor component of the Hookean material model},
    unit={\si{\pascal}}
}

\newglossaryentry{cylinderInnerRadius}{
    type=latin,
    name={\ensuremath{R_i}},
    description={Cylinder inner radius},
    unit={\si{\meter}}
}

\newglossaryentry{cylinderOuterRadius}{
    type=latin,
    name={\ensuremath{R_o}},
    description={Cylinder outer radius},
    unit={\si{\meter}}
}
\newglossaryentry{transmuralPressure}{
    type=latin,
    name={\ensuremath{\gls{pressure}_{tr}}},
    description={Transmural pressure},
    unit={\si{\pascal}}
}
\newglossaryentry{firstPrincipalCurvature}{
    type=greek,
    name={\ensuremath{\kappa_{1}}},
    description={First principal curvature of a surface},
    unit={\si{\per\meter}}
}
\newglossaryentry{secondPrincipalCurvature}{
    type=greek,
    name={\ensuremath{\kappa_{2}}},
    description={Second principal curvature of a surface},
    unit={\si{\per\meter}}
}
\newglossaryentry{meanCurvature}{
    type=latin,
    name={\ensuremath{H}},
    description={Mean curvature of a surface},
    unit={\si{\per\meter}}
}
\newglossaryentry{GaussianCurvature}{
    type=latin,
    name={\ensuremath{K}},
    description={Gaussian curvature of a surface},
    unit={\si{\per\square\meter}}
}

\newglossaryentry{membraneTension}{
    type=latin,
    name={\ensuremath{T}},
    description={Membrane tension},
    unit={\si{\pascal\meter}}
}
\newglossaryentry{volumetric}{
    type=subscript,
    name={\ensuremath{vol}},
    description={%
        Indicates the volumetric part of the strain-energy function
    }
}
\newglossaryentry{isochoric}{
    type=subscript,
    name={\ensuremath{iso}},
    description={%
        Indicates the isochoric or volume-preserving part of the strain-energy
        function
    }
}
\newglossaryentry{materialProjectionTensor}{
    type=latin,
    name={\ensuremath{\fourthTensor{P}}},
    description={%
        Projection tensor that maps the isochoric stress tensors to the
        physical stress tensor in Lagrangian coordinates
    },
    unit={-}
}
\newglossaryentry{YeohCoeffs}{
    type=hidden,
    name={\ensuremath{c}},
    description={%
        Material coefficients symbol of Yeoh's strain energy function
    },
    unit={\si{\pascal}}
}
\newglossaryentry{FungCoeffs}{
    type=hidden,
    name={\ensuremath{k}},
    description={%
        Material coefficients symbol of Fung's isotropic strain energy function
        ($c_1$ is given in \si{\pascal} whereas the others are dimensioneless)
    },
    unit={\si{\pascal}}
}
\newglossaryentry{axialWaveSpeed}{
    type=hidden,
    name={\ensuremath{U_n}},
    description={%
        Axial or longitudinal wave speed used in the advective boundary
        condition
    },
    unit={\si{\meter\per\second}}
}
\newglossaryentry{genericRelativeDifference}{
    type=hidden,
    name={\ensuremath{D}},
    description={%
        Relative difference between a wall mechanics metric between different
        modeling choices
    },
    unit={-}
}
\newglossaryentry{genericAbsoluteDifference}{
    type=hidden,
    name={\ensuremath{D}},
    description={%
        Absolute difference between a wall mechanics metric between different
        modeling choices
    },
    unit={-}
}
\newglossaryentry{meshCellsNumber}{
    type=hidden,
    name={\ensuremath{N_M}},
    description={%
        Number of cells in a computational mesh
    },
    unit={-}
}
\newglossaryentry{meshSurfaceCellDensity}{
    type=hidden,
    name={\ensuremath{\rho_{S}}},
    description={%
        Cells surface density in a computational mesh
    },
    unit={\si{cells\per\square\milli\meter}}
}
\newglossaryentry{maxEdgeSize}{
    type=hidden,
    name={\ensuremath{l_e}},
    description={%
        Maximum edge length size in a computational mesh
    },
    unit={\si{\milli\meter}}
}
\newglossaryentry{meshRefinementRatio}{
    type=latin,
    name={\ensuremath{r}},
    description={%
        Ration between element size of systematically refined meshes
    },
    unit={-}
}
\newglossaryentry{fvmOrder}{
    type=latin,
    name={\ensuremath{p}},
    description={%
        Formal order of accuracy of the Finite Volume Method
    },
    unit={-}
}
\newglossaryentry{discretizationError}{
    type=latin,
    name={\ensuremath{e}},
    description={%
        Discretization error of a particular mesh computed using Richardson's
        extrapolation
    },
    unit={-}
}
\newglossaryentry{pValue}{
    type=hidden,
    name={\ensuremath{p}},
    description={P-value of statistical tests},
    unit={-}
}
\newglossaryentry{significanceLevel}{
    type=hidden,
    name={\ensuremath{\alpha}},
    description={Significance level for statistical tests},
    unit={-}
}
\newglossaryentry{standardDeviation}{
    type=hidden,
    name={\ensuremath{\mathit{SD}}},
    description={Standard-deviation of a sample's distribution},
    unit={-}
}
\newglossaryentry{ruptureProbability}{
    type=latin,
    name={\ensuremath{P_{r}^{\textsc{ia}}}},
    description={Rupture probability of an intracranial aneurysm},
    unit={-}
}
\newglossaryentry{Yeoh}{
    type=superscript,
    name={\ensuremath{\mathrm{Y}}},
    description={%
        Indicates variables defined for the Yeoh hyperelastic law
    }
}
\newglossaryentry{Fung}{
    type=superscript,
    name={\ensuremath{\mathrm{F}}},
    description={%
        Indicates variables defined for the Fung hyperelastic law
    }
}
\newglossaryentry{MooneyRivlin}{
    type=superscript,
    name={\ensuremath{\mathrm{MR}}},
    description={%
        Indicates variables defined for the Mooney-Rivlin hyperelastic law
    }
}
\newglossaryentry{lumen}{
    type=superscript,
    name={\ensuremath{\mathrm{l}}},
    description={%
        Indicates variables defined on the lumen surface
    }
}
\newglossaryentry{ablumen}{
    type=superscript,
    name={\ensuremath{\mathrm{abl}}},
    description={%
        Indicates variables defined on the ablumen surface
    }
}
\newglossaryentry{mid}{
    type=superscript,
    name={\ensuremath{\mathrm{m}}},
    description={%
        Indicates variables defined on the mid surface of the vessel
    }
}
\newglossaryentry{iaWallMetric}{
    type=latin,
    name={\ensuremath{\mathcal{M}}},
    description={%
        Generic representation of a metric of the stress or stretch fields over
        the wall of an aneurysm
    }
}
\newglossaryentry{rupturedGroup}{
    type=latin,
    name={\ensuremath{\mathcal{R}}},
    description={%
        The group of ruptured aneurysms
    }
}
\newglossaryentry{unrupturedGroup}{
    type=latin,
    name={\ensuremath{\mathcal{U}}},
    description={%
        The group of unruptured aneurysms
    }
}
\newglossaryentry{typeIPatch}{
    type=subscript,
    name={\ensuremath{I}},
    description={%
        Indicates variables defined on the type-I patches of the aneurysm sac
    }
}
\newglossaryentry{typeIIPatch}{
    type=subscript,
    name={\ensuremath{II}},
    description={%
        Indicates variables defined on the type-II patches of the aneurysm sac
    }
}
\newglossaryentry{normalPatch}{
    type=subscript,
    name={\ensuremath{R}},
    description={%
        Indicates variables defined on the normal patches of the aneurysm sac
        (i.e., not type I or type II)
    }
}
\newglossaryentry{neckPatch}{
    type=subscript,
    name={\ensuremath{N}},
    description={%
        Indicates the neck patch of an \gls{ia} sac, defined by the
        physician-oriented patching
    }
}
\newglossaryentry{bodyPatch}{
    type=subscript,
    name={\ensuremath{B}},
    description={%
        Indicates the body patch of an \gls{ia} sac, defined by the
        physician-oriented patching
    }
}
\newglossaryentry{domePatch}{
    type=subscript,
    name={\ensuremath{D}},
    description={%
        Indicates the dome patch of an \gls{ia} sac, defined by the
        physician-oriented patching
    }
}

%% file: tex/elsevier-header.tex
\begin{frontmatter}

    \title{%
         A numerical investigation of the mechanics of intracranial aneurysms
         walls: Assessing the influence of tissue hyperelastic laws and
         heterogeneous properties on the stress and stretch fields
    }

    \author[feis-long]{I. L. Oliveira\corref{cor}}
    \ead{iago.oliveira@unesp.br}

    \author[ucd]{P. Cardiff}
    \ead{philip.cardiff@ucd.ie}

    \author[harvard]{C.E. Baccin}
    \ead{cebaccin@gmail.com}

    \author[feis-short]{J.L. Gasche}
    \ead{jose.gasche@unesp.br}

    \cortext[cor]{Corresponding author}

    \address[feis-short]{%
        São Paulo State University (UNESP), School of Engineering,
        Mechanical Engineering Department
    }

    \address[ucd]{
        University College Dublin (UCD), School of Mechanical and
        Materials Engineering, Dublin, Ireland
    }

    \address[harvard]{%
        Interventional Neuroradiology/Endovascular Neurosurgery,
        Beth Israel Deaconess Medical Center,
        Harvard Medical School, Boston, MA, US
    }

    \address[feis-long]{%
        São Paulo State University (UNESP), School of Engineering,
        Mechanical Engineering Department, Thermal Sciences Building,
        Avenida Brasil, 56, Ilha Solteira - SP, Brazil
    }

    \begin{abstract}
        Numerical simulations have been extensively used in the past two
        decades for the study of \glspl{ia}, a dangerous disease that occurs in
        the arteries that reach the brain. They may affect up to
        \SI{10}{\percent} of the world's population, with up to
        \SI{50}{\percent} mortality rate, in case of rupture. Physically, the
        blood flow inside \glspl{ia} should be modeled as a fluid-solid
        interaction problem.  However, the large majority of those works have
        focused on the hemodynamics of the intra-aneurysmal flow, while
        ignoring the wall tissue's mechanical response entirely, through
        rigid-wall modeling, or using limited modeling assumptions for the
        tissue mechanics. One of the explanations is the scarce data on the
        properties of \glspl{ia} walls, thus limiting the use of better
        modeling options. Unfortunately, this situation is still the case, thus
        our present study investigates the effect of different modeling
        approaches to simulate the motion of an \gls{ia}. We used three
        hyperelastic laws --- the Yeoh law, the three-parameter \glsxtrlong{mr}
        law, and a Fung-like law with a single parameter --- and two different
        ways of modeling the wall thickness and tissue mechanical properties
        --- one assumed that both were uniform while the other accounted for
        the heterogeneity of the wall by using a \enquote{hemodynamics-driven}
        approach in which both thickness and material constants varied
        spatially with the cardiac-cycle-averaged hemodynamics. Pulsatile
        numerical simulations, with patient-specific vascular geometries
        harboring \glspl{ia}, were carried out using the \glsxtrlong{1wfsi}
        solution strategy implemented in solids4foam, an extension of
        OpenFOAM\R, in which the blood flow is solved and applied as the
        driving force of the wall motion.  We found that different wall
        morphology models yield smaller absolute differences in the mechanical
        response than different hyperelastic laws. Furthermore, the stretch
        levels of \glspl{ia} walls were more sensitive to the hyperelastic and
        material constants than the stress.  These findings could be used to
        guide modeling decisions on \gls{ia} simulations, since the
        computational behavior of each law was different, for example, with the
        Yeoh law yielding the smallest computational time.  \end{abstract}

    \begin{keyword}
        intracranial aneurysms \sep%
        hyperelasticity \sep%
        wall morphology \sep%
        mechanical response \sep%
        numerical simulations
    \end{keyword}
\end{frontmatter}

%% file: tex/introduction.tex
\section{Introduction} \label{sec:introduction}

    \Glspl{ia} are pathological dilatations of the human vascular system
    normally found in the bifurcations of the cerebral arteries tree. The most
    common form has a saccular shape, with a prevalence of up to
    \SI{90}{\percent} in the brain arteries \citep{Diagbouga2018}, being a
    dangerous disease that may affect up to \SI{10}{\percent} of the world's
    population \citep{isuia1998} and with up to \SI{50}{\percent} mortality
    rate, in case of rupture \citep{Vlak2013,Saqr2019}. This pathology has,
    in the past three decades, been investigated experimentally
    \citep{Meng2007a,Metaxa2010}, which, for example, led to the understanding
    of the importance of hemodynamics on its development, but also numerically
    through \gls{cfd}, which provided detailed information on the hemodynamics
    \citep{Liang2019} --- although still being a debatable topic in the
    clinical practice \citep{Kallmes2012,Cebral2012}.

    Due to the nature of the pathology, better modeling is continuously sought
    to allow for more reliable numerical simulations, for example, through the
    use of \gls{fsi} modeling \citep{Bazilevs2010b,Lee2013a}, although
    numerical techniques to solve \gls{fsi} problems pose challenging numerical
    difficulties \citep{Causin2005,Forster2007}. Additionally, using an
    \gls{fsi} strategy employed for patient-specific \gls{ia} geometries
    requires the modeling of their wall tissue that also poses difficulties
    hard to overcome, such as the lack of patient-specific data of the wall
    thickness, the constitutive behavior of the tissue and its material
    properties. This is particularly important due to the large variability of
    the disease.

    Previous experimental works showed that \glspl{ia} walls are more likely
    to have wall thickness and mechanical and failure properties varying
    spatially \citep{Kadasi2013,Signorelli2018}. This local morphology is
    caused by the natural history of a particular \gls{ia}
    \citep{Frosen2019,Soldozy2019}. \citet{Meng2014}, for example,
    hypothesized two biological pathways, dependent on different local
    hemodynamic conditions, that would lead to different wall phenotypes ---
    the authors name these two phenotypes as \enquote{type-I}, comprising small
    \glspl{ia} with thin and translucent walls, and \enquote{type-II},
    encompassing large \glspl{ia} with thick, white or yellow, atherosclerotic
    walls. Moreover, a spectrum of morphologies would exist between these
    broad phenotypes, as investigated by \citet{Kadasi2013}, for example,
    who found that \SI{27}{\percent} of \glspl{ia} are type-I,
    \SI{8}{\percent} are type-II, and \SI{65}{\percent} contain both
    patch types.

    How a patient-specific \gls{ia} grows also influences its mechanical
    behavior, classically considered to be well described by hyperelastic laws
    \citep{Humphrey2000} --- even though particular laws to suitably represent
    it do not exist \citep{Parshin2019}. In the last decade, a few works
    mechanically characterized samples of \gls{ia} tissue using uniaxial tests
    to failure, obtaining the values of the material constants that appear in
    hyperelastic laws classically associated with artery tissue behavior
    \citep{Holzapfel2010}. Typical examples are the \gls{mr} law
    \citep{Costalat2011}, the Yeoh law \cite{Brunel2018}, and an isotropic
    exponential Fung-like quadratic law \citep{Robertson2015}.  Apart from the
    mechanical constants, other properties of \glspl{ia} tissue have also been
    reported by \citet{Costalat2011}, for example, who found that the tissue of
    unruptured \glspl{ia} is stiffer than ruptured \glspl{ia} tissue. Finally,
    in possession of \gls{ia} tissue samples, these works have also measured
    their average thickness, further confirming that an \gls{ia} is globally
    thinner than its surrounding arteries.

    Although it is still a challenge to measure the local morphology, i.e. the
    local wall thickness and tissue material properties, for a large number of
    patient-specific \glspl{ia}, some works on the subject exist.
    \citet{Signorelli2018}, for example, used an \enquote{indentation device}
    to measure, in a point-wise manner, the elasticity modulus of an \gls{ia}
    sac sample with a resolution of \SI{1}{\square\milli\meter}. Their findings
    suggest that the rupture site is less stiff, i.e. with a smaller elasticity
    modulus than the rest of the sac, where they found that stiff regions were
    mixed with thinner regions. The technique has the same drawbacks as
    classical uniaxial tests, though, because it still requires the aneurysm
    tissue to be collected, hence \textit{in vivo} measurements are unfeasible.
    In this regard, imaging techniques are thought to be a promising
    alternative to measuring the local thickness of a patient-specific \gls{ia}
    sac, as performed by \citet{Kleinloog2014} through an experimental study in
    which the wall thickness of \glspl{ia} was measured using a 7T \gls{mri}.

    \citet{Cebral2019} used \gls{cfd} to investigate an \gls{ia} sac's local
    morphological heterogeneity. The authors subdivided the wall of a sample of
    \glspl{ia} into five regions with specific phenotypes: atherosclerotic,
    hyperplastic, thin, the rupture site, and \enquote{normal-appearing} by
    intraoperative observation and correlated each of them with local
    hemodynamics. They found a similar relationship between the local
    hemodynamics conditions investigated by \citet{Meng2014}, so-called
    \enquote{low-flow} and \enquote{high-flow} effects, and the wall
    phenotypes.  Their study is a good example of how numerical simulations
    could be used to predict the \gls{ia} sac heterogeneity on a
    patient-by-patient basis.

    Therefore, accounting for all these modeling requirements makes the
    modeling of a patient-specific \gls{ia} wall a challenge due to both the
    scarce experimental data to feed numerical models and the large variability
    of the disease, which prevents patient-specific computations. This is
    reflected in the modeling choices used by the few numerical works that
    investigated the mechanical response of \glspl{ia}. For example, in terms
    of the approach to estimating the wall thickness, we found a majority that
    employed uniform thickness throughout the \gls{ia} sac and branches
    \citep{Torii2007,Torii2008,Lee2013b,Valencia2009},  and a small amount that
    employed a uniform thickness for the aneurysm sac and a different one on
    the branches \citep{Sanchez2013}, or a lumen-diameter thickness
    \citep{Bazilevs2010b}. Finally, only a single work obtained the
    patient-specific thickness distribution of the \gls{ia} sac
    \citep{Voss2016} and compared it with a whole uniform wall model,
    nonetheless the authors used micro-computed tomography to scan the aneurysm
    sac, a technique that is difficult to apply in a larger cohort of
    \glspl{ia}.

    Regarding the selection of constitutive law, works that numerically solved
    the \gls{fsi} problem with patient-specific \gls{ia} subjects have used
    several different ones. Surprisingly, we found a majority that has chosen
    the small-strain Hookean law that, rigorously, should not be used in
    finite-deformation motions \citep{Torii2006,Valencia2009,Lee2013a,Cho2020}.
    Other works employed the classic neo-Hookean law \citep{Bazilevs2010a} or
    more specialized ones, such as exponential laws \citep{Torii2008} and the
    \gls{mr} law \citep{Sanchez2014}, although in a smaller number, and not
    always using the material properties of patient-specific \gls{ia} tissue.

    Despite the uncertainty about which law to choose, the assessment of the
    impact of different material laws on the mechanics of \glspl{ia} walls has
    been the subject of even fewer studies. \citet{Torii2008} performed
    \gls{fsi} simulations for one \gls{ia} case, by assuming, first, the rigid
    wall assumption --- thus \enquote{pure} \gls{cfd} simulations ---, and
    three elastic laws: the Hookean law (thus, assuming small strains), the
    \gls{svk} law, and another hyperelastic law using the exponential
    strain-energy function proposed by \citet{Demiray1972}. Their findings
    showed that the displacement profiles were qualitatively similar among all
    the elastic laws, although the maximum displacement with the exponential
    hyperelastic law was \SI{36}{\percent} smaller than that for the \gls{svk}
    law.  Unfortunately, their results on the wall mechanics were limited to
    the displacement field, no stresses or strain were analyzed because their
    focus was on the hemodynamics.

    An earlier trial to assess different constitutive laws in \glspl{ia}'
    mechanical response was conducted by \citet{Ramachandran2012}, with
    patient-specific \glspl{ia} geometries. The authors assumed them to be
    statically determined, i.e. their mechanical response was independent of
    the material properties, and investigated the impact of different
    constitutive laws on the wall stresses and strains by numerically
    simulating only the aneurysm sac with \gls{csd} by using a numerical
    modeling similar to inflation experiments. They used both anisotropic and
    isotropic versions of Fung-like laws, the Yeoh law with three parameters,
    the \gls{svk} law, and Hooke's law too. Their results suggested that the
    aneurysm sac may indeed be statically determined regarding different
    material laws.  However, they only studied the aneurysm sac, i.e. they
    removed the surrounding arteries portions that may have had an impact on
    the aneurysms sac stresses, and their pressure-inflation model employed
    static \glspl{bc}, which limits their conclusions. Indeed, the authors
    highlighted that these conclusions may not stand when the full vasculature
    would be simulated with dynamical \glspl{bc} that realistically reflect the
    cardiac cycle forces.

    In this current scenario, it is clear that it remains broadly unknown what
    is the average impact of the use of different hyperelastic laws and wall
    morphology models in the mechanics of \glspl{ia}, i.e. in the stress and
    strain fields of the \gls{ia} sac. Therefore, the aim of this work was to
    assess what could be the impact of choosing different material laws and
    different morphology models to numerically obtain the mechanical response
    of \glspl{ia}. More specifically, we investigated whether a wall model with
    uniform thickness and material constants, for example, would be acceptable
    to be used, given the dominant heterogeneity existing for this disease.
    This is essential in investigations of \gls{ia} rupture while promising
    tools that could extract the heterogeneity of the wall more accurately are
    not ready to do that for a large cohort of patient-specific \glspl{ia} and,
    also, because ultimately the rupture event depends on the stress and strain
    levels on the wall.

%% file: tex/methodology.tex
\section{Numerical Methodology} \label{sec:numericalMethodology}

    \subsection{Sample Selection and Geometry Preparation}
    \label{sec:sampleSelection}

    We selected twelve vascular geometries from \gls{dsa} examinations
    collected retrospectively. Nine were collected in the Albert Einstein
    Israelite Hospital, São Paulo, and approved to be used by the institution's
    Research Ethics Committee as also by the Research Ethics Committee of
    the Faculty of Medicine of \gls{unesp}, Campus of Botucatu. The
    additional three vascular geometries were obtained from the Aneurisk
    dataset repository \citep{aneurisk}, which provides a set of \glspl{ia}
    geometries used during the Aneurisk project and are available under the
    \enquote{CC BY-NC 3.0} license. We used these additional geometries due to
    the lack of sufficient ruptured cases in the original dataset to build a
    representative sample.

    The twelve vasculatures harbored thirteen bifurcation \glspl{ia}, all of
    them originating from the more common bifurcation spots of \gls{ia}
    occurrences in the brain vessels (the \gls{ica} and \gls{mca}). Seven were
    unruptured and six ruptured with maximum dome diameter ranging from,
    roughly, \SI{3}{\milli\meter} to \SI{7}{\milli\meter}, thus categorizing
    them as small- or medium-sized \glspl{ia} (mean $\pm$ \gls{sd} equals
    \SI{4.93 \pm 1.59}{\milli\meter} for the ruptured group and \SI{6.28 \pm
    1.30}{\milli\meter} for the unruptured group). For reference, they were
    labeled by appending their rupture status, prefix \enquote{r} for ruptured
    and \enquote{ur} for unruptured, to their parent artery. For example, an
    unruptured case in the \gls{ica} bifurcation is labeled \enquote{urICA},
    followed by a natural number in case of repetition.

    The \gls{dsa} images were segmented using the \gls{vmtk}\R library
    \citep{vmtk} with the level-set segmentation method \citep{Piccinelli2009}.
    The selection of the \gls{voi} was not defined \emph{a priori} but chosen
    large enough to ensure that it enclosed only the aneurysm and its closest
    surrounding vessels. Subsequently, a triangulated surface was generated
    with the Marching Cubes algorithm \citep{Antiga2002,Antiga2008}. Inlet and
    outlet profiles were artificially created in the surfaces to impose
    \glspl{bc} for the numerical simulations.

    \subsection{%
        Physical and Mathematical Modeling and Boundary Conditions
    }
    \label{sec:mechanicalModeling}

    In the flow domain, blood was assumed to be a weakly compressible Newtonian
    fluid flowing in an isothermal laminar regime, hence the continuity,
    momentum, and an equation of state were solved for the fields of flow
    velocity, \gsuper{velocity}{fluid}, pressure \gsuper{pressure}{fluid}, and
    blood density \gsuper{density}{fluid}. For a moving control volume in the
    fluid domain, \gsuper{volume}{fluid}\functionOf{\gls{time}}, and using the
    \gls{ale} framework, where the reference coordinate system is given by
    \gls{meshCoord}, the continuity equation is written as:
        \begin{equation} \label{eq:continuityEquationIntegral}
            \ddt{}
            {
                \bigg(
                    \int\limits_{
                        \gsuper{volume}{fluid}
                        \functionOf{\gls{time}}
                    }{
                        \gsuper{density}{fluid}
                    }\diff{\gls{volume}}
                \bigg)
            }\biggr\rvert_{\gls{meshCoord}}
            +
            \oint\limits_{
                \gsuper{surface}{fluid}
                \functionOf{\gls{time}}
            }{
                \gsuper{density}{fluid}
                \left(
                    \gsuper{velocity}{fluid}
                    -
                    \gls{meshVelocity}
                \right)
                \dprod
                \gsuper{sNormalVector}{fluid}
            }\diff{\gls{surface}}
            =
            0\,,
        \end{equation}
    \noindent where \gls{meshVelocity} is the velocity of the referential
    system of the \gls{ale} formulation, and \gsuper{sNormalVector}{fluid} is
    the outward normal vector to the control surface \gsuper{surface}{fluid};
    additionally, the momentum equation in integral form is given by:
        \begin{equation} \label{eq:momentumEquationIntegral}
            \ddt{}
            {
                \bigg(
                    \int\limits_{
                        \gsuper{volume}{fluid}
                        \functionOf{\gls{time}}
                    }{
                        \gsuper{density}{fluid}
                        \gsuper{velocity}{fluid}
                    }\diff{\gls{volume}}
                \bigg)
            }\biggr\rvert_{\gls{meshCoord}}
            +
            \oint\limits_{
                \gsuper{surface}{fluid}
                \functionOf{\gls{time}}
            }{
                \gsuper{density}{fluid}
                \gsuper{velocity}{fluid}
                \left(
                    \gsuper{velocity}{fluid}
                    -
                    \gls{meshVelocity}
                \right)
                \dprod
                \gsuper{sNormalVector}{fluid}
            }\diff{\gls{surface}}
            =
            \oint\limits_{
                \gsuper{surface}{fluid}
                \functionOf{\gls{time}}
            }{
                \gsuper{CauchyStress}{fluid}
                \dprod
                \gsuper{sNormalVector}{fluid}
            }\diff{\gls{surface}}\,,
        \end{equation}
    where \gsuper{CauchyStress}{fluid} is the Cauchy stress tensor, given by:
        \begin{equation} \label{eq:NewtonianFluidContitutiveEq}
            \gsuper{CauchyStress}{fluid}
            =
            -
            \gsuper{pressure}{fluid}
            \gls{identityTensor}
            +
            \gls{dynamicViscosity}
            \left[
                \grad{\gsuper{velocity}{fluid}}
                +
                \trans{(\grad{\gsuper{velocity}{fluid}})}
            \right]
            -
            \frac{2}{3}
            \gls{dynamicViscosity}
            \left(
                \div{
                    \gsuper{velocity}{fluid}
                }
            \right)
            \gls{identityTensor}\,,
        \end{equation}
    and \gls{identityTensor} is the second-order identity tensor. Blood
    dynamic viscosity was assumed to be $\gls{dynamicViscosity} =
    \SI{3.3e-3}{\pascal\second}$.

    Finally, blood compressibility was assumed to be governed by the
    barotropic equation of state, which has already been used to model the
    behavior of blood in \gls{fsi} problems
    \citep{Kanyanta2009,Kanyanta2009a,Tandis2019}. It can be applied for
    liquids at \enquote{low pressures} when the change in density is linearly
    related to the pressure change according to the definition of the bulk
    modulus of the fluid, \gsuper{bulkModulus}{fluid}:
        \begin{equation} \label{eq:bulkModulusFluidDefinition}
            \gsuper{bulkModulus}{fluid}
            =
            \gsuper{density}{fluid}
            \ddx{
                \gsuper{pressure}{fluid}
            }{
                \gsuper{density}{fluid}
            }\,.
        \end{equation}
    In this case, the pressure and the fluid density are related by:
        \begin{equation} \label{eq:eqStateWeaklyCompressibleFluid}
            \gsuper{density}{fluid}
            =
            \gsupsub{density}{fluid}{zeroTime}
            +
            \frac{
                \gsupsub{density}{fluid}{zeroTime}
            }{
                \gsuper{bulkModulus}{fluid}
            }
            \left(
                \gsuper{pressure}{fluid}
                -
                \gsupsub{pressure}{fluid}{zeroTime}
            \right)\,,
        \end{equation}
    \noindent where the subscript \enquote{\gls{zeroTime}} indicates a
    reference state of the fluid, assumed to be blood at an average cardiac
    cycle pressure, \SI{100}{\mmHg}, with $\gsupsub{density}{fluid}{zeroTime} =
    \SI{1000.0}{\kilo\gram\per\cubic\meter}$. Blood bulk modulus was assumed
    to be \SI{2.2e9}{\pascal} \citep{Kanyanta2009}.

    The solid domain, i.e. the \gls{ia} and artery walls, was assumed as
    isotropic and represented by a hyperelastic constitutive law. In the
    finite-deformation regime, the momentum equation was solved in the total
    Lagrangian formulation and integral form, written as:
        \begin{equation} \label{eq:momentumEquationTotalLag}
            \int\limits_{\gsupsub{volume}{solid}{zeroTime}}{
                \gsupsub{density}{solid}{zeroTime}
                \dsdts{\gls{solidDisplacement}}
                \functionOf{
                    \gls{LagrangeCoord},
                    \gls{time}
                }
            }\diff{\gsub{volume}{zeroTime}}
            =
            \int\limits_{\gsupsub{volume}{solid}{zeroTime}}{
                \mdiv{
                    \left[
                        \gls{Jacobian}
                        \inv{\gls{deformationGradient}}
                        \dprod
                        \left(
                            \gsuper{CauchyStress}{solid}
                            +
                            \gsupsub{CauchyStress}{solid}{zeroTime}
                        \right)
                    \right]
                }
            }\diff{\gsub{volume}{zeroTime}}\,,
        \end{equation}
    where \gls{solidDisplacement} is the motion displacement,
    \gsupsub{density}{solid}{zeroTime} is the tissue density at the reference
    configuration, and $\gls{deformationGradient} = \gls{identityTensor} +
    \trans{(\mgrad{\gls{solidDisplacement}})}$ is the deformation gradient of
    the motion, with $\gls{Jacobian} = \det{(\gls{deformationGradient})}$. The
    subscript \enquote{\gls{zeroTime}} indicates any property or derivative
    that was evaluated with respect to the coordinates system of the reference
    configuration, indicated by \gls{LagrangeCoord}. The reference
    configuration was assumed to be the domain configuration at the time zero
    and to be pre-stressed with a Cauchy prestress field calculated by
    $\gsupsub{CauchyStress}{solid}{zeroTime} = \inv{\gls{Jacobian}}
    \gls{deformationGradient} \dprod \gsub{2PKstress}{zeroTime} \dprod
    \gls{deformationGradient}$. The second Piola-Kirchhoff prestress tensor,
    \gsub{2PKstress}{zeroTime}, was computed using the same approach employed
    by \citet{Bazilevs2010b}.

    Finally, the Cauchy stress of the solid, \gsuper{CauchyStress}{solid}, was
    calculated using three different hyperelastic laws that are defined based
    on their strain-energy function, \gls{strainEnergyFunction}, as follows:
    \begin{itemize}
        \item the \gls{mr} law with 3 material constants
            \citep{Mooney1940}:
            \begin{equation} \label{eq:MooneyRivlin3CoeffsIsochoric}
                \gls{strainEnergyFunction}
                \functionOf{
                    \invariant{1}{},
                    \invariant{2}{}
                }
                =
                \materialcoeff{MooneyCoeffs}{1}{0}
                \left(
                    \invariant{1}{} - 3
                \right)
                +
                \materialcoeff{MooneyCoeffs}{0}{1}
                \left(
                    \invariant{2}{} - 3
                \right)
                +
                \materialcoeff{MooneyCoeffs}{1}{1}
                \left(
                    \invariant{1}{} - 3
                \right)
                \left(
                    \invariant{2}{} - 3
                \right)\,;
            \end{equation}

        \item the Yeoh law with 3 material constants
            \begin{equation} \label{eq:YeohModelIsochoric}
                \gls{strainEnergyFunction}
                \functionOf{
                    \invariant{1}{}
                }
                =
                \materialcoeff{YeohCoeffs}{1}{0}
                \left(
                    \invariant{1}{} - 3
                \right)
                +
                \materialcoeff{YeohCoeffs}{2}{0}
                \left(
                    \invariant{1}{} - 3
                \right)^2
                +
                \materialcoeff{YeohCoeffs}{3}{0}
                \left(
                    \invariant{1}{} - 3
                \right)^3\,;
            \end{equation}

        \item and an \enquote{isotropic} version of the exponential Fung-like
            law, originally proposed by \citet{Demiray1972}:
            \begin{equation} \label{eq:isotropicFungModelIsochoric}
                \gls{strainEnergyFunction}
                \functionOf{
                    \invariant{1}{}
                }
                =
                \frac{
                    \materialcoeff{FungCoeffs}{1}{}
                }{
                    \materialcoeff{FungCoeffs}{2}{}
                }
                \left[
                    \exp{
                        \frac{\materialcoeff{FungCoeffs}{2}{}}{2}
                        \left(
                            \invariant{1}{}
                            -
                            3
                        \right)
                    }
                    - 1
                \right]\,,
            \end{equation}
    \end{itemize}
    where \invariant{1}{} and \invariant{2}{} are the first and second
    invariants of the right Cauchy-Green deformation tensor. The complete
    constitutive model employed a volumetric-decomposition
    approach as explained in \citet{Holzapfel2000}.  In this framework, the
    volumetric part of the tissue motion, dependent on the tissue
    compressibility, was ultimately measured by the Poisson's ratio of the
    tissue, \gls{PoissonRatio} through its bulk modulus,
    \gsuper{bulkModulus}{solid}, as computed by the linearized theory:
    \begin{equation} \label{eq:solidBulkModulus}
        \gsuper{bulkModulus}{solid}
        =
        \frac{
            \gls{YoungModulus}
            }{
            3(1 - 2\gls{PoissonRatio})
        }\,,
    \end{equation}
    For the simulations in this work, the Poisson ratio was assumed to be
    \num{0.48} --- we performed a parametric study of its influence on the
    stresses and stretch on an \gls{ia} sac surface and found that by
    increasing it above \num{0.48}, both average fields over the sac changed by
    less than \SI{2}{\percent}, on average, for all the hyperelastic laws,
    while still yielding reasonable computational times. The linearized Young's
    modulus for the hyperelastic laws was also calculated based on the uniaxial
    deformation, resulting in %
    $
        \gls{YoungModulus}
        =
        6
        \left(
            \ddx{\gls{strainEnergyFunction}}{\invariant{1}{}}
            +
            \ddx{\gls{strainEnergyFunction}}{\invariant{2}{}}
        \right)
    $ \citep{Holzapfel2000a}.

    \subsubsection*{\Glsfmtfullpl{bc}}

\begin{figure}[!htb]
        \includegraphics[%
            width=0.6\textwidth
        ]   {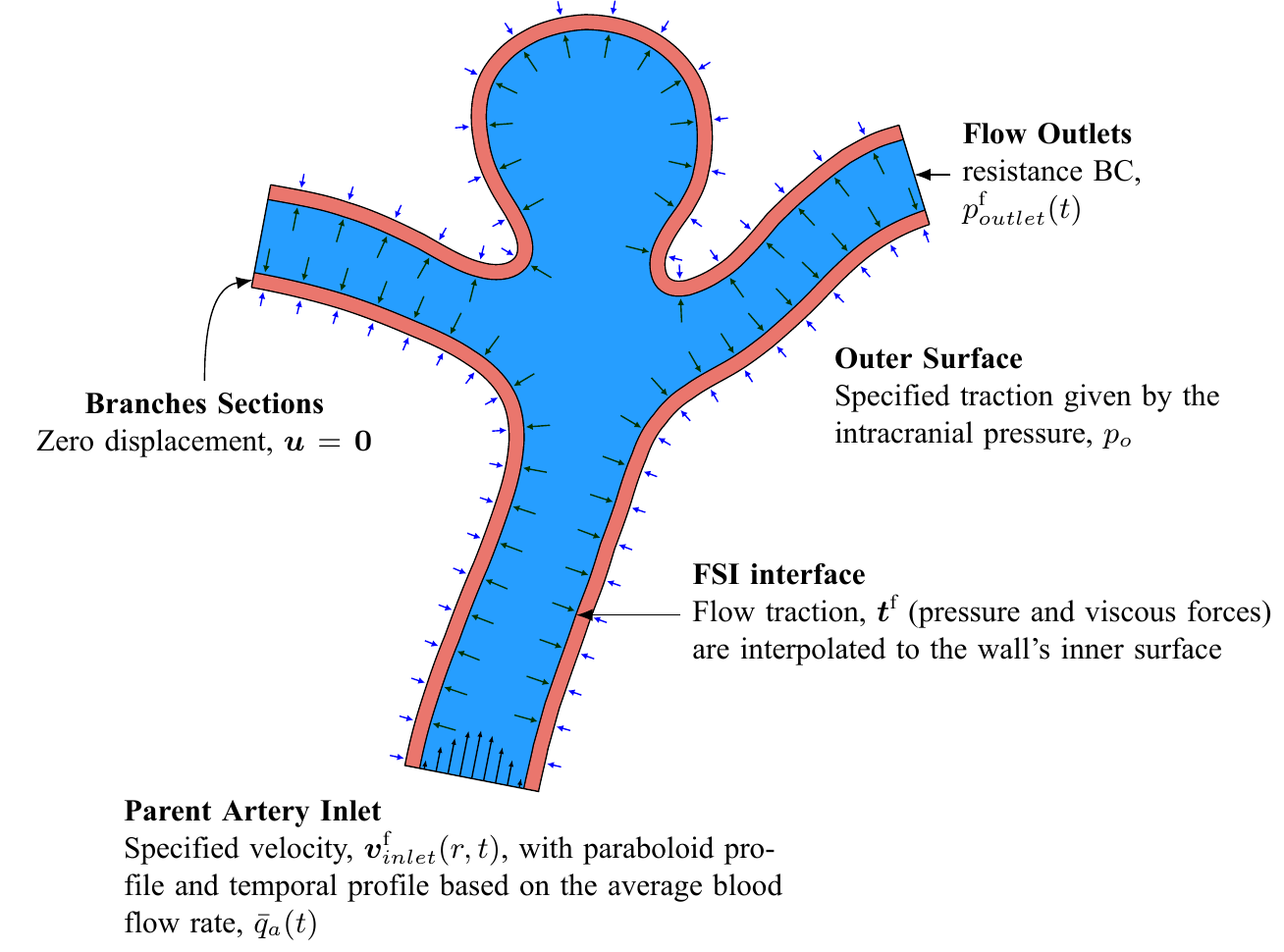}

        \caption{
            Schematic \gls{2d} representation of the \glspl{bc} applied in the
            aneurysms \glsxtrshort{1wfsi} simulations and the domains of artery
            and \gls{ia} wall tissue and blood flow.
        }

        \label{fig:icaAneurysmFsiDomain}
    \end{figure}

\begin{figure}[!htb]
        \includegraphics[%
            width=\linewidth
        ]   {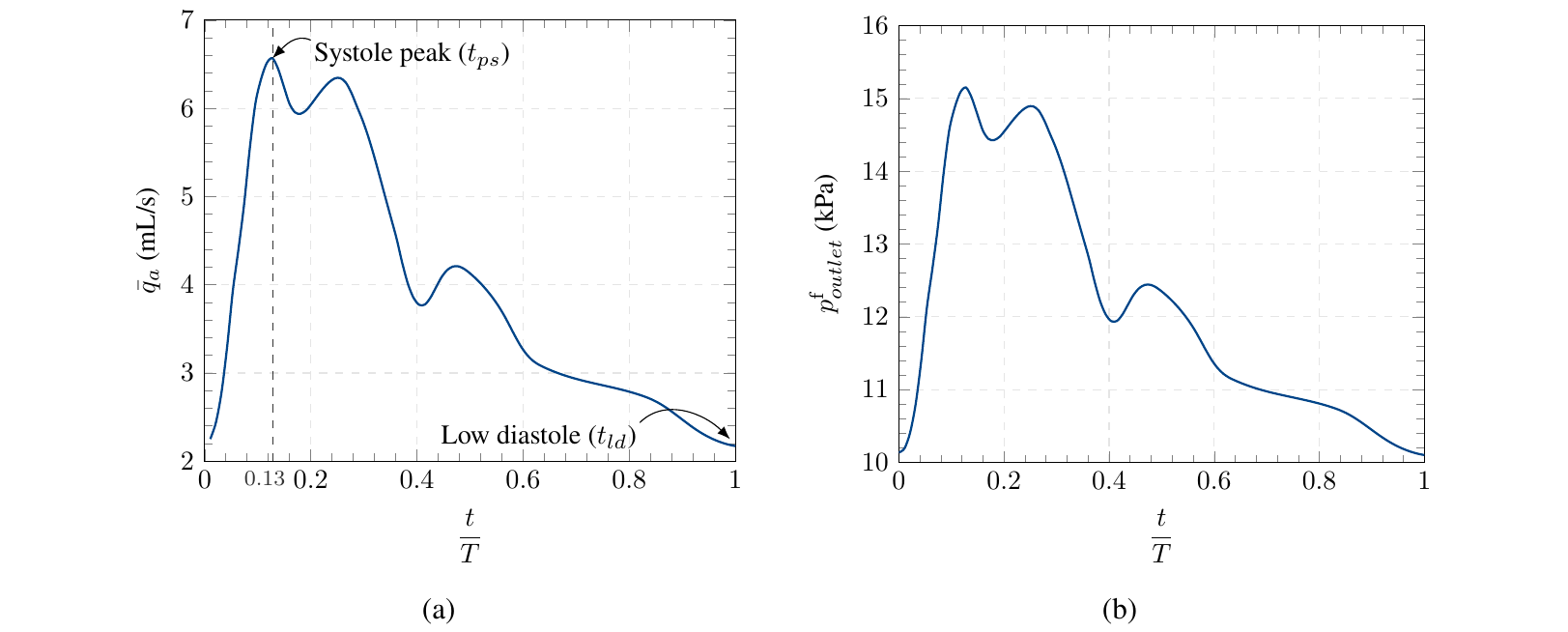}

        \caption{
            (a) Flow rate profile waveform for
            older adults in a cardiac cycle from \citet{Hoi2010}
            (non-dimensional time axis by dividing per the cardiac period,
            $\gls{cardiacPeriod} = \SI{0.94}{\second}$) and
            (b) pressure waveform used at
            the flow outlets computed as being proportional to the flow rate
            and between the normal range \SIrange{80}{120}{\mmHg}.
        }

        \label{fig:flowAndPressureProfiles}
    \end{figure}

    At the flow inlet (see \cref{fig:icaAneurysmFsiDomain}), a time-varying
    pulsatile velocity profile was imposed varying spatially according to the
    fully-developed laminar flow in a pipe:
        \begin{equation} \label{eq:parabolicInletCondition}
            \gsuper{velocity}{fluid}_{inlet}
            \functionOf{
                \gls{radial},
                \gls{time}
            }
            =
            2\frac{
                \gls{meanBloodFlowRate}
                \functionOf{\gls{time}}
            }{
                \gls{surfaceArea}_{inlet}
            }
            \left[
                1
                -
                \frac{
                    4\gls{radial}^2
                }{
                    \gls{arteryDiameter}^2
                }
            \right]\,,
        \end{equation}
    where $\gls{surfaceArea}_{inlet}$ is the cross-sectional area of the inlet
    artery, \gls{arteryDiameter} is its diameter, and \gls{radial} is the
    radial coordinate of the circular inlet section --- an artificial
    circular-section extension, with a length equal to twice the diameter
    \gls{arteryDiameter}, was added to the artery inlet to impose this inlet
    flow condition. The blood flow rate,
    \gls{meanBloodFlowRate}\functionOf{\gls{time}}, corresponding to the flow
    pulse from the beginning of systole until the end of the diastole, was
    obtained by multiplying the normalized flow rate reported by
    \citet{Hoi2010}, for older adults, by the mean blood flow rate in the
    respective \gls{ia} parent artery reported by \citet{Zarrinkoob2015}.  This
    population-averaged rate was employed because the patient-specific blood
    flow rate waveform at the \gls{ica} was not available. An example of the
    profile used for \glspl{ia} on the \gls{ica} is shown in
    \cref{fig:flowAndPressureProfiles}a. Moreover, the pressure gradient was set to
    zero at the inlet.

    At the outlets, a flux-corrected velocity and a resistance \gls{bc} for
    pressure were imposed. The resistance \gls{bc} is defined as being
    proportional to the blood flow rate profile, but with levels ranging
    between the normal cardiac cycle pressure levels, i.e. from
    \SIrange{80}{120}{\mmHg} (approximately \SIrange{10}{15}{\kilo\pascal}, see
    \cref{fig:flowAndPressureProfiles}b). Since the distances between the
    outlets and the aneurysm were not sufficiently long and we did not add
    cylindrical extensions at the outlets to avoid large computational meshes,
    reports show that it is important to use this \gls{bc} \citep{Chnafa2018}.

    On the outer surface of the solid wall (the abluminal surface), a pressure
    of $\gls{intracranialPressure} = \SI{5}{\mmHg}$, corresponding to the
    intracranial pressure, was imposed. Although the intracranial pressure
    seems to vary among patients, we have found similar values in related
    studies \citep{Valencia2013,Sanchez2014}. On branch \enquote{sections}
    that were artificially created due to the segmentation process
    (indicated in \cref{fig:icaAneurysmFsiDomain}), we imposed
    zero displacements, i.e.:
        \begin{equation} \label{eq:bcZeroFixedDisplacement}
            \gls{solidDisplacement}
            =
            \vec{0}\,.
        \end{equation}

    \subsection{%
        Modeling the Heterogeneity of the \glsfmtshort{ia} Wall
    }
    \label{sec:modelingAneurysmHeterogeneity}

    The modeling of the thickness, \gls{aneurysmThickness}, and material
    constants of the \gls{ia} sac wall --- represented here as
    \materialcoeff{MooneyCoeffs}{i}{j} as they correspond to the constants
    appearing in the laws given by
    \cref{eq:MooneyRivlin3CoeffsIsochoric,eq:YeohModelIsochoric,%
    eq:isotropicFungModelIsochoric} --- will be referenced as \enquote{wall
    morphology modeling} and two approaches were employed.  In the
    \emph{\gls{uwm}}, we assumed that both \gls{aneurysmThickness} and
    \materialcoeff{MooneyCoeffs}{i}{j} were uniform over the \gls{ia} sac,
    while in the \gls{awm} we assumed that the two wall properties'
    distributions are governed by the adjacent \gls{tawss} and \gls{osi} fields
    of the intra-aneurysmal flow. Both approaches used the same modeling for
    the \gls{ia}'s surrounding branches.

\begin{figure}[!htb]
        \includegraphics[%
            width=0.7\textwidth
        ]   {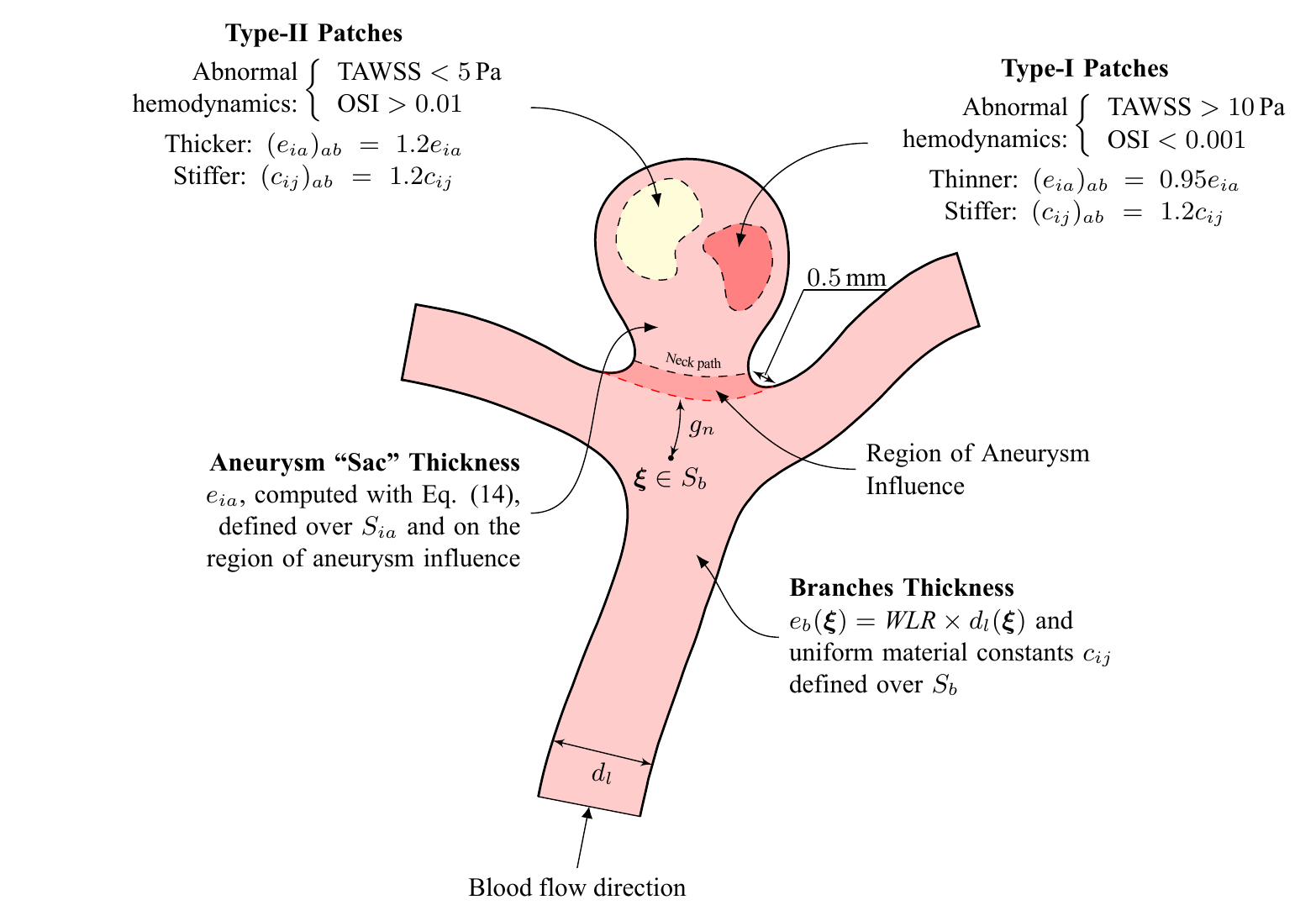}

        \caption{%
            Schematic illustration of the morphology model's components used in
            this work: the branches surface, \gsub{surface}{branch}, has a
            lumen diameter-dependent thickness and uniform material constants;
            with the \gls{uwm}, the aneurysm sac has a \enquote{sac} uniform
            thickness and material constants, while with the \gls{awm}, this
            prior uniform assumption was changed according to
            \enquote{abnormal-hemodynamics} patches defined based on the local
            \gls{tawss} and the \gls{osi}.
        }
        \label{fig:icaWithAbnormalWallRegions}
    \end{figure}

    Computationally, both were built with scripts in \gls{vmtk}\R and an
    in-house code based on the \gls{vtk}\R. The thickness field defined by each
    model was used to build the computational mesh of the solid domain (see
    \cref{sec:computationalMeshes}), whereas the material constants fields
    were computationally created in these meshes. The starting point was the
    manual delineation of the \emph{neck contour}, over the triangulated
    surface extracted from the \gls{dsa} exams, by interactively marking its
    path. Mathematically, the neck contour or path, \gsub{contour}{aneurysm},
    separates the surface of the branches, \gsub{surface}{branch}, from the
    \gls{ia} sac, \gsub{surface}{aneurysm} (see
    \cref{fig:icaWithAbnormalWallRegions} for a schematic example of
    \gsub{contour}{aneurysm} and other terms used in this section).

    \subsubsection*{Morphology of the Branches Wall}

    The thickness of \gsub{surface}{branch}, \gls{branchesThickness}, was based
    on established evidence that arteries thickness is dependent on the vessel
    lumen diameter \citep{Fung1993} and used the \gls{wlr}
    \citep{Nakagawa2016} to estimate the branch thickness. The \gls{wlr} is
    defined as %
    $
        \gls{wallLumenRatio}
        \equiv
        \dfrac{
            \text{artery wall thickness}
        }{
            \text{artery lumen diameter}
        }
        =
        \dfrac{
            \gls{branchesThickness}
        }{
            \gls{lumenDiameter}
        }
    $, where \gls{lumenDiameter} is the artery lumen diameter at a
    specified position along the vasculature. Therefore, the local branch
    thickness was calculated as %
    $
        \gls{branchesThickness}
        \functionOf{
            \gls{LagrangeCoord}
        }
        =
        \gls{wallLumenRatio}
        \times
        \gls{lumenDiameter}
        \functionOf{
            \gls{LagrangeCoord}
        }
    $ (note that \gls{branchesThickness} was defined at the reference
    configuration, hence defined in material coordinates, \gls{LagrangeCoord}).

    The lumen diameter of each geometry was estimated by using \gls{vmtk}\R, by
    computing the distance between the vasculature \emph{centerlines} and its
    surface \citep{Piccinelli2009}. Empirical values of the \gls{wlr} in the
    cerebral arteries were reported by \citet{Nakagawa2016} and used to
    define a functional form of \gls{wlr} according to the lumen diameter, as
    follows:
        \begin{equation} \label{eq:wlrFunction}
            \gls{wallLumenRatio}
            =
            \left\{
                \begin{array}{l l}
                    0.070
                &   \gls{lumenDiameter} < \SI{2}{\milli\meter} \\
                    0.070 + 0.018 \, (\gls{lumenDiameter} - 2 )
                &   \SI{2}{\milli\meter} <
                    \gls{lumenDiameter}  <
                    \SI{3}{\milli\meter} \\
                    0.088
                &   \gls{lumenDiameter} > \SI{3}{\milli\meter} \\

                \end{array}
            \right.\,.
        \end{equation}

    \subsubsection*{\Glsfmtlong{uwm}}

    We estimated the uniform thickness of the \gls{ia} sac,
    \gls{aneurysmThickness}, as a weighted average of the surrounding branches
    thickness field, \gls{branchesThickness}\functionOf{\gls{LagrangeCoord}} as
    follows:
        \begin{equation} \label{eq:aneurysmWallThickness}
            \gls{aneurysmThickness}
            =
            \gls{aneurysmThicknessFactor}
            \frac{
                \displaystyle
                \int_{\gsub{surface}{branch}}{
                    \gls{distanceToNeckLine}
                    \functionOf{
                        \gls{LagrangeCoord}
                    }
                    \gls{branchesThickness}
                    \functionOf{
                        \gls{LagrangeCoord}
                    }
                }\diff{\gsub{surface}{branch}}
            }{
                \displaystyle
                \int_{\gsub{surface}{branch}}{
                    \gls{distanceToNeckLine}
                    \functionOf{
                        \gls{LagrangeCoord}
                    }
                }\diff{\gsub{surface}{branch}}
            }\,,
        \end{equation}
    \noindent where \gls{aneurysmThicknessFactor} is a factor to control how
    much thinner the aneurysm wall was compared to the vasculature;
    $\gls{aneurysmThicknessFactor} = 0.75$ was used consistently for all
    geometries, consequently \gls{aneurysmThickness} was within the range that
    agrees with measured values available in the literature. The weight
    function, \gls{distanceToNeckLine}, was the minimum \emph{geodesic} distance
    between each point of the surrounding branches, \gsub{surface}{branch}, and
    the line that separates the sac and what was named the \emph{region of
    aneurysm influence} (see \cref{fig:icaWithAbnormalWallRegions}). This line
    may not be coincident with the aneurysm \emph{neck contour}, because,
    depending on its morphology and location, there might exist regions that
    neither belong to the healthy vasculature nor the aneurysm sac. Therefore,
    it may be imagined as a separation between the \emph{hypothetical healthy
    vasculature} and the region of aneurysm influence.  This line was computed
    automatically as being \SI{0.5}{\milli\meter} apart from the neck contour.

    Finally, the computational procedure used to build these fields created a
    discontinuity between the thickness on \gsub{surface}{aneurysm} and the
    branches thickness distribution defined over \gsub{surface}{branch}. To
    correct this biologically unrealistic discontinuity, the resulting
    thickness was smoothed out using the array smoothing script provided by
    \gls{vmtk}\R with 15 iterations.

\begin{figure}[!htb]
        \includegraphics[%
            width=\textwidth
        ]   {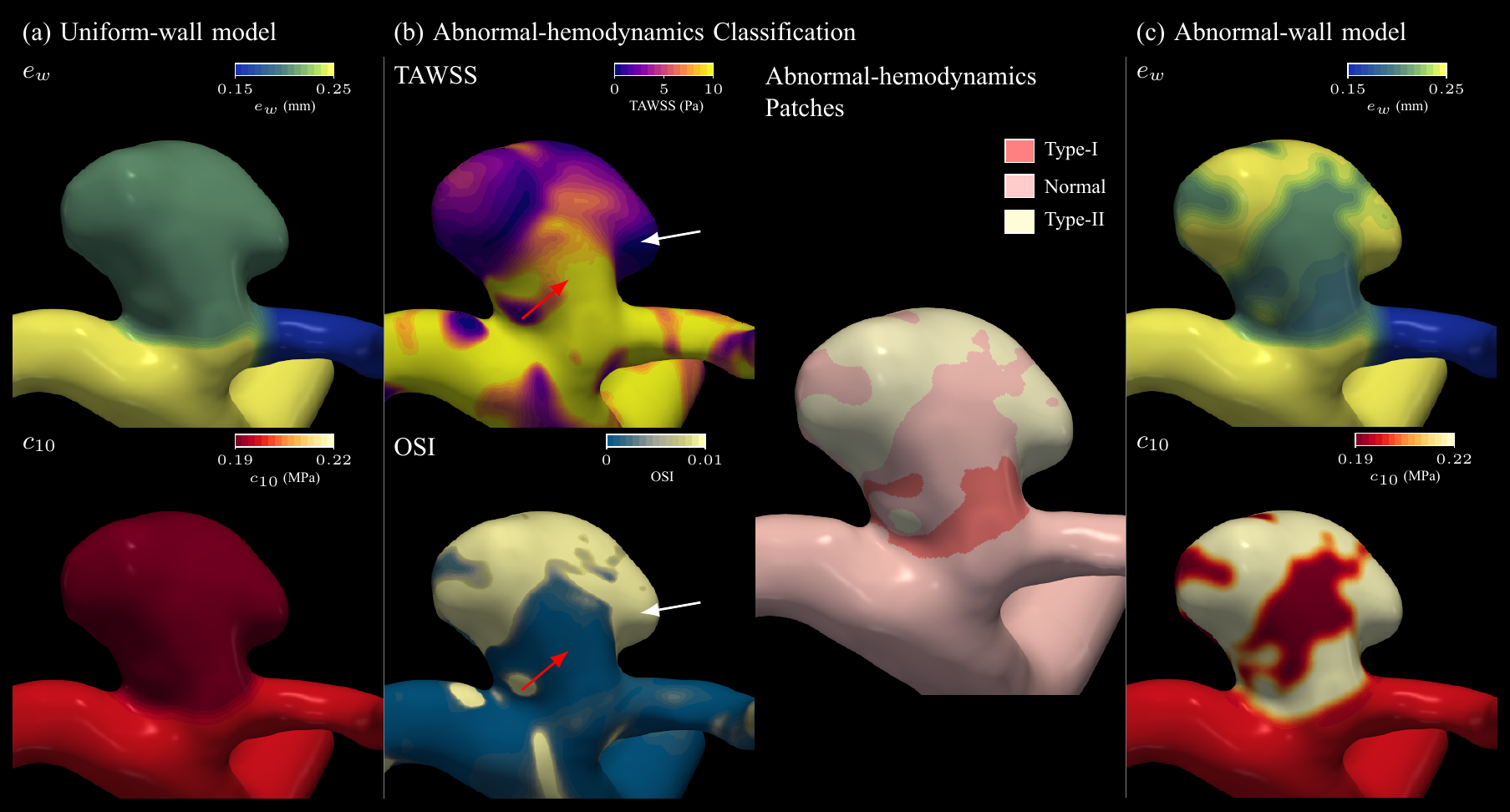}

        \caption{%
            Example of the morphology models construction for an example of
            a ruptured \gls{ia} case: (a) Thickness and material constant
            ($\gls{MooneyCoeffs}_{10}$ as an example) fields defined \emph{a
            priori} on each \gls{ia} for the \gls{uwm}; (b) identification of
            abnormal-hemodynamics patches on the \gls{ia} sac surface:
            \gls{tawss} and \gls{osi} fields over the vasculature computed with
            \gls{cfd} simulations of the blood flow (left); the red and white
            arrows point to examples of regions where abnormal hemodynamics
            occur and identification of the aneurysm sac patches with abnormal
            hemodynamics (right); \enquote{normal} are patches that are neither
            type-I nor type-II. (c) Finally, fields of thickness and
            $\gls{MooneyCoeffs}_{10}$ resulted by accounting for abnormal
            intra-aneurysmal hemodynamics (see
            \cref{fig:icaWithAbnormalWallRegions} too).
        }

        \label{fig:wallMorphologyModeling}
\end{figure}

    We assumed the material constants, \materialcoeff{MooneyCoeffs}{i}{j}, as
    uniform on both \gsub{surface}{branch} and \gsub{surface}{aneurysm}, but
    with different values according to rupture status, following experimental
    evidence that ruptured \glspl{ia} are less stiff than unruptured ones.
    Hence, the values for each constant (see
    \cref{tab:materialConstantsRupturedAndUnruptured}) were based on averages
    of experimental data, with ruptured and unruptured \gls{ia}, provided by
    \citet{Costalat2011} for the \gls{mr} law and \citet{Brunel2018} for the
    Yeoh law. We were not able to find any work that fit mechanical data of
    \gls{ia} tissue to the isotropic Fung law, hence we based the values used
    here on the constants used by \citet{Torii2010},
    $\materialcoeff{FungCoeffs}{1}{} = \SI{0.3536}{\mega\pascal}$ and
    $\materialcoeff{FungCoeffs}{2}{} = \num{16.7}$, based on experiments with
    porcine carotid arteries. See \cref{fig:wallMorphologyModeling}a for an
    example of the resulting fields of
    \gls{wallThickness}\functionOf{\gls{LagrangeCoord}} and
    \materialcoeff{MooneyCoeffs}{1}{0}\functionOf{\gls{LagrangeCoord}}.

    \begin{table}[!htb]
        \caption{%
            Material constants selected for arteries branches,
            \gsub{surface}{branch}, and the \gls{ia} sac,
            \gsub{surface}{aneurysm}, according to rupture status based on
            experimental works by \citet{Costalat2011} and
            \citet{Brunel2018}, for the material laws employed here.
        }
        \small
        \begin{tabular}{c c S S S}
            \toprule
                \multirow{2}{*}{Law}
            &   \multirow{2}{*}{Constant}
            &   {\multirow{2}{*}{\gsub{surface}{branch}}}
            &   \multicolumn{2}{c}{\gsub{surface}{aneurysm}} \\
            \cmidrule(lr){4-5}
                \multicolumn{2}{c}{}
            &
            &   {Ruptured}
            &   {Unruptured} \\
            \midrule
                \multirow{3}{*}{\gls{mr}}
            &   \materialcoeff{MooneyCoeffs}{1}{0} (\si{\mega\pascal})
            &   0.1966
            &   0.19
            &   0.19       \\

            &   \materialcoeff{MooneyCoeffs}{0}{1} (\si{\mega\pascal})
            &   0.0163
            &   0.026
            &   0.023      \\

            &   \materialcoeff{MooneyCoeffs}{1}{1} (\si{\mega\pascal})
            &   7.837
            &   1.377
            &   11.780     \\
            \midrule
                \multirow{3}{*}{Yeoh}
            &   \materialcoeff{YeohCoeffs}{1}{0} (\si{\mega\pascal})
            &   0.1067
            &   0.07
            &   0.12          \\

            &   \materialcoeff{YeohCoeffs}{2}{0} (\si{\mega\pascal})
            &   5.1602
            &   2.10
            &   6.80       \\

            &   \materialcoeff{YeohCoeffs}{3}{0} (\si{\mega\pascal})
            &   0.0
            &   0.0
            &   0.0        \\
            \midrule
                \multirow{2}{*}{Fung}
            &   \materialcoeff{FungCoeffs}{1}{} (\si{\mega\pascal})
            &   0.3536
            &   0.1768
            &   0.7072        \\

            &   \materialcoeff{FungCoeffs}{2}{} (-)
            &   16.7
            &   16.7
            &   16.7       \\
            \bottomrule
        \end{tabular}%
        \\[\baselineskip]
        \label{tab:materialConstantsRupturedAndUnruptured}
\end{table}

    \subsubsection*{\Glsfmtlong{awm}}

    We selected two \enquote{abnormal-hemodynamics} conditions adjacent to the
    aneurysm wall that were already associated with phenotypic changes, more
    specifically, with type-I and type-II regions as proposed by
    \citet{Meng2014}. The type-I phenotype is more likely to be caused by the
    occurrence of high \gls{tawss}, $> \SI{10}{\pascal}$, and low \gls{osi}, $<
    \num{0.001}$ (\enquote{high-flow} effects) and cause local wall thinning
    and stiffening. On the other hand, the type-II phenotype is more likely to
    be caused by disturbed flow characterized by low \gls{tawss}, $<
    \SI{5}{\pascal}$, and high \gls{osi}, $> \num{0.01}$ (\enquote{low-flow
    effects}), causing local stiffening and thickening with atherosclerosis.
    Those specific thresholds for \gls{tawss} and \gls{osi} were chosen based
    on the averaged values reported by \citet{Furukawa2018} and
    \citet{Cebral2019}.

    For each patient-specific \gls{ia} geometry, these regions were identified
    after performing \gls{cfd} simulations using the rigid-wall assumption ---
    the specific methodology of these simulations can be found elsewhere
    \citep{Oliveira2021}. The \gls{tawss} and \gls{osi} were computed and the
    identification of the abnormal-hemodynamics patches was performed
    automatically by using an in-house code implemented with the \gls{vtk}\R
    library (see \cref{fig:wallMorphologyModeling}b).  Patches that were
    neither identified as type-I or II were labeled \enquote{normal}.

    The fields of aneurysm thickness, \gls{aneurysmThickness}, and of each
    material constants on \gsub{surface}{aneurysm} were computationally
    \enquote{updated} according to the illustration in
    \cref{fig:icaWithAbnormalWallRegions}. If a portion of the aneurysm is
    identified as a type-I patch, the thickness was decreased by
    \SI{5}{\percent} and the material constants were increased by
    \SI{20}{\percent}. On type-II patches, both thickness and material
    constants were increased by \SI{20}{\percent}. The properties on normal
    patches remained unaltered, i.e. with the same properties as defined in the
    \gls{uwm} (see \cref{fig:wallMorphologyModeling}c). Finally, these fields
    were also smoothed to avoid unrealistic discontinuities introduced during
    the computation procedure.

    \subsection{%
        Computational Meshes and Numerical Strategies
    }
    \label{sec:computationalMeshes}

    The numerical simulations were performed in solids4foam
    \citep{Cardiff2018}, an extension of the foam-extend library
    \citep{foam-extend,Weller1998}, version 4.0, which uses the
    second-order-accurate \gls{fvm} as the discretization method. The
    \gls{1wfsi} solution strategy implemented in solids4foam was
    used, in which the blood flow is first solved assuming a rigid wall and, at
    each time-step, the wall traction forces are applied as the driving force
    of the wall motion, whereas the solid deformation is not passed back to the
    fluid domain. This strategy was chosen to avoid the numerical instabilities
    that arise in the numerical solution of flow in arterial geometries. That
    allowed us to simulate a relatively large number of patient-specific
    \gls{ia} geometries.  Furthermore, in this work we were interested in the
    mechanical response of the wall, hence only simulating it with the blood
    flow forces is sufficient for this comparative analysis.

    The discretized version of the \cref{eq:momentumEquationTotalLag}, with the
    different hyperelastic laws and morphology models, was solved by using the
    \emph{segregated} algorithm in solids4foam, while the flow
    governing equations were solved with the \gls{piso} algorithm
    \citep{Issa1986}, adapted for the flow of a compressible fluid. In the
    solution of both sub-problems, we selected second-order interpolation
    profiles for the spatial discretization to maintain the second-order
    accuracy level of the \gls{fvm}. The central differences scheme was used
    for all the Laplacian terms, with non-orthogonal and skewness corrections
    \citep{JasakThesis1996}. Particularly for the flow's momentum and
    pressure equations, the second-order upwind scheme was used for the
    advective term. Moreover, all the gradients in the equations were
    discretized with the least-squares scheme.  The temporal discretization was
    performed by using the implicit first-order Euler approach for the solid
    momentum equation and the implicit second-order Euler approach for the
    flow's momentum equation. Finally, the normalized residual convergence
    criteria were: \num{1e-6} for the flow 's pressure equation, \num{1e-08}
    for the flow's momentum equation, and \num{1e-9} for the outer iterations
    of the solid's momentum.

    The fluid and solid computational meshes were built with the triangulated
    surfaces extracted with \gls{vmtk}\R. The mesh of the flow domain was
    generated using the utility \command{cartesianMesh} provided by
    foam-extend as part of the cfMesh library. This utility
    automatically creates polyhedral meshes that are predominantly composed of
    hexahedral cells, i.e. the interior of the mesh consists of cells close in
    shape to hexahedra. To fit the curved boundary, the cells adjacent to the
    wall are, generically, polyhedra with a prismatic boundary-layer-refined
    region composed of five layers. Subsequently, the \gls{fsi} interface of
    this fluid mesh, a quadrilateral surface, was \enquote{re-meshed}, i.e. the
    structure of it was modified to only contain triangular faces by using
    \gls{vmtk}\R. This re-meshing procedure was necessary because of the
    possibility of pressure oscillations infecting the solid displacement
    solution when using quadrilateral cells with the segregated approach and
    the \gls{fvm} \citep{Wheel1999}. The solid mesh was, then, created by
    extruding that re-meshed surface in the outward direction using
    \gls{vmtk}\R with the thickness field defined in
    \cref{sec:modelingAneurysmHeterogeneity}. Six layers were used in the
    extrusion.

    For the type of fluid meshes used in this work, mesh-sensitivity studies
    were carried out extensively with different \glspl{ia} geometries and were
    not included here. They yielded a volume density of cells in the range of
    \SIrange{3000}{4000}{cells\per\milli\cubic\meter} (an example of this study
    can be found in \citet{Oliveira2021}). This level of refinement was
    assured for the meshes used in this work. We carried out a separate
    mesh-independence study of the solid meshes by also using the \gls{1wfsi}
    strategy with three systematically-refined meshes. The result was a mesh
    with a surface-cells density of approximately
    \SI{240.0}{cells\per\milli\square\meter}, with 6 layers of cells along with
    the thickness. We also carried out a time-step refinement study yielding a
    time-step of \SI{1e-4}{\second}. Finally, two cardiac cycles were solved in
    each simulation, but only the second one was used for the analysis.

    \subsection{Data Analysis} \label{sec:dataAnalysis}

    \subsubsection*{Physical Variables of Analysis And Metrics Employed}

    We selected the largest principal Cauchy stress field of the solid wall,
    $\gls{maxPrincipalCauchyStress} \equiv
    \gsuper{CauchyStressComp}{solid}_{1}$\functionOf{\gls{EulerCoord},
    \gsub{time}{peakSystole}} (we drop the superscript \enquote{\gls{solid}}
    henceforth), and its largest principal stretch,
    \gls{maxStretch}\functionOf{\gls{EulerCoord}, \gsub{time}{peakSystole}}
    (given by the square root of the largest principal value of
    \gls{rightCauchyGreenDeformation}) as the main subjects of our analysis.
    Both were taken at  the \emph{deformed} configuration at the peak-systole,
    i.e. $\gls{time} = \gsub{time}{peakSystole}$ (see
    \cref{fig:flowAndPressureProfiles}a).

    Both fields were analyzed only on the luminal and abluminal sides of the
    aneurysmal region, i.e. on \gsupsub{surface}{lumen}{aneurysm} and
    \gsupsub{surface}{ablumen}{aneurysm}. To perform the statistical analysis,
    we computed two metrics of the fields. The \nth{99} percentile of each
    field, $(\gls{maxPrincipalCauchyStress})_{\gls{percentil99}}$ and
    $(\gls{maxStretch})_{\gls{percentil99}}$, were used as \enquote{proxies} of
    the fields' maximum specifically for the statistical analysis, due to
    possible mesh-induced differences in the absolute maximum. The second
    metric was the surface average of a field over a surface, \gls{surface},
    defined as follows, for \gls{maxPrincipalCauchyStress}:
    \begin{equation} \label{eq:surfaceAvgSigmaMax}
        \savg{\gls{maxPrincipalCauchyStress}}_{\gls{surface}}
        =
        \surfaceavg{
            \gls{surface}
        }{
            \gls{maxPrincipalCauchyStress}
            \functionOf{
                \gls{EulerCoord}
            }
        },
        \,\,\,\,
        \gls{EulerCoord} \in \gls{surface}\,,
    \end{equation}
    \noindent where $\gls{surfaceArea}\functionOf{}$ is the area operator and
    \gls{surface} is either \gsupsub{surface}{lumen}{aneurysm} or
    \gsupsub{surface}{ablumen}{aneurysm}.  For \gls{maxPrincipalCauchyStress},
    the surface-average metric gives a measure of the total traction on
    \gls{surface} caused by this stress. A similar physical meaning cannot be
    said about \savg{\gls{maxStretch}} although the same definition was used
    regardless.

    Computationally, the \nth{99} percentile was computed with Python's NumPy
    library \citep{numpy}, whereas the surface average was computed using an
    in-house code that computes the integral in \cref{eq:surfaceAvgSigmaMax}
    with first-order accuracy

    \subsubsection*{Statistical Analysis}

    Statistical tests were performed with n = 13, for the \nth{99} percentiles
    and the surface averages, by using the SciPy library \citep{scipy} with
    a significance level of $\gls{significanceLevel} = 0.05$ (hence,
    \SI{95}{\percent} confidence interval). All distributions were tested for
    normality by using the Shapiro-Wilk test.

    To compare the two wall morphology models, for each hyperelastic law, the
    paired t-test and the Wilcoxon signed-rank tests were used for normal and
    non-normal distributions, respectively. Similarly, to compare the three
    distributions of the hyperelastic laws, the ANOVA test and the
    Kruskal-Wallis test were used \emph{a priori} for normal and for non-normal
    distributions, respectively. Subsequently, pair-wise posthoc analyses were
    performed to test the distributions. The t-test and Dunn's posthoc methods
    were employed, in this case, for normal and non-normal distributions,
    respectively, via Python's scikit-posthoc library.

    \subsubsection*{Relative Comparison Among Different Models}

    To quantify the differences among the metrics, we computed the
    \emph{absolute} differences between the mean of the distributions obtained
    with each modeling approach. If \tavg{\gls{iaWallMetric}} is the mean of
    the distribution of a metric \gls{iaWallMetric}, for a fixed hyperelastic
    law, the mean difference between the uniform and abnormal wall morphologies
    for a sample was defined as:
    \begin{equation} \label{eq:absoluteDifference}
        \gsub{genericAbsoluteDifference}{iaWallMetric}^{\textsc{WM}}
        =
        \left|
            \tavg{\gls{iaWallMetric}}_{\mathit{\glsxtrshort{awm}}}
            -
            \tavg{\gls{iaWallMetric}}_{\mathit{\glsxtrshort{uwm}}}
        \right|\,,
    \end{equation}
    where the superscript \enquote{$\textsc{WM}$} stands for \enquote{wall
    morphology}, and, consequentely, \enquote{\glsxtrshort{uwm}} and
    \enquote{\glsxtrshort{awm}} indicate the uniform and abnormal-wall models.

    Because three hyperelastic laws were employed and none can be assumed the
    \enquote{gold standard} to represent \gls{ia} tissue, first, a difference
    similar to \cref{eq:absoluteDifference}, but in a pair-wise manner
    between the three laws, \gls{mr}, Yeoh, and isotropic Fung, was computed.
    Then, the maximum of these values was found, or mathematically:
    \begin{equation} \label{eq:absoluteDifferenceHyperelasticLaws}
        \gsub{genericAbsoluteDifference}{iaWallMetric}^{\textsc{HL}}
        =
        \underset{%
            i,j
        }{\max}\,
        {\left(
            \left|
                \tavg{\gls{iaWallMetric}}_{i}
                -
                \tavg{\gls{iaWallMetric}}_{j}
            \right|
        \right)
        }\,,
    \end{equation}
    where the superscript \enquote{$\textsc{HL}$} stands for
    \enquote{hyperelastic laws} and the pair $(i, j)$ assumes the values in the
    set of permutations among the three hyperelastic laws.

%% file: tex/results.tex
\section{Results} \label{sec:results}

\begin{figure}[!hp]
        \includegraphics[%
            width=0.9\textwidth
        ]   {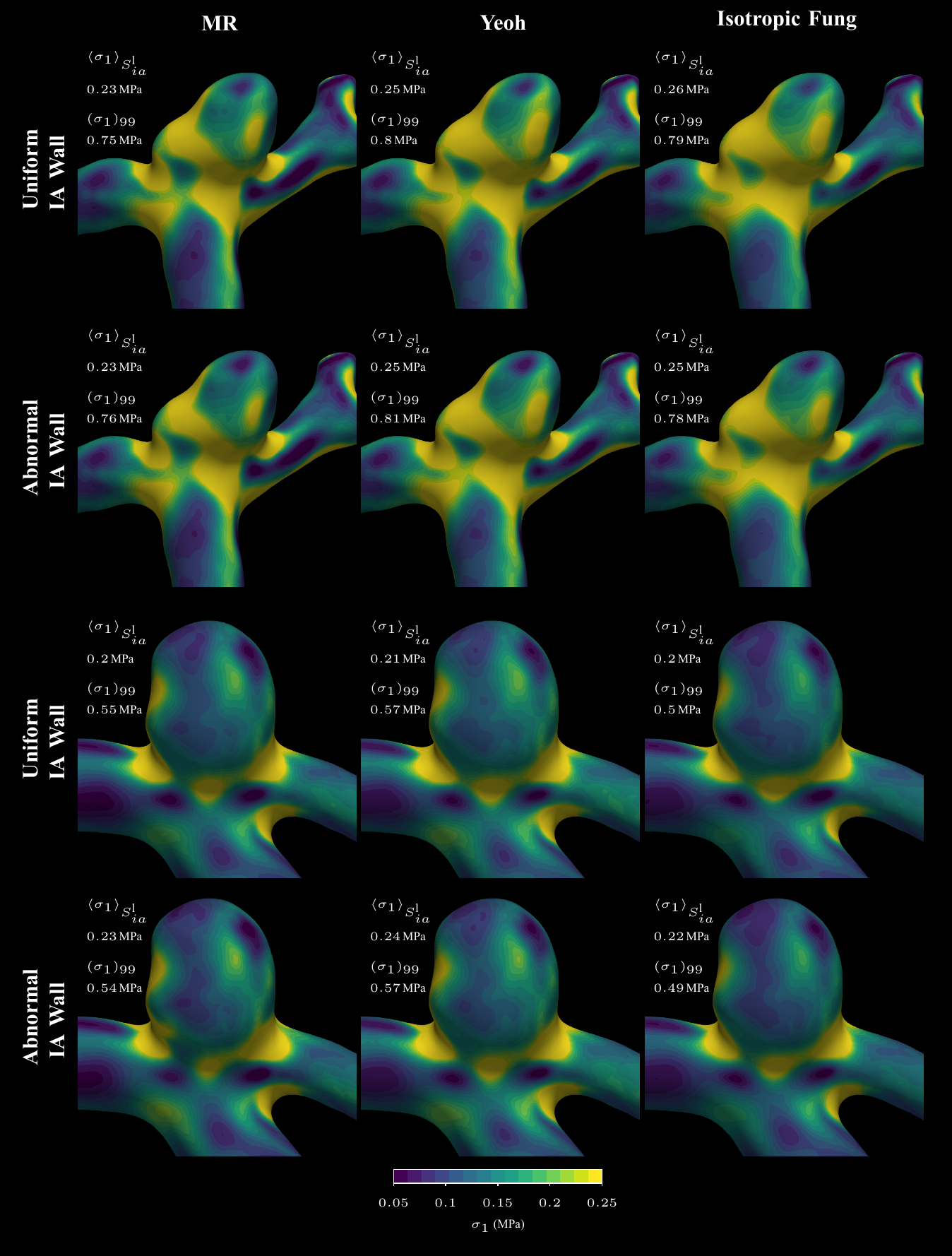}

        \caption{%
            Largest principal Cauchy stress, \gls{maxPrincipalCauchyStress},
            at the peak systole, on \gsupsub{surface}{lumen}{aneurysm},
            for cases \case{case50} (top panel) and \case{case74} (bottom
            panel), with different hyperelastic laws (columns), and wall
            morphology models (rows).
        }

        \label{fig:sigmaMaxCompMatLawAndAneurysmWallLumen}
    \end{figure}

\begin{figure}[!hp]

        \includegraphics[%
            width=0.9\textwidth
        ]   {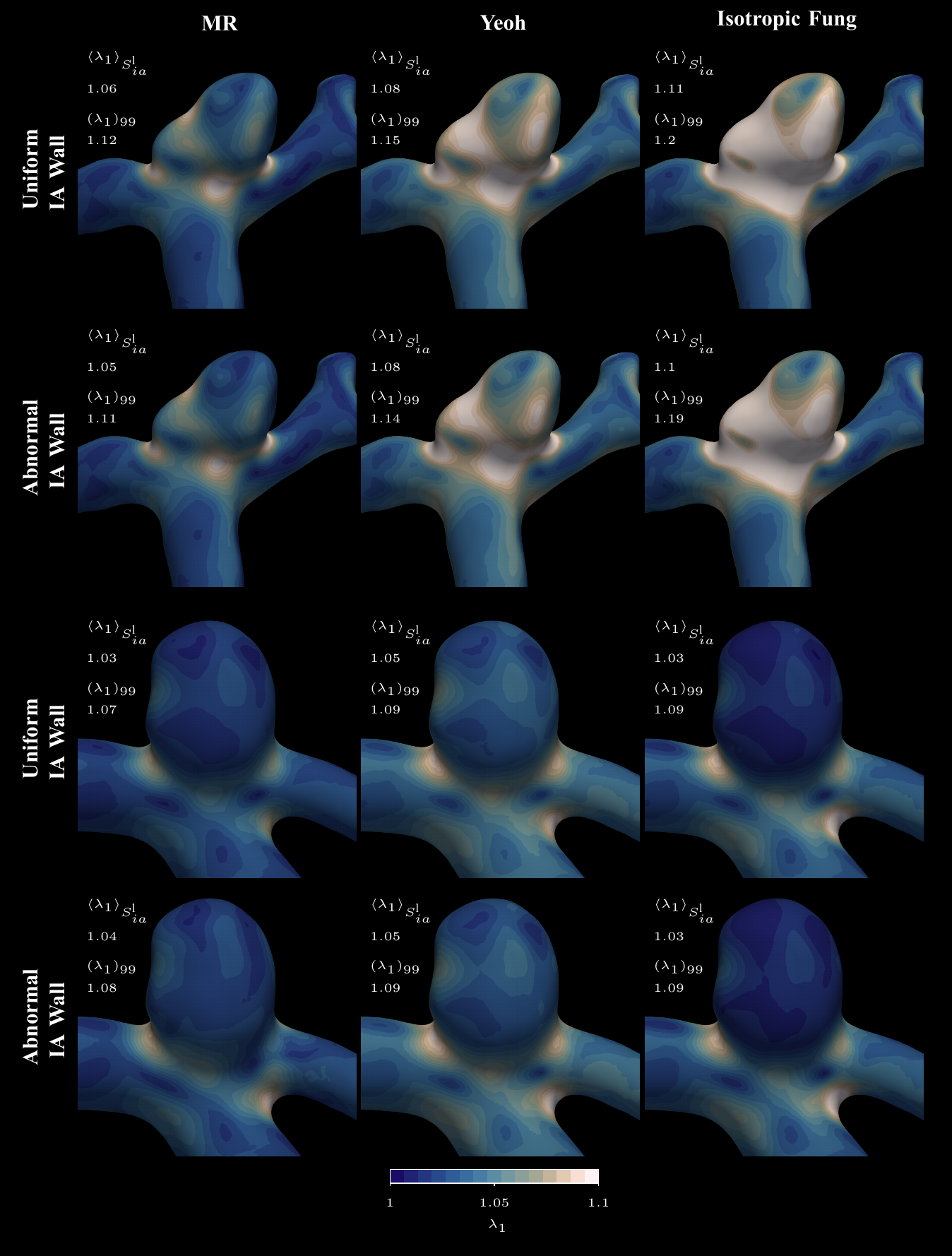}

        \caption{%
            Largest principal stretch, \gls{maxStretch}, at the peak
            systole, on \gsupsub{surface}{lumen}{aneurysm}, for cases
            \case{case50} (top panel) and \case{case74} (bottom panel), with
            different hyperelastic laws (columns), and wall morphology models
            (rows).
        }

        \label{fig:stretchMaxCompMatLawAndAneurysmWallLumen}
    \end{figure}

    A preliminary qualitative analysis of the \gls{maxPrincipalCauchyStress}
    and \gls{maxStretch} fields in two representative cases of the ruptured and
    unruptured groups (labeled from now on as \gls{rupturedGroup} and
    \gls{unrupturedGroup}, respectively), namely, cases \case{case50} and
    \case{case74} suggests that the different hyperelastic laws exert less
    influence on \gls{maxPrincipalCauchyStress} than on \gls{maxStretch} and
    this seems to be true for both \glspl{ia} (see
    \cref{fig:sigmaMaxCompMatLawAndAneurysmWallLumen}). The \gls{maxStretch}
    field in case \case{case50} has a different overall magnitude on its sac
    depending on the law, whereas the qualitative differences in \case{case74}
    are less apparent (see on the top panel of
    \cref{fig:stretchMaxCompMatLawAndAneurysmWallLumen} how the Fung and Yeoh
    laws produce higher levels of stretch in much larger areas than the
    \gls{mr} law). On the other hand, for both \gls{maxPrincipalCauchyStress}
    and \gls{maxStretch}, the two morphology models produce similar fields,
    regardless of the particular hyperelastic law or the rupture status.
    Similar trends were found when inspecting the abluminal surface,
    \gsupsub{surface}{ablumen}{aneurysm}.

\begin{figure}[!htb]
        \includegraphics[%
            width=\textwidth
        ]   {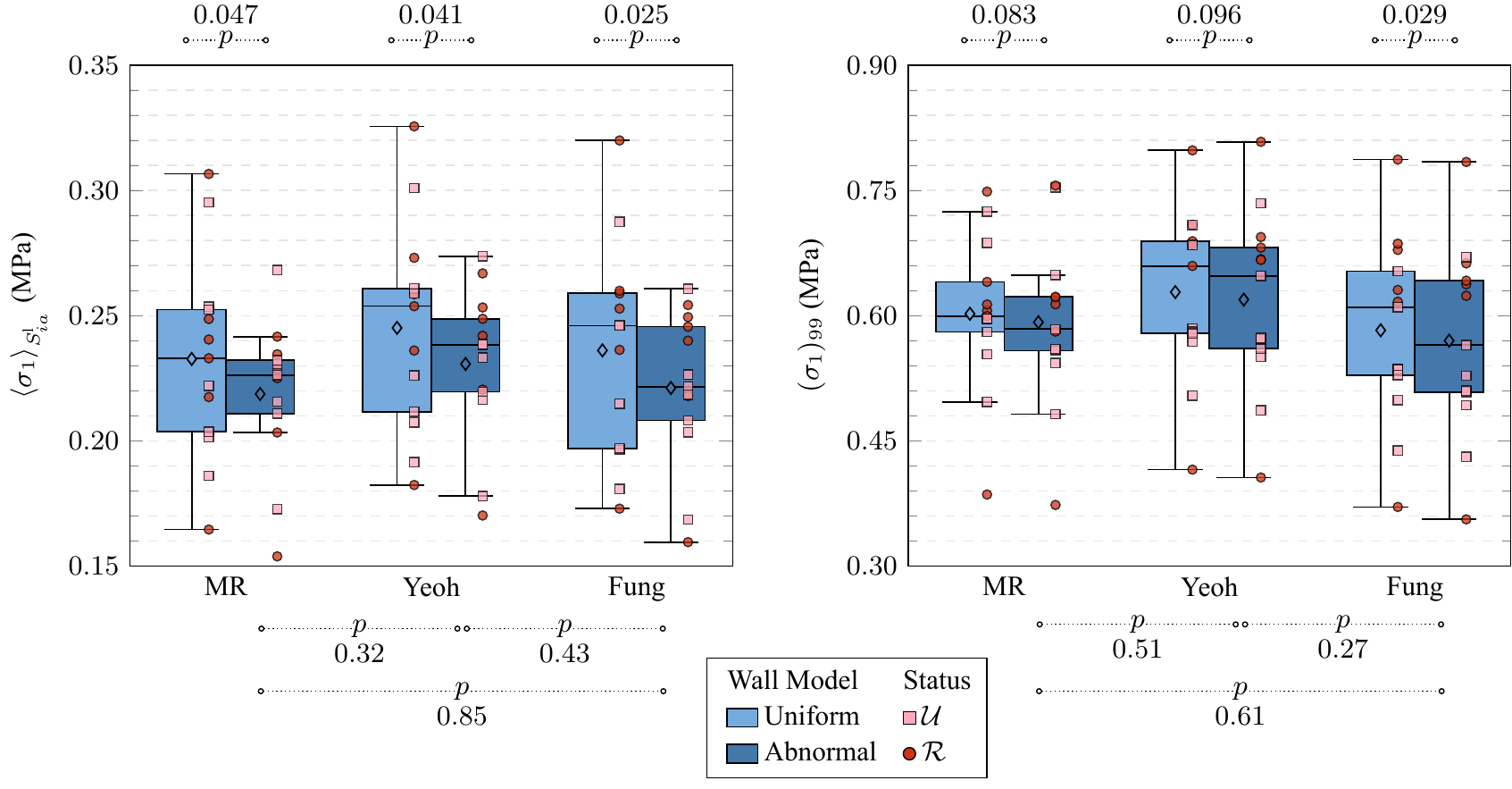}

        \caption{%
            Box plots and data points (placed on top of their respective box
            plot) of the distributions of
            $\savg{\gls{maxPrincipalCauchyStress}}_{\gsupsub{surface}{lumen}{aneurysm}}$
            (left) and $(\gls{maxPrincipalCauchyStress})_{\gls{percentil99}}$
            (right) for the \gls{ia} sample simulated with the three
            hyperelastic laws (along the x-axis) and segregated by morphology
            model and rupture status. The diamond shape indicates the mean of
            each distribution. The p-values resulting from hypothesis tests
            comparing the distributions are shown below and above the plot's
            area through the special symbol where the two circles visually
            indicate the two distributions that were compared and the resulting
            p-value.
        }

        \label{fig:boxPlotsSigmaMaxDistPerHyperelasticLaw}
    \end{figure}

    These two \gls{ia} cases were chosen as representative examples because
    they yielded the largest absolute difference
    $\gsub{genericAbsoluteDifference}{iaWallMetric}^{\textsc{HL}}$ in the
    groups \gls{rupturedGroup} and \gls{unrupturedGroup}, respectively.
    Nevertheless, the patterns indicated by them seem to be examples of larger
    trends. By comparing the sample's distributions for the three hyperelastic
    laws (see \cref{fig:boxPlotsSigmaMaxDistPerHyperelasticLaw}),
    \gls{maxPrincipalCauchyStress}'s metrics did not reach statistical
    significance ($\gls{pValue} = 0.73$ and $\gls{pValue} = 0.57$, with the
    \gls{uwm} and \gls{awm}, respectively, for
    $\savg{\gls{maxPrincipalCauchyStress}}_{\gsupsub{surface}{lumen}{aneurysm}}$;
    similar values were found for
    $(\gls{maxPrincipalCauchyStress})_{\gls{percentil99}}$). Furthermore, a
    posthoc analysis using the t-test in a pair-wise manner yielded that the
    three distributions are not significantly different (p-values depicted in
    \cref{fig:boxPlotsSigmaMaxDistPerHyperelasticLaw} with the symbol
    \enquote{\tikz{\pic {picStatsTest=north/{}};}}, for the \gls{awm} only;
    similar p-values were found with the \gls{uwm}).

\begin{figure}[!h]
        \includegraphics[%
            width=\textwidth
        ]   {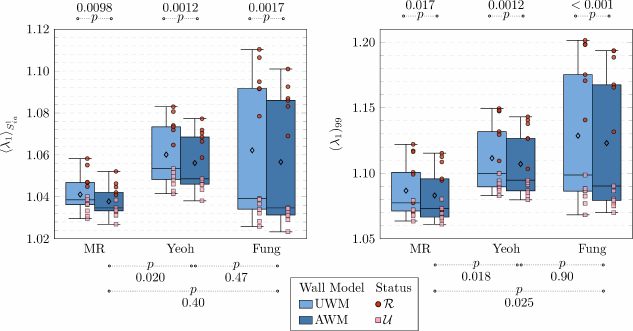}

        \caption{%
            Box plots and data points (placed on top of their respective box
            plot) of the distributions of
            $\savg{\gls{maxStretch}}_{\gsupsub{surface}{lumen}{aneurysm}}$
            (left) and $(\gls{maxStretch})_{\gls{percentil99}}$
            (right) for the \gls{ia} sample simulated with the three
            hyperelastic laws (along the x-axis) and segregated by morphology
            model and rupture status. The diamond shape indicates the mean of
            each distribution. The p-values resulting from hypothesis tests
            comparing the distributions are shown below and above the plot's
            area through the special symbol where the two circles visually
            indicate the two distributions that were compared and the resulting
            p-value.
        }

        \label{fig:boxPlotsStretchMaxDistPerHyperelasticLaw}
\end{figure}

    The opposite was the case for the distribution of \gls{maxStretch}'s
    (\cref{fig:boxPlotsStretchMaxDistPerHyperelasticLaw}), in which case the
    three distributions were significantly different ($\gls{pValue} = 0.033$
    with the \gls{uwm} and $\gls{pValue} = 0.026$ with the \gls{awm} for
    $\savg{\gls{maxStretch}}_{\gsupsub{surface}{lumen}{aneurysm}}$, with
    similar values found for $(\gls{maxStretch})_{\gls{percentil99}}$).
    Nevertheless, a posthoc analysis, using Dunn's test, indicated that only
    the differences between the pairs (\gls{mr}, Yeoh) and (\gls{mr}, Fung)
    were indeed significant (see p-values below the plots in
    \cref{fig:boxPlotsStretchMaxDistPerHyperelasticLaw}).

    Furthermore, the different behavior of \gls{maxStretch} between the
    ruptured and unruptured cases evidenced above is part of a trend by
    noting a visible and consistent \enquote{separation} of
    \gls{maxStretch} levels according to rupture status (see the data points in
    \cref{fig:boxPlotsStretchMaxDistPerHyperelasticLaw}), independently of
    hyperelastic law or morphology model. That gap is the largest for the
    isotropic Fung law, explaining the large dispersion of the distributions of
    \gls{maxStretch}'s metrics, although the same can also be perceived with
    the Yeoh and \gls{mr} models, to a smaller extent.  This particular feature
    did not occur for \gls{maxPrincipalCauchyStress}'s metrics --- note that
    they spread over a range between, approximately, \SI{0.15}{\mega\pascal} to
    \SI{0.30}{\mega\pascal} but are not visibly segregated by rupture status
    (see the data points in \cref{fig:boxPlotsSigmaMaxDistPerHyperelasticLaw}).

    By comparing the distributions between the \gls{uwm} and \gls{awm}, they
    reached statistical significance for both \gls{maxStretch} metrics,
    irrespective of the hyperelastic law (see the p-values above the plots of
    \cref{fig:boxPlotsStretchMaxDistPerHyperelasticLaw}). The same did not
    occur for the \gls{maxPrincipalCauchyStress}, for which statistical
    significance was only reached by \savg{\gls{maxPrincipalCauchyStress}}'s
    distributions with all laws, whereas the distribution of
    $(\gls{maxPrincipalCauchyStress})_{\gls{percentil99}}$ was only reached for
    the Fung law.

    The absolute differences between the means of
    \gls{maxPrincipalCauchyStress} and \gls{maxStretch} were larger when the
    hyperelastic laws were compared than when comparing the wall morphology
    models (\cref{tab:absolutDifferencesModelings}), especially on the luminal
    surface and irrespective of the morphology model, confirming that the
    mechanical response is more sensitive to the choice of hyperelastic law.
    More specifically, it is possible to note that the influence of the
    hyperelastic laws is more drastic for \gls{maxStretch} and depends on the
    rupture status --- note how the mean absolute difference is larger for the
    ruptured group than the unruptured one in
    \cref{tab:absolutDifferencesModelings}. Finally, as verified above, that
    difference for \gls{maxStretch} reached statistical significance, whereas
    it did not with \gls{maxPrincipalCauchyStress}.

    \begin{table}[!h]
        \caption{%
            Maximal absolute difference for the selected metrics
            (surface average and \nth{99} percentile) of the
            \gls{maxPrincipalCauchyStress} and \gls{maxStretch}
            fields on different surfaces of the \gls{ia} wall, among the set of
            pair-wise comparisons of the different hyperelastic laws used for
            both wall morphology models
            (\cref{eq:absoluteDifferenceHyperelasticLaws}).  The differences
            were averaged over each rupture-status group.
        }

        \sisetup{%
            group-digits=false,
            round-mode=figures
        }
        \small
        \begin{tabular}{
            c
            l
            S[round-precision=3]
            S[round-precision=3]
            S[round-precision=3]
            S[round-precision=3]
            S[round-precision=3]
            S[round-precision=3]
            S[round-precision=3]
            S[round-precision=3]
            S[round-precision=3]
            S[round-precision=3]
        }
            \toprule
                    \multirow{4}{*}{%
                        \parbox{0.4cm}{%
                            \centering
                            \scriptsize
                            Side
                        }
                    }
                &   \multirow{4}{*}{%
                        \parbox{2cm}{%
                            \centering
                            Field Metric, \gls{iaWallMetric}
                        }
                    }
                &   \multicolumn{4}{c}{%
                        $\gsub{genericAbsoluteDifference}{iaWallMetric}^{\textsc{HL}}$
                    }
                &   \multicolumn{6}{c}{%
                    $\gsub{genericAbsoluteDifference}{iaWallMetric}^{\textsc{WM}}$
                    } \\
            \cmidrule(lr){3-6}
            \cmidrule(lr){7-12}
                &
                &   \multicolumn{2}{c}{Abnormal}
                &   \multicolumn{2}{c}{Uniform}
                &   \multicolumn{2}{c}{\glsxtrshort{mr}}
                &   \multicolumn{2}{c}{Yeoh}
                &   \multicolumn{2}{c}{Fung} \\
            \cmidrule(lr){3-4}
            \cmidrule(lr){5-6}
            \cmidrule(lr){7-8}
            \cmidrule(lr){9-10}
            \cmidrule(lr){11-12}
                &
                &   \gls{rupturedGroup} & \gls{unrupturedGroup}
                &   \gls{rupturedGroup} & \gls{unrupturedGroup}
                &   \gls{rupturedGroup} & \gls{unrupturedGroup}
                &   \gls{rupturedGroup} & \gls{unrupturedGroup}
                &   \gls{rupturedGroup} & \gls{unrupturedGroup}  \\
            \midrule
                    \multirow{4}{*}{%
                        \gsupsub{surface}{lumen}{aneurysm}
                    }
                &   \savg{\gls{maxPrincipalCauchyStress}}
                    (\si{\kilo\pascal})
                &    19.0536 &    12.9331 &  19.7834 &    12.6261
                &      20.6107 &     8.2461 &  21.3405 &     8.4856 &       22.3513 &     8.7926 \\
                &   $(\gls{maxPrincipalCauchyStress})_{\gls{percentil99}}$
                    (\si{\kilo\pascal})
                &    59.3431 &    60.5537 &  60.3936 &    61.9273
                &       4.2327 &    15.1203 &   5.2833 &    12.2384 &       10.7635 &    13.7467 \\
            \cmidrule(lr){2-12}
                &   \savg{\gls{maxStretch}}
                    ($\times 10^{-3}$)
                &    43.2804 &    14.1925 &  47.3691 &    14.0591
                &       5.0279 &     1.7363 &   5.7444 &     2.4678 &        9.1166 &     2.6013 \\
                &   $(\gls{maxStretch})_{\gls{percentil99}}$
                    ($\times 10^{-3}$)
                &    73.6272 &    18.8659 &  75.3114 &    19.9087
                &       5.7180 &     1.9019 &   6.2974 &     2.9447 &        7.4023 &     4.1654 \\
            \cmidrule(lr){1-12}
            \cmidrule(lr){2-12}
                    \multirow{4}{*}{%
                        \gsupsub{surface}{ablumen}{aneurysm}
                    }
                &   \savg{\gls{maxPrincipalCauchyStress}}
                    (\si{\kilo\pascal})
                &     3.7949 &     3.4943 &   4.7853 &     4.0638
                &       6.2821 &     6.5107 &   6.7077 &     7.9658 &        7.6981 &     9.3516 \\
                &   $(\gls{maxPrincipalCauchyStress})_{\gls{percentil99}}$
                    (\si{\kilo\pascal})
                &     8.8968 &    55.1615 &   6.7427 &    46.8756
                &       0.1238 &     6.1629 &   1.8555 &    13.1996 &        4.0096 &    14.4488 \\
            \cmidrule(lr){2-12}
                &   \savg{\gls{maxStretch}}
                    ($\times 10^{-3}$)
                &    37.0968 &    12.2291 &  40.3033 &    12.2952
                &       3.0638 &     1.4654 &   4.0682 &     2.2405 &        6.2703 &     2.1743 \\
                &   $(\gls{maxStretch})_{\gls{percentil99}}$
                    ($\times 10^{-3}$)
                &    62.5605 &    14.7802 &  64.9294 &    17.0586
                &       4.4509 &     0.6471 &   5.0477 &     2.9255 &        6.8198 &     3.8671 \\
            \bottomrule
        \end{tabular}
        \\[\baselineskip]
        \label{tab:absolutDifferencesModelings}
\end{table}

    It is important to note that similar plots of
    \cref{fig:boxPlotsSigmaMaxDistPerHyperelasticLaw,%
    fig:boxPlotsStretchMaxDistPerHyperelasticLaw} were inspected for the
    metrics calculated on the abluminal surface and the same trends were found
    (as also indicated by \cref{tab:absolutDifferencesModelings}). This
    suggests that the trends reported here occur on the whole wall of the
    \gls{ia}.

\begin{figure}[!htb]
        \includegraphics[%
            width=\textwidth
        ]   {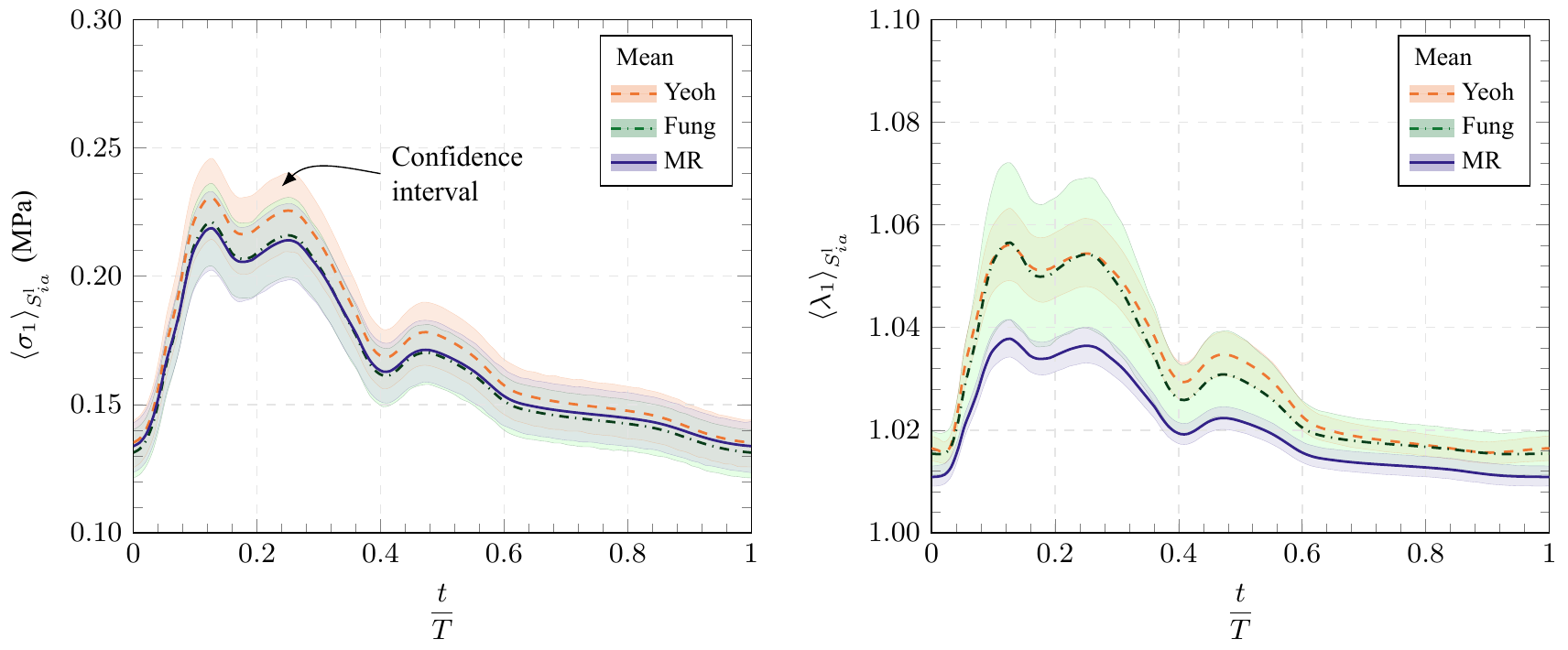}

        \caption{%
            Central tendency plots (mean and confidence interval based on the
            95-percentile) along a cardiac cycle (dimensionless time,
            normalized by the cardiac period \gls{cardiacPeriod}) of
            \savg{\gls{maxPrincipalCauchyStress}} and \savg{\gls{maxStretch}}
            and \gls{maxStretch}, computed over
            \gsupsub{surface}{lumen}{aneurysm}, for the whole \gls{ia} sample
            for the three hyperelastic laws and the \gls{awm}.
        }

        \label{fig:temporalProfiles}
\end{figure}

    Finally, it is important to note that we performed the same analysis
    presented above for the \gls{maxPrincipalCauchyStress} and \gls{maxStretch}
    fields but taken at the instant of low diastole, \gsub{time}{lowDiastole}
    (see \cref{fig:flowAndPressureProfiles}a), and in general the same trends still
    hold for the \gls{maxPrincipalCauchyStress}.  Regarding \gls{maxStretch}'s
    metrics, the same trends above hold too, although the values of the
    absolute differences between the models tend to decrease uniformly. These
    trends can be perceived by inspecting \savg{\gls{maxPrincipalCauchyStress}}
    and \savg{\gls{maxStretch}} along a cardiac cycle in
    \cref{fig:temporalProfiles}, where it is clear that the largest differences
    among the models occur near the peak-systole and any difference tends to
    decrease after the systolic period towards the diastolic period. The large
    differences among the hyperelastic laws during the systole probably occur
    as a manifestation of the nonlinearities of the modeling depending on the
    level of forces driving the motion, which are higher during the systole.

%% file: tex/discussion.tex
\section{Discussion} \label{sec:discussion}

    The use of numerical simulations to study \glspl{ia} has increased
    substantially in the last two decades, although the majority of the works
    were focused on the hemodynamics inside the \gls{ia} \citep{Saqr2019},
    thus primarily using \gls{cfd} as their main computational tool.
    Investigations on the hemodynamics were helpful to a more complete
    understanding of the role played by the blood flow in an \gls{ia}'s natural
    history \citep{Castro2013}. Nevertheless, investigations on the
    mechanical response of \gls{ia} geometries under physiologically realistic
    conditions, which seems to account for a small percentage of all the works
    on the subject, are of foremost importance and can be helpful to further
    understand both their mechanics and, more importantly, the rupture event,
    which is ultimately a phenomenon occurring in the wall tissue. As argued in
    the introduction of this work, the lack of that kind of study is
    understandable due to the complex modeling requirements to realistically
    represent a patient-specific \gls{ia} wall tissue. In this regard, to the
    authors' knowledge, our study is the first attempt to thoroughly assess the
    impact of both hyperelastic laws and wall morphology models --- i.e.
    thickness and material constants --- on the mechanical response of
    \glspl{ia} with a relatively large population's sample.

    We employed a realistic modeling of the wall morphology --- what we named
    the \enquote{\gls{awm}} --- and the technique to create it allows for
    patient-specific prediction of the wall morphology based on the underlying
    hemodynamics. Nonetheless, our results suggest that the \gls{awm} predicts
    a similar mechanical response to a model where the aneurysm wall has
    uniform thickness and material constants, the \gls{uwm} --- a common
    alternative used in previous works when patient-specific data is missing,
    for example ---, with absolute differences smaller than when different
    hyperelastic laws were compared and consistent for the whole cardiac cycle.
    These findings indicate that the \gls{uwm} may be used to find the mechanical
    response in populational studies of \glspl{ia}. Care must be taken, though,
    depending on the particular application of the numerical results because
    \citet{Cebral2015}, for example, using similar modeling to the \gls{awm}
    to investigate the rupture site of \glspl{ia}, found that a wall with both
    uniform thickness and stiffness was not able to predict the rupture point
    specifically, compared to models with focal properties changes.

    Furthermore, according to our results, when numerically simulating the
    geometry of an \gls{ia}, the mechanical response in terms of the stretches,
    \gls{maxStretch}, was more sensitive to the constitutive law chosen and,
    additionally, yielded clear differences between ruptured and unruptured
    \glspl{ia}. In that regard, the comparative behavior among each law is most
    likely to be related to their response given the stiffer
    material properties of the unruptured group. Therefore, as stiffer an
    \gls{ia} tissue is, all laws tend to yield the same, or at least
    similar, mechanical response, in terms of stresses, whereas as less stiff
    it is, i.e.  closer to a ruptured condition, the \gls{mr} deviates from
    both Yeoh and Fung results.

    We found few studies that investigated the impact of different modeling
    choices on the mechanical response of \glspl{ia}. \citet{Torii2008}
    assessed the impact of different materials laws mainly on the hemodynamics,
    using a single \gls{ia} geometry, thus limiting the possibility of
    comparison with our results (regarding the wall motion the authors only
    reported the maximum displacement on the \gls{ia} sac and only used one law
    that we also employed, the Fung law). \citet{Ramachandran2012} also
    directly compared different material laws in patient-specific \gls{ia}
    geometries, although they assumed static \glspl{bc} and simulated only the
    \gls{ia} sac, i.e.  without the branches walls. They employed anisotropic
    and isotropic versions of the Fung law, the Yeoh law, and both small and
    finite strain versions of Hooke's law. Although the material constants they
    have employed were different, their conclusions broadly agree with the
    findings in this work, i.e. that the material laws predicted similar
    responses in terms of the wall stresses. Although, it is important to note
    that their study did not use the \gls{mr} law, the one that presented the
    most divergent response compared to the Yeoh and Fung responses.  The
    authors recognized, though, that their results could not hold as more
    patient-specific modeling features were added to the whole \gls{ia} wall
    model in populational studies by avoiding the \enquote{uniform modeling
    choices applied across the patient population}. The comparison performed
    here between the \gls{uwm} and \gls{awm} suitably addresses their concerns
    by showing that the absolute differences between a uniform wall \gls{ia}
    and a realistic one, in terms of their mechanical response as given by
    \gls{maxPrincipalCauchyStress} and \gls{maxStretch}, are smaller than when
    comparing different material laws.

    From a practical perspective, by assessing the impact of different
    materials laws and morphology models, suitable modeling can be chosen
    depending on the particular goals of the simulations. In large cohort
    studies, for example, simpler models could be used to assess average
    quantities over the \glspl{ia} sacs, while keeping the computational times
    low --- for example, the Yeoh law yielded the fastest results while the
    \gls{uwm} and \gls{awm} yielded similar computational times.

    \subsection*{Limitations}

    There exist some controversy on the \glspl{bc} applied on the artificial
    sections of the vascular branches because it is difficult to predict which
    numerical \gls{bc} can be realistically applied in there.  The majority
    of works we found that have simulated \gls{csd} or \gls{fsi} in vascular
    geometries commonly employed zero-displacements too
    \citep{Lee2013a,Valencia2013}, although a few \citep{Bazilevs2010b} applied
    a \gls{bc} that allows the arterial branch to slide along the section's
    tangential direction, but constrains the displacement along with its normal
    --- hence a \enquote{zero-shear traction} \gls{bc}. We performed a
    numerical study comparing both and found that, although the displacement
    field was affected by different \glspl{bc}, both
    \gls{maxPrincipalCauchyStress} and \gls{maxStretch} were unaltered on the
    aneurysm sacs, with qualitative differences on these fields only near the
    locations of the sections, as expected. Furthermore, we found that the
    sections should be made at least two local diameters away from the
    \glspl{ia} neck lines to safely assume that the zero-displacement \gls{bc}
    would not influence the stress and stretch distribution on the \gls{ia}.
    For all the geometries used here, this length was assured for all the
    branches, including the parent artery.

    Additionally, it is important to note that we only employed isotropic laws,
    despite the tissue of arteries and \glspl{ia} being anisotropic, based on
    evidence that anisotropic laws yield similar mechanical responses compared
    to their isotropic versions \citep{Ramachandran2012}, although it is
    important to further confirm that in larger studies.

    The values by which thickness and material constants were altered in the
    assumptions made for the \gls{awm} (see
    \cref{fig:icaWithAbnormalWallRegions}) were somewhat arbitrary, even if
    based on scarce data available in the literature on this subject. For
    example, we could not find quantitative data on how much thinner the type-I
    patches are compared to normal walls. In atherosclerotic walls, hence
    type-II patches, data gathered by \citet{Holzapfel2003}, based on earlier
    studies, show that arteries with atherosclerotic plaques can be up to
    \num{1.5} times thicker than when they are healthy. Similarly, their
    elastic modulus can be 4 times larger in the later stages of plaque
    development and in the presence of calcification, but much smaller than
    \num{1.5} times in earlier stages. Due to this lack of data, the factors
    that correctly represented the histological observations for each patch
    were selected and used consistently among all geometries. Nevertheless,  we
    carried out a parametric study by incrementally changing these scale
    factors of thickness and material constants separately on type-I and
    type-II patches (see supplementary material). By then computing
    $\savg{\gls{maxPrincipalCauchyStress}}_{\gsupsub{surface}{lumen}{aneurysm}}$
    and $\savg{\gls{maxStretch}}_{\gsupsub{surface}{lumen}{aneurysm}}$, we
    found a mean maximum absolute difference of \SI{20.9}{\kilo\pascal} and
    \num{5.47e-3}, respectively, computed between the extremes of the
    parametric intervals, when the thickness of type-II patches was varied ---
    all the other yield smaller differences. Most importantly, the variation
    induced by focal changes in thickness and material constants were
    consistent among all \glspl{ia}, as expected, therefore our main
    conclusions would likely withstand if other scale factors were chosen.

    Finally, a further limitation was due to the use of the \gls{1wfsi}
    numerical strategy that ignores the two-way coupling between the fluid and
    solid domains. Nevertheless, given the comparative nature of this work, it
    is sufficient to set a baseline modeling for all cases simulated, in such a
    way that a \enquote{full} \gls{2wfsi} modeling would alter the
    \gls{maxPrincipalCauchyStress} and \gls{maxStretch} consistently and, thus
    not altering our main conclusions.

\section{Conclusions} \label{sec:conclusions}

   The challenges involved in the experimental acquisition of mechanical
   properties of patient-specific \glspl{ia} prevent the use of better modeling
   to study this disease, especially if numerical simulations are to be used to
   help in this endeavor. This scenario is particularly pressing because a
   better understanding of the rupture, and eventual prediction of it, passes
   through the capacity to compute the mechanical response of an \gls{ia} wall.
   In this scenario, this study help to assess the likely impact of certain
   modeling choices on the mechanical response of \glspl{ia}.

   We found that different wall morphology models --- i.e. different thickness
   and material properties --- yield smaller absolute differences than when
   comparing different constitutive laws, in terms of both stress and stretches
   of the mechanical response. Furthermore, different hyperelastic laws
   produced significantly different stretch fields, explained by the likely
   higher sensitivity of stretch to the material constants of the law,
   indicating that ruptured aneurysms had much larger stretches than unruptured
   ones. The same behavior was not encountered for the stress fields in the
   \gls{ia} sac. These findings may help future studies to choose more
   suitable modeling to investigate other aspects of the mechanical response of
   \glspl{ia}.

%% file: tex/supplementaryMaterial.tex
\pagestyle{empty}
\counterwithin{equation}{section}
\renewcommand{\thefigure}{S\arabic{figure}}
\setcounter{figure}{0}

\section*{Supplementary Material} \label{sec:supplementaryMaterial}

    \section*{Influence of Local Thickness and Material Constants}
    \label{sec:parametricThicknessAndElasticity}

    Two parametric studies were carried out by changing the scale factors
    applied to build the \gls{awm} by scaling the thickness and material
    constants of type-I and type-II patches. A simpler modeling of the flow was
    applied, though, by using a steady-state blood flow model at peak-systole
    conditions as the driven force of the arterial and \gls{ia} wall motion and
    numerically solving it with the \gls{1wfsi} strategy with the same
    \glspl{bc} presented in main study.  Only the Yeoh hyperelastic
    law was used --- it is reasonable to assume that the same behavior would be
    found for the \gls{mr} and Fung laws ---, with
    $\gls{PoissonRatio} = 0.48$.  The study was carried out in two stages:
    first, by varying \gls{aneurysmThickness}, locally, and, in the sequence,
    changing the fields of the the material constants,
    \materialcoeff{YeohCoeffs}{1}{} and \materialcoeff{YeohCoeffs}{2}{}.

    In the first stage, uniform elasticities over the aneurysm,
    \materialcoeff{YeohCoeffs}{1}{} and \materialcoeff{YeohCoeffs}{2}{}, were
    assumed to isolate the effect of the varying \gls{aneurysmThickness} on
    each patch type, calculated as follows:

    \begin{enumerate}
        \item on type-I patches, \gls{aneurysmThickness} was first kept
            unchanged and then decreased by \num{5}, \num{10}, \num{20}, and
            \SI{30}{\percent}, while keeping the type-II patches
            \SI{20}{\percent} thicker;

        \item on type-II patches, \gls{aneurysmThickness} was first kept
            unchanged and then increased by \num{5}, \num{10}, \num{20}, and
            \SI{30}{\percent}, while keeping the type-I patches
            \SI{5}{\percent} thinner.
    \end{enumerate}

    The computational meshes were created similarly as in the main study, but,
    in this case, for each new thickness field, a new mesh was generated,
    whereas a single fluid computational mesh was used.

    In the second stage, uniform \gls{aneurysmThickness} was assumed over
    \gsupsub{surface}{lumen}{aneurysm} --- thus, a single mesh was used --- to
    isolate the effect of varying the material constants,
    \materialcoeff{YeohCoeffs}{1}{} and \materialcoeff{YeohCoeffs}{2}{}, which
    were both, first, kept uniform and, then, decreased by \num{10}, \num{20},
    and \SI{30}{\percent}, on type-I and type-II patches at the same time.

\begin{figure}[!htb]
            \includegraphics[%
                width=0.8\textwidth
            ]   {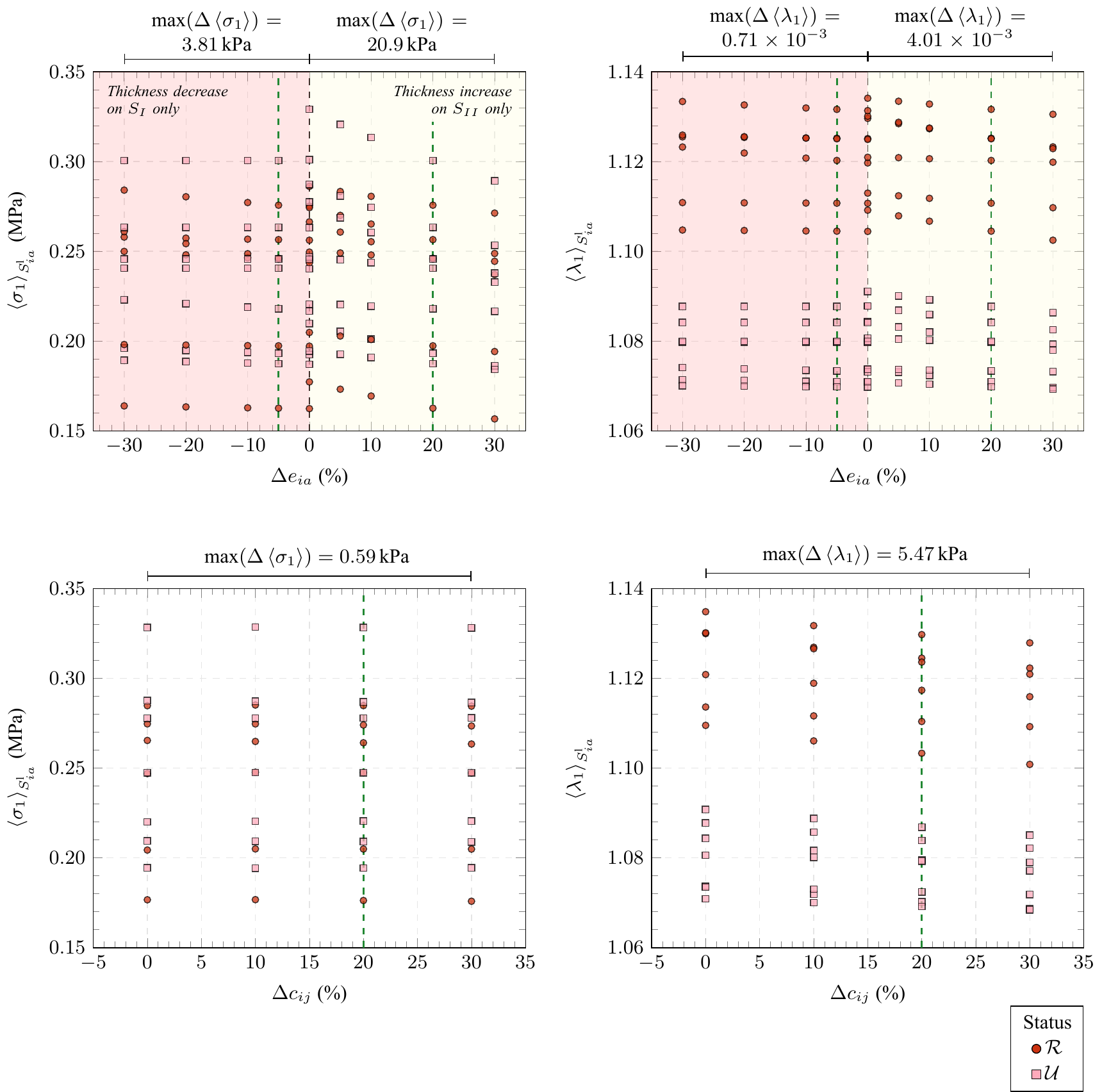}

            \caption{
                Surface-averaged \gls{maxPrincipalCauchyStress} (left column)
                and \gls{maxStretch} (right column) over
                \gsupsub{surface}{lumen}{aneurysm}, grouped by rupture status,
                versus the percentual variation of \gls{aneurysmThickness}
                (first row) and material constants,
                $\gls{YeohCoeffs}_{10}$ and $\gls{YeohCoeffs}_{20}$,
                (second row) applied separately on type-I and type-II patches.
                Above each plots is annotated the mean maximum absolute
                difference among all \glspl{ia}.
            }

            \label{fig:paramMechVarsVsWallMorphology}
    \end{figure}

    Note that the parametric intervals employed
    ($\interval{\SI{5}{\percent}}{\SI{30}{\percent}}$) included the actual
    values selected for the abnormal wall morphology model. The
    surface-averages of \gls{maxPrincipalCauchyStress} and \gls{maxStretch}
    were computed over \gsupsub{surface}{lumen}{aneurysm} at the deformed
    configuration. The results are shown in
    \cref{fig:paramMechVarsVsWallMorphology}.

    The plots show how
    $\savg{\gls{maxPrincipalCauchyStress}}_{\gsupsub{surface}{lumen}{aneurysm}}$
    and $\savg{\gls{maxStretch}}_{\gsupsub{surface}{lumen}{aneurysm}}$ changed
    with both local changes in \gls{aneurysmThickness} and
    \materialcoeff{YeohCoeffs}{i}{j}. The absolute difference between the
    extremes of the parametric intervals was computed for each \gls{ia} case,
    and its average annotated above each plot.

%% file: main.bbl
\begin{thebibliography}{69}
\providecommand{\natexlab}[1]{#1}
\providecommand{\url}[1]{\texttt{#1}}
\providecommand{\urlprefix}{URL }
\expandafter\ifx\csname urlstyle\endcsname\relax
  \providecommand{\doi}[1]{doi:\discretionary{}{}{}#1}\else
  \providecommand{\doi}[1]{doi:\discretionary{}{}{}\begingroup
  \urlstyle{rm}\url{#1}\endgroup}\fi
\providecommand{\bibinfo}[2]{#2}

\bibitem[{Diagbouga et~al.(2018)Diagbouga, Morel, Bijlenga, and
  Kwak}]{Diagbouga2018}
\bibinfo{author}{M.~R. Diagbouga}, \bibinfo{author}{S.~Morel},
  \bibinfo{author}{P.~Bijlenga}, \bibinfo{author}{B.~R. Kwak},
  \bibinfo{title}{Role of Hemodynamics in Initiation/Growth of Intracranial
  Aneurysms}, \bibinfo{journal}{European Journal of Clinical Investigation}
  \bibinfo{volume}{48} (\bibinfo{year}{2018}) \bibinfo{pages}{1--12},
  \doi{\bibinfo{doi}{10.1111/eci.12992}}.

\bibitem[{{The International Study of Unruptured Intracranial Aneurysms
  Investigators}(1998)}]{isuia1998}
\bibinfo{author}{{The International Study of Unruptured Intracranial Aneurysms
  Investigators}}, \bibinfo{title}{Unruptured Intracranial Aneurysms - Risk of
  Rupture and Risks of Surgical Intervention}, \bibinfo{journal}{The New
  England Journal of Medicine} \bibinfo{volume}{339}~(\bibinfo{number}{24})
  (\bibinfo{year}{1998}) \bibinfo{pages}{1725--1733}.

\bibitem[{Vlak et~al.(2013)Vlak, Rinkel, Greebe, and Algra}]{Vlak2013}
\bibinfo{author}{M.~H.~M. Vlak}, \bibinfo{author}{G.~J.~E. Rinkel},
  \bibinfo{author}{P.~Greebe}, \bibinfo{author}{A.~Algra}, \bibinfo{title}{Risk
  of Rupture of an Intracranial Aneurysm Based on Patient Characteristics:
  {{A}} Case-Control Study}, \bibinfo{journal}{Stroke}
  \bibinfo{volume}{44}~(\bibinfo{number}{5}) (\bibinfo{year}{2013})
  \bibinfo{pages}{1256--1259}, ISSN \bibinfo{issn}{00392499},
  \doi{\bibinfo{doi}{10.1161/STROKEAHA.111.000679}}.

\bibitem[{Saqr et~al.(2019)Saqr, Rashad, Tupin, Niizuma, Hassan, Tominaga, and
  Ohta}]{Saqr2019}
\bibinfo{author}{K.~M. Saqr}, \bibinfo{author}{S.~Rashad},
  \bibinfo{author}{S.~Tupin}, \bibinfo{author}{K.~Niizuma},
  \bibinfo{author}{T.~Hassan}, \bibinfo{author}{T.~Tominaga},
  \bibinfo{author}{M.~Ohta}, \bibinfo{title}{What Does Computational Fluid
  Dynamics Tell Us about Intracranial Aneurysms? {{A}} Meta-Analysis and
  Critical Review}, \bibinfo{journal}{Journal of Cerebral Blood Flow and
  Metabolism} \bibinfo{volume}{0}~(\bibinfo{number}{0}) (\bibinfo{year}{2019})
  \bibinfo{pages}{1--19}, \doi{\bibinfo{doi}{10.1177/0271678X19854640}}.

\bibitem[{Meng et~al.(2007)Meng, Wang, Hoi, Gao, Metaxa, Swartz, Kolega,
  Swartz, and Kolega}]{Meng2007a}
\bibinfo{author}{H.~Meng}, \bibinfo{author}{Z.~Wang}, \bibinfo{author}{Y.~Hoi},
  \bibinfo{author}{L.~Gao}, \bibinfo{author}{E.~Metaxa}, \bibinfo{author}{D.~D.
  Swartz}, \bibinfo{author}{J.~Kolega}, \bibinfo{author}{D.~D. Swartz},
  \bibinfo{author}{J.~Kolega}, \bibinfo{title}{Complex {{Hemodynamics}} at the
  {{Apex}} of an {{Arterial Bifurcation Induces Vascular Remodeling Resembling
  Cerebral Aneurysm Initiation}}}, \bibinfo{journal}{Stroke}
  \bibinfo{volume}{38} (\bibinfo{year}{2007}) \bibinfo{pages}{1924--1931},
  \doi{\bibinfo{doi}{10.1161/STROKEAHA.106.481234}}.

\bibitem[{Metaxa et~al.(2010)Metaxa, Tremmel, Natarajan, Xiang, Paluch,
  Mandelbaum, Siddiqui, Kolega, Mocco, and Meng}]{Metaxa2010}
\bibinfo{author}{E.~Metaxa}, \bibinfo{author}{M.~Tremmel},
  \bibinfo{author}{S.~K. Natarajan}, \bibinfo{author}{J.~Xiang},
  \bibinfo{author}{R.~A. Paluch}, \bibinfo{author}{M.~Mandelbaum},
  \bibinfo{author}{A.~H. Siddiqui}, \bibinfo{author}{J.~Kolega},
  \bibinfo{author}{J.~Mocco}, \bibinfo{author}{H.~Meng},
  \bibinfo{title}{Characterization of Critical Hemodynamics Contributing to
  Aneurysmal Remodeling at the Basilar Terminus in a Rabbit Model},
  \bibinfo{journal}{Stroke} \bibinfo{volume}{41}~(\bibinfo{number}{8})
  (\bibinfo{year}{2010}) \bibinfo{pages}{1774--1782}, ISSN
  \bibinfo{issn}{00392499}, \doi{\bibinfo{doi}{10.1161/STROKEAHA.110.585992}}.

\bibitem[{Liang et~al.(2019)Liang, Steinman, Brina, Chnafa, Cancelliere, and
  Pereira}]{Liang2019}
\bibinfo{author}{L.~Liang}, \bibinfo{author}{D.~A. Steinman},
  \bibinfo{author}{O.~Brina}, \bibinfo{author}{C.~Chnafa},
  \bibinfo{author}{N.~M. Cancelliere}, \bibinfo{author}{V.~M. Pereira},
  \bibinfo{title}{Towards the {{Clinical}} Utility of {{CFD}} for Assessment of
  Intracranial Aneurysm Rupture - {{A}} Systematic Review and Novel
  Parameter-Ranking Tool}, \bibinfo{journal}{Journal of NeuroInterventional
  Surgery} \bibinfo{volume}{11} (\bibinfo{year}{2019})
  \bibinfo{pages}{153--158}, ISSN \bibinfo{issn}{17598486},
  \doi{\bibinfo{doi}{10.1136/neurintsurg-2018-014246}}.

\bibitem[{Kallmes(2012)}]{Kallmes2012}
\bibinfo{author}{D.~F. Kallmes}, \bibinfo{title}{Point:
  {{CFD}}\textemdash{{Computational Fluid Dynamics}} or {{Confounding Factor
  Dissemination}}}, \bibinfo{journal}{American Journal of Neuroradiology}
  \bibinfo{volume}{33} (\bibinfo{year}{2012}) \bibinfo{pages}{393--398},
  \doi{\bibinfo{doi}{10.3174/ajnr.A2993}}.

\bibitem[{Cebral and Meng(2012)}]{Cebral2012}
\bibinfo{author}{J.~R. Cebral}, \bibinfo{author}{H.~Meng},
  \bibinfo{title}{Counterpoint: {{Realizing}} the {{Clinical Utility}} of
  {{Computational Fluid Dynamics}}\textemdash{{Closing}} the {{Gap}}},
  \doi{\bibinfo{doi}{10.3174/ajnr.a2993}}, \bibinfo{year}{2012}.

\bibitem[{Bazilevs et~al.(2010{\natexlab{a}})Bazilevs, Hsu, Zhang, Wang,
  Kvamsdal, Hentschel, and Isaksen}]{Bazilevs2010b}
\bibinfo{author}{Y.~Bazilevs}, \bibinfo{author}{M.~C. Hsu},
  \bibinfo{author}{Y.~Zhang}, \bibinfo{author}{W.~Wang},
  \bibinfo{author}{T.~Kvamsdal}, \bibinfo{author}{S.~Hentschel},
  \bibinfo{author}{J.~G. Isaksen}, \bibinfo{title}{Computational Vascular
  Fluid-Structure Interaction: {{Methodology}} and Application to Cerebral
  Aneurysms}, \bibinfo{journal}{Biomechanics and Modeling in Mechanobiology}
  \bibinfo{volume}{9} (\bibinfo{year}{2010}{\natexlab{a}})
  \bibinfo{pages}{481--498}, ISSN \bibinfo{issn}{16177959},
  \doi{\bibinfo{doi}{10.1007/s10237-010-0189-7}}.

\bibitem[{Lee et~al.(2013{\natexlab{a}})Lee, Zhang, Takao, Murayama, and
  Qian}]{Lee2013a}
\bibinfo{author}{C.~J. Lee}, \bibinfo{author}{Y.~Zhang},
  \bibinfo{author}{H.~Takao}, \bibinfo{author}{Y.~Murayama},
  \bibinfo{author}{Y.~Qian}, \bibinfo{title}{The Influence of Elastic Upstream
  Artery Length on Fluid-Structure Interaction Modeling: {{A}} Comparative
  Study Using Patient-Specific Cerebral Aneurysm}, \bibinfo{journal}{Medical
  Engineering and Physics} \bibinfo{volume}{35}~(\bibinfo{number}{9})
  (\bibinfo{year}{2013}{\natexlab{a}}) \bibinfo{pages}{1377--1384}, ISSN
  \bibinfo{issn}{13504533},
  \doi{\bibinfo{doi}{10.1016/j.medengphy.2013.03.009}}.

\bibitem[{Causin et~al.(2005)Causin, Gerbeau, and Nobile}]{Causin2005}
\bibinfo{author}{P.~Causin}, \bibinfo{author}{J.~F. Gerbeau},
  \bibinfo{author}{F.~Nobile}, \bibinfo{title}{Added-Mass Effect in the Design
  of Partitioned Algorithms for Fluid-Structure Problems},
  \bibinfo{journal}{Computational Methods in Applied Mechanical Engineering}
  \bibinfo{volume}{194}~(\bibinfo{number}{42-44}) (\bibinfo{year}{2005})
  \bibinfo{pages}{4506--4527}, ISSN \bibinfo{issn}{00457825},
  \doi{\bibinfo{doi}{10.1016/j.cma.2004.12.005}}.

\bibitem[{F{\"o}rster et~al.(2007)F{\"o}rster, Wall, and Ramm}]{Forster2007}
\bibinfo{author}{C.~F{\"o}rster}, \bibinfo{author}{W.~A. Wall},
  \bibinfo{author}{E.~Ramm}, \bibinfo{title}{Artificial Added Mass
  Instabilities in Sequential Staggered Coupling of Nonlinear Structures and
  Incompressible Viscous Flows}, \bibinfo{journal}{Computer Methods in Applied
  Mechanics and Engineering} \bibinfo{volume}{196}~(\bibinfo{number}{7})
  (\bibinfo{year}{2007}) \bibinfo{pages}{1278--1293}, ISSN
  \bibinfo{issn}{00457825}, \doi{\bibinfo{doi}{10.1016/j.cma.2006.09.002}}.

\bibitem[{Kadasi et~al.(2013)Kadasi, Dent, and Malek}]{Kadasi2013}
\bibinfo{author}{L.~M. Kadasi}, \bibinfo{author}{W.~C. Dent},
  \bibinfo{author}{A.~M. Malek}, \bibinfo{title}{Colocalization of Thin-Walled
  Dome Regions with Low Hemodynamic Wall Shear Stress in Unruptured Cerebral
  Aneurysms}, \bibinfo{journal}{Journal of Neurosurgery}
  \bibinfo{volume}{119}~(\bibinfo{number}{1}) (\bibinfo{year}{2013})
  \bibinfo{pages}{172--179}, ISSN \bibinfo{issn}{0022-3085},
  \doi{\bibinfo{doi}{10.3171/2013.2.jns12968}}.

\bibitem[{Signorelli et~al.(2018)Signorelli, {Pailler-Mattei}, Gory, Larquet,
  Robinson, Vargiolu, Zahouani, Labeyrie, Guyotat, {Pelissou-Guyotat},
  Berthiller, and Turjman}]{Signorelli2018}
\bibinfo{author}{F.~Signorelli}, \bibinfo{author}{C.~{Pailler-Mattei}},
  \bibinfo{author}{B.~Gory}, \bibinfo{author}{P.~Larquet},
  \bibinfo{author}{P.~Robinson}, \bibinfo{author}{R.~Vargiolu},
  \bibinfo{author}{H.~Zahouani}, \bibinfo{author}{P.-E. Labeyrie},
  \bibinfo{author}{J.~Guyotat}, \bibinfo{author}{I.~{Pelissou-Guyotat}},
  \bibinfo{author}{J.~Berthiller}, \bibinfo{author}{F.~Turjman},
  \bibinfo{title}{Biomechanical {{Characterization}} of {{Intracranial Aneurysm
  Wall}}: {{A Multiscale Study}}}, \bibinfo{journal}{World Neurosurgery}
  \bibinfo{volume}{119} (\bibinfo{year}{2018}) \bibinfo{pages}{e882--e889},
  ISSN \bibinfo{issn}{18788750},
  \doi{\bibinfo{doi}{10.1016/j.wneu.2018.07.290}}.

\bibitem[{Fr{\"o}sen et~al.(2019)Fr{\"o}sen, Cebral, Robertson, and
  Aoki}]{Frosen2019}
\bibinfo{author}{J.~Fr{\"o}sen}, \bibinfo{author}{J.~Cebral},
  \bibinfo{author}{A.~M. Robertson}, \bibinfo{author}{T.~Aoki},
  \bibinfo{title}{Flow-Induced, Inflammation-Mediated Arterial Wall Remodeling
  in the Formation and Progression of Intracranial Aneurysms},
  \bibinfo{journal}{Neurosurgical Focus}
  \bibinfo{volume}{47}~(\bibinfo{number}{1}) (\bibinfo{year}{2019})
  \bibinfo{pages}{E21}, ISSN \bibinfo{issn}{1092-0684},
  \doi{\bibinfo{doi}{10.3171/2019.5.FOCUS19234}}.

\bibitem[{Soldozy et~al.(2019)Soldozy, Norat, Elsarrag, Chatrath, Costello,
  Sokolowski, Tvrdik, Kalani, and Park}]{Soldozy2019}
\bibinfo{author}{S.~Soldozy}, \bibinfo{author}{P.~Norat},
  \bibinfo{author}{M.~Elsarrag}, \bibinfo{author}{A.~Chatrath},
  \bibinfo{author}{J.~S. Costello}, \bibinfo{author}{J.~D. Sokolowski},
  \bibinfo{author}{P.~Tvrdik}, \bibinfo{author}{M.~Y.~S. Kalani},
  \bibinfo{author}{M.~S. Park}, \bibinfo{title}{The Biophysical Role of
  Hemodynamics in the Pathogenesis of Cerebral Aneurysm Formation and Rupture},
  \bibinfo{journal}{Neurosurgical Focus}
  \bibinfo{volume}{47}~(\bibinfo{number}{1}) (\bibinfo{year}{2019})
  \bibinfo{pages}{1--9}, \doi{\bibinfo{doi}{10.3171/2019.4.focus19232}}.

\bibitem[{Meng et~al.(2014)Meng, Tutino, Xiang, and Siddiqui}]{Meng2014}
\bibinfo{author}{H.~Meng}, \bibinfo{author}{V.~M. Tutino},
  \bibinfo{author}{J.~Xiang}, \bibinfo{author}{A.~Siddiqui},
  \bibinfo{title}{High {{WSS}} or {{Low WSS}}? {{Complex}} Interactions of
  Hemodynamics with Intracranial Aneurysm Initiation, Growth, and Rupture:
  {{Toward}} a Unifying Hypothesis}, \bibinfo{journal}{American Journal of
  Neuroradiology} \bibinfo{volume}{35}~(\bibinfo{number}{7})
  (\bibinfo{year}{2014}) \bibinfo{pages}{1254--1262}, ISSN
  \bibinfo{issn}{1936959X}, \doi{\bibinfo{doi}{10.3174/ajnr.A3558}}.

\bibitem[{Humphrey and Canham(2000)}]{Humphrey2000}
\bibinfo{author}{J.~D. Humphrey}, \bibinfo{author}{P.~B. Canham},
  \bibinfo{title}{Structure, {{Mechanical Properties}}, and {{Mechanics}} of
  {{Intracranial Saccular Aneurysms}}}, \bibinfo{journal}{Journal of
  Elasticity} \bibinfo{volume}{61} (\bibinfo{year}{2000})
  \bibinfo{pages}{49--81}.

\bibitem[{Parshin et~al.(2019)Parshin, Lipovka, Yunoshev, Ovsyannikov, Dubovoy,
  and Chupakhin}]{Parshin2019}
\bibinfo{author}{D.~V. Parshin}, \bibinfo{author}{A.~I. Lipovka},
  \bibinfo{author}{A.~S. Yunoshev}, \bibinfo{author}{K.~S. Ovsyannikov},
  \bibinfo{author}{A.~V. Dubovoy}, \bibinfo{author}{A.~P. Chupakhin},
  \bibinfo{title}{On the Optimal Choice of a Hyperelastic Model of Ruptured and
  Unruptured Cerebral Aneurysm}, \bibinfo{journal}{Scientific Reports}
  \bibinfo{volume}{9}~(\bibinfo{number}{1}) (\bibinfo{year}{2019})
  \bibinfo{pages}{15865}, ISSN \bibinfo{issn}{2045-2322},
  \doi{\bibinfo{doi}{10.1038/s41598-019-52229-y}}.

\bibitem[{Holzapfel et~al.(2010)Holzapfel, Ogden, and Olzapfel}]{Holzapfel2010}
\bibinfo{author}{G.~A. Holzapfel}, \bibinfo{author}{R.~W. Ogden},
  \bibinfo{author}{B.~Y. G. E. A.~H. Olzapfel}, \bibinfo{title}{Constitutive
  Modelling of Arteries}, \bibinfo{journal}{Proceedings of the Royal Society A:
  Mathematical, Physical and Engineering Sciences}
  \bibinfo{volume}{466}~(\bibinfo{number}{2118}) (\bibinfo{year}{2010})
  \bibinfo{pages}{1551--1597}, ISSN \bibinfo{issn}{1364-5021},
  \doi{\bibinfo{doi}{10.1098/rspa.2010.0058}}.

\bibitem[{Costalat et~al.(2011)Costalat, Sanchez, Ambard, Thines, Lonjon,
  Nicoud, Brunel, Lejeune, Dufour, Bouillot, Lhaldky, Kouri, Segnarbieux,
  Maurage, Lobotesis, {Villa-Uriol}, Zhang, Frangi, Mercier, Bonaf{\'e}, Sarry,
  and Jourdan}]{Costalat2011}
\bibinfo{author}{V.~Costalat}, \bibinfo{author}{M.~Sanchez},
  \bibinfo{author}{D.~Ambard}, \bibinfo{author}{L.~Thines},
  \bibinfo{author}{N.~Lonjon}, \bibinfo{author}{F.~Nicoud},
  \bibinfo{author}{H.~Brunel}, \bibinfo{author}{J.~P. Lejeune},
  \bibinfo{author}{H.~Dufour}, \bibinfo{author}{P.~Bouillot},
  \bibinfo{author}{J.~P. Lhaldky}, \bibinfo{author}{K.~Kouri},
  \bibinfo{author}{F.~Segnarbieux}, \bibinfo{author}{C.~A. Maurage},
  \bibinfo{author}{K.~Lobotesis}, \bibinfo{author}{M.~C. {Villa-Uriol}},
  \bibinfo{author}{C.~Zhang}, \bibinfo{author}{A.~F. Frangi},
  \bibinfo{author}{G.~Mercier}, \bibinfo{author}{A.~Bonaf{\'e}},
  \bibinfo{author}{L.~Sarry}, \bibinfo{author}{F.~Jourdan},
  \bibinfo{title}{Biomechanical Wall Properties of Human Intracranial Aneurysms
  Resected Following Surgical Clipping ({{IRRAs Project}})},
  \bibinfo{journal}{Journal of Biomechanics}
  \bibinfo{volume}{44}~(\bibinfo{number}{15}) (\bibinfo{year}{2011})
  \bibinfo{pages}{2685--2691},
  \doi{\bibinfo{doi}{10.1016/j.jbiomech.2011.07.026}}.

\bibitem[{Brunel et~al.(2018)Brunel, Ambard, Dufour, Roche, Costalat, and
  Jourdan}]{Brunel2018}
\bibinfo{author}{H.~Brunel}, \bibinfo{author}{D.~Ambard},
  \bibinfo{author}{H.~Dufour}, \bibinfo{author}{P.~Roche},
  \bibinfo{author}{V.~Costalat}, \bibinfo{author}{F.~Jourdan},
  \bibinfo{title}{Rupture Limit Evaluation of Human Cerebral Aneurysms Wall:
  {{Experimental}} Study}, \bibinfo{journal}{Journal of Biomechanics}
  \bibinfo{volume}{77} (\bibinfo{year}{2018}) \bibinfo{pages}{76--82}, ISSN
  \bibinfo{issn}{00219290},
  \doi{\bibinfo{doi}{10.1016/j.jbiomech.2018.06.016}}.

\bibitem[{Robertson et~al.(2015)Robertson, Duan, Aziz, Hill, Watkins, and
  Cebral}]{Robertson2015}
\bibinfo{author}{A.~M. Robertson}, \bibinfo{author}{X.~Duan},
  \bibinfo{author}{K.~M. Aziz}, \bibinfo{author}{M.~R. Hill},
  \bibinfo{author}{S.~C. Watkins}, \bibinfo{author}{J.~R. Cebral},
  \bibinfo{title}{Diversity in the {{Strength}} and {{Structure}} of
  {{Unruptured Cerebral Aneurysms}}}, \bibinfo{journal}{Annals of Biomedical
  Engineering} \bibinfo{volume}{43}~(\bibinfo{number}{7})
  (\bibinfo{year}{2015}) \bibinfo{pages}{1502--1515},
  \doi{\bibinfo{doi}{10.1007/s10439-015-1252-4}}.

\bibitem[{Kleinloog et~al.(2014)Kleinloog, Korkmaz, Zwanenburg, Kuijf, Visser,
  Blankena, Post, Ruigrok, Luijten, Regli, Rinkel, and Verweij}]{Kleinloog2014}
\bibinfo{author}{R.~Kleinloog}, \bibinfo{author}{E.~Korkmaz},
  \bibinfo{author}{J.~J.~M. Zwanenburg}, \bibinfo{author}{H.~J. Kuijf},
  \bibinfo{author}{F.~Visser}, \bibinfo{author}{R.~Blankena},
  \bibinfo{author}{J.~A. Post}, \bibinfo{author}{Y.~M. Ruigrok},
  \bibinfo{author}{P.~R. Luijten}, \bibinfo{author}{L.~Regli},
  \bibinfo{author}{G.~J.~E. Rinkel}, \bibinfo{author}{B.~H. Verweij},
  \bibinfo{title}{Visualization of the Aneurysm Wall: A 7.0-Tesla Magnetic
  Resonance Imaging Study}, \bibinfo{journal}{Neurosurgery}
  \bibinfo{volume}{75}~(\bibinfo{number}{December}) (\bibinfo{year}{2014})
  \bibinfo{pages}{614--622}, \doi{\bibinfo{doi}{10.1227/NEU.0000000000000559}}.

\bibitem[{Cebral et~al.(2019)Cebral, Detmer, Chung, {Choque-Velasquez}, Rezai,
  Lehto, Tulamo, Hernesniemi, Niemela, Yu, Williamson, Aziz, Sakur,
  {Amin-Hanjani}, Charbel, Tobe, Robertson, and Fr{\"o}sen}]{Cebral2019}
\bibinfo{author}{J.~R. Cebral}, \bibinfo{author}{F.~Detmer},
  \bibinfo{author}{B.~J. Chung}, \bibinfo{author}{J.~{Choque-Velasquez}},
  \bibinfo{author}{B.~Rezai}, \bibinfo{author}{H.~Lehto},
  \bibinfo{author}{R.~Tulamo}, \bibinfo{author}{J.~Hernesniemi},
  \bibinfo{author}{M.~Niemela}, \bibinfo{author}{A.~Yu},
  \bibinfo{author}{R.~Williamson}, \bibinfo{author}{K.~Aziz},
  \bibinfo{author}{S.~Sakur}, \bibinfo{author}{S.~{Amin-Hanjani}},
  \bibinfo{author}{F.~Charbel}, \bibinfo{author}{Y.~Tobe},
  \bibinfo{author}{A.~Robertson}, \bibinfo{author}{J.~Fr{\"o}sen},
  \bibinfo{title}{Local Hemodynamic Conditions Associated with Focal Changes in
  the Intracranial Aneurysm Wall}, \bibinfo{journal}{American Journal of
  Neuroradiology} \bibinfo{volume}{40}~(\bibinfo{number}{3})
  (\bibinfo{year}{2019}) \bibinfo{pages}{510--516}, ISSN
  \bibinfo{issn}{1936959X}, \doi{\bibinfo{doi}{10.3174/ajnr.A5970}}.

\bibitem[{Torii et~al.(2007)Torii, Oshima, Kobayashi, Takagi, and
  Tezduyar}]{Torii2007}
\bibinfo{author}{R.~Torii}, \bibinfo{author}{M.~Oshima},
  \bibinfo{author}{T.~Kobayashi}, \bibinfo{author}{K.~Takagi},
  \bibinfo{author}{T.~E. Tezduyar}, \bibinfo{title}{Influence of Wall
  Elasticity in Patient-Specific Hemodynamic Simulations},
  \bibinfo{journal}{Computers and Fluids} \bibinfo{volume}{36}
  (\bibinfo{year}{2007}) \bibinfo{pages}{160--168}, ISSN
  \bibinfo{issn}{00457930},
  \doi{\bibinfo{doi}{10.1016/j.compfluid.2005.07.014}}.

\bibitem[{Torii et~al.(2008)Torii, Oshima, Kobayashi, Takagi, and
  Tezduyar}]{Torii2008}
\bibinfo{author}{R.~Torii}, \bibinfo{author}{M.~Oshima},
  \bibinfo{author}{T.~Kobayashi}, \bibinfo{author}{K.~Takagi},
  \bibinfo{author}{T.~E. Tezduyar}, \bibinfo{title}{Fluid-Structure Interaction
  Modeling of a Patient-Specific Cerebral Aneurysm: {{Influence}} of Structural
  Modeling}, \bibinfo{journal}{Computational Mechanics} \bibinfo{volume}{43}
  (\bibinfo{year}{2008}) \bibinfo{pages}{151--159}, ISSN
  \bibinfo{issn}{01787675}, \doi{\bibinfo{doi}{10.1007/s00466-008-0325-8}}.

\bibitem[{Lee et~al.(2013{\natexlab{b}})Lee, Zhang, Takao, Murayama, and
  Qian}]{Lee2013b}
\bibinfo{author}{C.~J. Lee}, \bibinfo{author}{Y.~Zhang},
  \bibinfo{author}{H.~Takao}, \bibinfo{author}{Y.~Murayama},
  \bibinfo{author}{Y.~Qian}, \bibinfo{title}{A Fluid-Structure Interaction
  Study Using Patient-Specific Ruptured and Unruptured Aneurysm: {{The}} Effect
  of Aneurysm Morphology, Hypertension and Elasticity},
  \bibinfo{journal}{Journal of Biomechanics}
  \bibinfo{volume}{46}~(\bibinfo{number}{14})
  (\bibinfo{year}{2013}{\natexlab{b}}) \bibinfo{pages}{2402--2410}, ISSN
  \bibinfo{issn}{00219290},
  \doi{\bibinfo{doi}{10.1016/j.jbiomech.2013.07.016}}.

\bibitem[{Valencia et~al.(2009)Valencia, Mu{\~n}oz, Arayaa, Rivera, and
  Bravo}]{Valencia2009}
\bibinfo{author}{A.~Valencia}, \bibinfo{author}{F.~Mu{\~n}oz},
  \bibinfo{author}{S.~Arayaa}, \bibinfo{author}{R.~Rivera},
  \bibinfo{author}{E.~Bravo}, \bibinfo{title}{Comparison between Computational
  Fluid Dynamics, Fluid-Structure Interaction and Computational Structural
  Dynamics Predictions of Flow-Induced Wall Mechanics in an Anatomically
  Realistic Cerebral Aneurysm Model}, \bibinfo{journal}{International Journal
  of Computational Fluid Dynamics} \bibinfo{volume}{23}~(\bibinfo{number}{9})
  (\bibinfo{year}{2009}) \bibinfo{pages}{649--666}, ISSN
  \bibinfo{issn}{10618562}, \doi{\bibinfo{doi}{10.1080/10618560903476386}}.

\bibitem[{Sanchez et~al.(2013)Sanchez, Ambard, Costalat, Mendez, Jourdan, and
  Nicoud}]{Sanchez2013}
\bibinfo{author}{M.~Sanchez}, \bibinfo{author}{D.~Ambard},
  \bibinfo{author}{V.~Costalat}, \bibinfo{author}{S.~Mendez},
  \bibinfo{author}{F.~Jourdan}, \bibinfo{author}{F.~Nicoud},
  \bibinfo{title}{Biomechanical Assessment of the Individual Risk of Rupture of
  Cerebral Aneurysms: {{A}} Proof of Concept}, \bibinfo{journal}{Annals of
  Biomedical Engineering} \bibinfo{volume}{41}~(\bibinfo{number}{1})
  (\bibinfo{year}{2013}) \bibinfo{pages}{28--40}, ISSN
  \bibinfo{issn}{00906964}, \doi{\bibinfo{doi}{10.1007/s10439-012-0632-2}}.

\bibitem[{Vo{\ss} et~al.(2016)Vo{\ss}, Gla{\ss}er, Hoffmann, Beuing, Weigand,
  Jachau, Preim, Th{\'e}venin, Janiga, and Berg}]{Voss2016}
\bibinfo{author}{S.~Vo{\ss}}, \bibinfo{author}{S.~Gla{\ss}er},
  \bibinfo{author}{T.~Hoffmann}, \bibinfo{author}{O.~Beuing},
  \bibinfo{author}{S.~Weigand}, \bibinfo{author}{K.~Jachau},
  \bibinfo{author}{B.~Preim}, \bibinfo{author}{D.~Th{\'e}venin},
  \bibinfo{author}{G.~Janiga}, \bibinfo{author}{P.~Berg},
  \bibinfo{title}{Fluid-{{Structure Simulations}} of a {{Ruptured Intracranial
  Aneurysm}}: {{Constant}} versus {{Patient-Specific Wall Thickness}}},
  \bibinfo{journal}{Computational and Mathematical Methods in Medicine}
  \bibinfo{volume}{2016}, \doi{\bibinfo{doi}{10.1155/2016/9854539}}.

\bibitem[{Torii et~al.(2006)Torii, Oshima, Kobayashi, Takagi, and
  Tezduyar}]{Torii2006}
\bibinfo{author}{R.~Torii}, \bibinfo{author}{M.~Oshima},
  \bibinfo{author}{T.~Kobayashi}, \bibinfo{author}{K.~Takagi},
  \bibinfo{author}{T.~E. Tezduyar}, \bibinfo{title}{Fluid-Structure Interaction
  Modeling of Aneurysmal Conditions with High and Normal Blood Pressures},
  \bibinfo{journal}{Computational Mechanics} \bibinfo{volume}{38}
  (\bibinfo{year}{2006}) \bibinfo{pages}{482--490}, ISSN
  \bibinfo{issn}{01787675}, \doi{\bibinfo{doi}{10.1007/s00466-006-0065-6}}.

\bibitem[{Cho et~al.(2020)Cho, Yang, Kim, Oh, and Kim}]{Cho2020}
\bibinfo{author}{K.-C. Cho}, \bibinfo{author}{H.~Yang}, \bibinfo{author}{J.-J.
  Kim}, \bibinfo{author}{J.~H. Oh}, \bibinfo{author}{Y.~B. Kim},
  \bibinfo{title}{Prediction of Rupture Risk in Cerebral Aneurysms by Comparing
  Clinical Cases with Fluid\textendash Structure Interaction Analyses},
  \bibinfo{journal}{Scientific Reports}
  \bibinfo{volume}{10}~(\bibinfo{number}{1}) (\bibinfo{year}{2020})
  \bibinfo{pages}{18237}, ISSN \bibinfo{issn}{2045-2322},
  \doi{\bibinfo{doi}{10.1038/s41598-020-75362-5}}.

\bibitem[{Bazilevs et~al.(2010{\natexlab{b}})Bazilevs, Hsu, Zhang, Wang, Liang,
  Kvamsdal, Brekken, and Isaksen}]{Bazilevs2010a}
\bibinfo{author}{Y.~Bazilevs}, \bibinfo{author}{M.~C. Hsu},
  \bibinfo{author}{Y.~Zhang}, \bibinfo{author}{W.~Wang},
  \bibinfo{author}{X.~Liang}, \bibinfo{author}{T.~Kvamsdal},
  \bibinfo{author}{R.~Brekken}, \bibinfo{author}{J.~G. Isaksen},
  \bibinfo{title}{A Fully-Coupled Fluid-Structure Interaction Simulation of
  Cerebral Aneurysms}, \bibinfo{journal}{Computational Mechanics}
  \bibinfo{volume}{46} (\bibinfo{year}{2010}{\natexlab{b}})
  \bibinfo{pages}{3--16}, ISSN \bibinfo{issn}{01787675},
  \doi{\bibinfo{doi}{10.1007/s00466-009-0421-4}}.

\bibitem[{Sanchez et~al.(2014)Sanchez, Ecker, Ambard, Jourdan, Nicoud, Mendez,
  Lejeune, Thines, Dufour, Brunel, Machi, Lobotesis, Bonafe, and
  Costalat}]{Sanchez2014}
\bibinfo{author}{M.~Sanchez}, \bibinfo{author}{O.~Ecker},
  \bibinfo{author}{D.~Ambard}, \bibinfo{author}{F.~Jourdan},
  \bibinfo{author}{F.~Nicoud}, \bibinfo{author}{S.~Mendez},
  \bibinfo{author}{J.~P. Lejeune}, \bibinfo{author}{L.~Thines},
  \bibinfo{author}{H.~Dufour}, \bibinfo{author}{H.~Brunel},
  \bibinfo{author}{P.~Machi}, \bibinfo{author}{K.~Lobotesis},
  \bibinfo{author}{A.~Bonafe}, \bibinfo{author}{V.~Costalat},
  \bibinfo{title}{Intracranial Aneurysmal Pulsatility as a New Individual
  Criterion for Rupture Risk Evaluation: {{Biomechanical}} and Numeric Approach
  ({{IRRAs Project}})}, \bibinfo{journal}{American Journal of Neuroradiology}
  \bibinfo{volume}{35} (\bibinfo{year}{2014}) \bibinfo{pages}{1765--1771},
  \doi{\bibinfo{doi}{10.3174/ajnr.A3949}}.

\bibitem[{Demiray(1972)}]{Demiray1972}
\bibinfo{author}{H.~Demiray}, \bibinfo{title}{A Note on the Elasticity of Soft
  Biological Tissues}, \bibinfo{journal}{Journal of Biomechanics}
  \bibinfo{volume}{5} (\bibinfo{year}{1972}) \bibinfo{pages}{309--311},
  \doi{\bibinfo{doi}{10.1016/0021-9290(72)90047-4}}.

\bibitem[{Ramachandran et~al.(2012)Ramachandran, Laakso, Harbaugh, and
  Raghavan}]{Ramachandran2012}
\bibinfo{author}{M.~Ramachandran}, \bibinfo{author}{A.~Laakso},
  \bibinfo{author}{R.~E. Harbaugh}, \bibinfo{author}{M.~L. Raghavan},
  \bibinfo{title}{On the Role of Modeling Choices in Estimation of Cerebral
  Aneurysm Wall Tension}, \bibinfo{journal}{Journal of Biomechanics}
  \bibinfo{volume}{45}~(\bibinfo{number}{16}) (\bibinfo{year}{2012})
  \bibinfo{pages}{2914--2919}, ISSN \bibinfo{issn}{00219290},
  \doi{\bibinfo{doi}{10.1016/j.jbiomech.2012.07.029}}.

\bibitem[{{T. Passerini, M. Piccinelli, A. Veneziani and L.
  Antiga}(2021)}]{aneurisk}
\bibinfo{author}{{T. Passerini, M. Piccinelli, A. Veneziani and L. Antiga}},
  \bibinfo{title}{{aneurisk}},
  \bibinfo{howpublished}{\url{http://ecm2.mathcs.emory.edu/aneuriskweb/index}},
  \bibinfo{year}{2021}.

\bibitem[{VMTK(2017)}]{vmtk}
\bibinfo{author}{VMTK}, \bibinfo{title}{{VMTK Website}},
  \bibinfo{howpublished}{\url{http://www.vmtk.org/}}, \bibinfo{note}{[Accessed
  19-June-2017]}, \bibinfo{year}{2017}.

\bibitem[{Piccinelli et~al.(2009)Piccinelli, Veneziani, Steinman, Remuzzi, and
  Antiga}]{Piccinelli2009}
\bibinfo{author}{M.~Piccinelli}, \bibinfo{author}{A.~Veneziani},
  \bibinfo{author}{D.~A. Steinman}, \bibinfo{author}{A.~Remuzzi},
  \bibinfo{author}{L.~Antiga}, \bibinfo{title}{A Framework for Geometric
  Analysis of Vascular Structures: {{Application}} to Cerebral Aneurysms},
  \bibinfo{journal}{IEEE Transactions on Medical Imaging}
  \bibinfo{volume}{28}~(\bibinfo{number}{8}) (\bibinfo{year}{2009})
  \bibinfo{pages}{1141--1155}, \doi{\bibinfo{doi}{10.1109/TMI.2009.2021652}}.

\bibitem[{Antiga et~al.(2002)Antiga, {Ene-Iordache}, Caverni, Cornalba, and
  Remuzzi}]{Antiga2002}
\bibinfo{author}{L.~Antiga}, \bibinfo{author}{B.~{Ene-Iordache}},
  \bibinfo{author}{L.~Caverni}, \bibinfo{author}{G.~P. Cornalba},
  \bibinfo{author}{A.~Remuzzi}, \bibinfo{title}{Geometric Reconstruction for
  Computational Mesh Generation of Arterial Bifurcations from {{CT}}
  Angiography}, \bibinfo{journal}{Computerized Medical Imaging and Graphics}
  \bibinfo{volume}{26} (\bibinfo{year}{2002}) \bibinfo{pages}{227--235}.

\bibitem[{Antiga et~al.(2008)Antiga, Piccinelli, Botti, {Ene-Iordache},
  Remuzzi, and Steinman}]{Antiga2008}
\bibinfo{author}{L.~Antiga}, \bibinfo{author}{M.~Piccinelli},
  \bibinfo{author}{L.~Botti}, \bibinfo{author}{B.~{Ene-Iordache}},
  \bibinfo{author}{A.~Remuzzi}, \bibinfo{author}{D.~A. Steinman},
  \bibinfo{title}{An Image-Based Modeling Framework for Patient-Specific
  Computational Hemodynamics}, \bibinfo{journal}{Medical and Biological
  Engineering and Computing} \bibinfo{volume}{46} (\bibinfo{year}{2008})
  \bibinfo{pages}{1097--1112}, \doi{\bibinfo{doi}{10.1007/s11517-008-0420-1}}.

\bibitem[{Kanyanta(2009)}]{Kanyanta2009}
\bibinfo{author}{V.~Kanyanta}, \bibinfo{title}{Towards Early Diagnosis of
  Atherosclerosis: Accurate Prediction of Wall Shear Stress}, Ph.D. thesis,
  \bibinfo{year}{2009}.

\bibitem[{Kanyanta et~al.(2009)Kanyanta, Ivankovic, and Karac}]{Kanyanta2009a}
\bibinfo{author}{V.~Kanyanta}, \bibinfo{author}{A.~Ivankovic},
  \bibinfo{author}{A.~Karac}, \bibinfo{title}{Validation of a Fluid-Structure
  Interaction Numerical Model for Predicting Flow Transients in Arteries},
  \bibinfo{journal}{Journal of Biomechanics}
  \bibinfo{volume}{42}~(\bibinfo{number}{11}) (\bibinfo{year}{2009})
  \bibinfo{pages}{1705--1712}, ISSN \bibinfo{issn}{00219290},
  \doi{\bibinfo{doi}{10.1016/j.jbiomech.2009.04.023}}.

\bibitem[{Tandis and Ashrafizadeh(2019)}]{Tandis2019}
\bibinfo{author}{E.~Tandis}, \bibinfo{author}{A.~Ashrafizadeh},
  \bibinfo{title}{A Numerical Study on the Fluid Compressibility Effects in
  Strongly Coupled Fluid\textendash Solid Interaction Problems},
  \bibinfo{journal}{Engineering with Computers} ISSN \bibinfo{issn}{14355663},
  \doi{\bibinfo{doi}{10.1007/s00366-019-00880-4}}.

\bibitem[{Mooney(1940)}]{Mooney1940}
\bibinfo{author}{M.~Mooney}, \bibinfo{title}{A {{Theory}} of {{Large Elastic
  Deformation}}}, \bibinfo{journal}{Journal of Applied Physics}
  \bibinfo{volume}{11}~(\bibinfo{number}{153}),
  \doi{\bibinfo{doi}{10.1063/1.1713863}}.

\bibitem[{Holzapfel et~al.(2000)Holzapfel, Gasser, and Ogden}]{Holzapfel2000}
\bibinfo{author}{G.~A. Holzapfel}, \bibinfo{author}{T.~C. Gasser},
  \bibinfo{author}{R.~W. Ogden}, \bibinfo{title}{A New Constitutive Framework
  for Arterial Wall Mechanics and a Comparative Study of Material Models},
  \bibinfo{journal}{Journal of Elasticity}
  \bibinfo{volume}{61}~(\bibinfo{number}{1-3}) (\bibinfo{year}{2000})
  \bibinfo{pages}{1--48}, ISSN \bibinfo{issn}{03743535},
  \doi{\bibinfo{doi}{10.1023/A:1010835316564}}.

\bibitem[{Holzapfel(2000)}]{Holzapfel2000a}
\bibinfo{author}{G.~A. Holzapfel}, \bibinfo{title}{Nonlinear {{Solid
  Mechanics}}}, \bibinfo{publisher}{{John Wiley \& Sons, Inc.}},
  \bibinfo{address}{{Chichester}}, \bibinfo{year}{2000}.

\bibitem[{Hoi et~al.(2010)Hoi, Wasserman, Xie, Najjar, Ferruci, Lakatta,
  Gerstenblith, and Steinman}]{Hoi2010}
\bibinfo{author}{Y.~Hoi}, \bibinfo{author}{B.~A. Wasserman},
  \bibinfo{author}{Y.~J. Xie}, \bibinfo{author}{S.~S. Najjar},
  \bibinfo{author}{L.~Ferruci}, \bibinfo{author}{E.~G. Lakatta},
  \bibinfo{author}{G.~Gerstenblith}, \bibinfo{author}{D.~A. Steinman},
  \bibinfo{title}{Characterization of Volumetric Flow Rate Waveforms at the
  Carotid Bifurcations of Older Adults}, \bibinfo{journal}{Physiological
  Measurement} \bibinfo{volume}{31}~(\bibinfo{number}{3})
  (\bibinfo{year}{2010}) \bibinfo{pages}{291--302}, ISSN
  \bibinfo{issn}{09673334}, \doi{\bibinfo{doi}{10.1088/0967-3334/31/3/002}}.

\bibitem[{Zarrinkoob et~al.(2015)Zarrinkoob, Ambarki, W{\aa}hlin, Birgander,
  Eklund, and Malm}]{Zarrinkoob2015}
\bibinfo{author}{L.~Zarrinkoob}, \bibinfo{author}{K.~Ambarki},
  \bibinfo{author}{A.~W{\aa}hlin}, \bibinfo{author}{R.~Birgander},
  \bibinfo{author}{A.~Eklund}, \bibinfo{author}{J.~Malm}, \bibinfo{title}{Blood
  Flow Distribution in Cerebral Arteries}, \bibinfo{journal}{Journal of
  Cerebral Blood Flow and Metabolism} \bibinfo{volume}{35}
  (\bibinfo{year}{2015}) \bibinfo{pages}{648--654}, ISSN
  \bibinfo{issn}{15597016}, \doi{\bibinfo{doi}{10.1038/jcbfm.2014.241}}.

\bibitem[{Chnafa et~al.(2018)Chnafa, Brina, Pereira, and Steinman}]{Chnafa2018}
\bibinfo{author}{X.~C. Chnafa}, \bibinfo{author}{X.~O. Brina},
  \bibinfo{author}{V.~M. Pereira}, \bibinfo{author}{X.~D.~A. Steinman},
  \bibinfo{title}{Better {{Than Nothing}}: {{A Rational Approach}} for
  {{Minimizing}} the {{Impact}} of {{Outflow Strategy}} on {{Cerebrovascular
  Simulations}}}, \bibinfo{journal}{American Journal of Neuroradiology}
  \bibinfo{volume}{39} (\bibinfo{year}{2018}) \bibinfo{pages}{337--343},
  \doi{\bibinfo{doi}{10.3174/ajnr.A5484}}.

\bibitem[{Valencia et~al.(2013)Valencia, Burdiles, Ignat, Mura, Bravo, Rivera,
  and Sordo}]{Valencia2013}
\bibinfo{author}{A.~Valencia}, \bibinfo{author}{P.~Burdiles},
  \bibinfo{author}{M.~Ignat}, \bibinfo{author}{J.~Mura},
  \bibinfo{author}{E.~Bravo}, \bibinfo{author}{R.~Rivera},
  \bibinfo{author}{J.~Sordo}, \bibinfo{title}{Fluid Structural Analysis of
  Human Cerebral Aneurysm Using Their Own Wall Mechanical Properties},
  \bibinfo{journal}{Computational and Mathematical Methods in Medicine}
  \bibinfo{volume}{2013} (\bibinfo{year}{2013}) \bibinfo{pages}{1--18}, ISSN
  \bibinfo{issn}{1748670X}, \doi{\bibinfo{doi}{10.1155/2013/293128}}.

\bibitem[{Fung(1993)}]{Fung1993}
\bibinfo{author}{Y.-C. Fung}, \bibinfo{title}{Biomechanics - {{Mechanical
  Properties}} of {{Living Tissues}}}, \bibinfo{publisher}{{Springer New
  York}}, \bibinfo{address}{{New York, NY}}, ISBN
  \bibinfo{isbn}{978-1-4419-3104-7 978-1-4757-2257-4},
  \doi{\bibinfo{doi}{10.1007/978-1-4757-2257-4}}, \bibinfo{year}{1993}.

\bibitem[{Nakagawa et~al.(2016)Nakagawa, Shojima, Yoshino, Kin, Imai, Nomura,
  Saito1, Nakatomi, Oyama1, and Saito}]{Nakagawa2016}
\bibinfo{author}{D.~Nakagawa}, \bibinfo{author}{M.~Shojima},
  \bibinfo{author}{M.~Yoshino}, \bibinfo{author}{T.~Kin},
  \bibinfo{author}{H.~Imai}, \bibinfo{author}{S.~Nomura},
  \bibinfo{author}{T.~Saito1}, \bibinfo{author}{H.~Nakatomi},
  \bibinfo{author}{H.~Oyama1}, \bibinfo{author}{N.~Saito},
  \bibinfo{title}{Wall-to-lumen Ratio of Intracranial Arteries Measured by
  Indocyanine Green Angiography}, \bibinfo{journal}{Asian Journal of
  Neurosurgery} \bibinfo{volume}{11} (\bibinfo{year}{2016})
  \bibinfo{pages}{361--364}, \doi{\bibinfo{doi}{10.4103/1793-5482.175637}}.

\bibitem[{Torii et~al.(2010)Torii, Oshima, Kobayashi, Takagi, and
  Tezduyar}]{Torii2010}
\bibinfo{author}{R.~Torii}, \bibinfo{author}{M.~Oshima},
  \bibinfo{author}{T.~Kobayashi}, \bibinfo{author}{K.~Takagi},
  \bibinfo{author}{T.~E. Tezduyar}, \bibinfo{title}{Influence of Wall Thickness
  on Fluid\textendash Structure Interaction Computations of Cerebral
  Aneurysms}, \bibinfo{journal}{International Journal for Numerical Methods in
  Biomedical Engineering} \bibinfo{volume}{26} (\bibinfo{year}{2010})
  \bibinfo{pages}{336--347}, ISSN \bibinfo{issn}{20407939},
  \doi{\bibinfo{doi}{10.1002/cnm}}.

\bibitem[{Furukawa et~al.(2018)Furukawa, Ishida, Tsuji, Miura, Kishimoto,
  Shiba, Tanemura, Umeda, Sano, Yasuda, Shimosaka, and Suzuki}]{Furukawa2018}
\bibinfo{author}{K.~Furukawa}, \bibinfo{author}{F.~Ishida},
  \bibinfo{author}{M.~Tsuji}, \bibinfo{author}{Y.~Miura},
  \bibinfo{author}{T.~Kishimoto}, \bibinfo{author}{M.~Shiba},
  \bibinfo{author}{H.~Tanemura}, \bibinfo{author}{Y.~Umeda},
  \bibinfo{author}{T.~Sano}, \bibinfo{author}{R.~Yasuda},
  \bibinfo{author}{S.~Shimosaka}, \bibinfo{author}{H.~Suzuki},
  \bibinfo{title}{Hemodynamic Characteristics of Hyperplastic Remodeling
  Lesions in Cerebral Aneurysms}, \bibinfo{journal}{PLOS ONE}
  \bibinfo{volume}{13} (\bibinfo{year}{2018}) \bibinfo{pages}{1--11}, ISSN
  \bibinfo{issn}{1932-6203}, \doi{\bibinfo{doi}{10.1371/journal.pone.0191287}}.

\bibitem[{Oliveira et~al.(2021)Oliveira, Santos, Militzer, Baccin, Tatit, and
  Gasche}]{Oliveira2021}
\bibinfo{author}{I.~L. Oliveira}, \bibinfo{author}{G.~B. Santos},
  \bibinfo{author}{J.~Militzer}, \bibinfo{author}{C.~E. Baccin},
  \bibinfo{author}{R.~T. Tatit}, \bibinfo{author}{J.~L. Gasche},
  \bibinfo{title}{A Longitudinal Study of a Lateral Intracranial Aneurysm:
  Identifying the Hemodynamic Parameters behind Its Inception and Growth Using
  Computational Fluid Dynamics}, \bibinfo{journal}{Journal of the Brazilian
  Society of Mechanical Sciences and Engineering} \bibinfo{volume}{43}
  (\bibinfo{year}{2021}) \bibinfo{pages}{138},
  \doi{\bibinfo{doi}{10.1007/s40430-021-02836-6}}.

\bibitem[{Cardiff et~al.(2018)Cardiff, Kara{\v c}, De~Jaeger, Jasak, Nagy,
  Ivankovi{\'c}, and Tukovi{\'c}}]{Cardiff2018}
\bibinfo{author}{P.~Cardiff}, \bibinfo{author}{A.~Kara{\v c}},
  \bibinfo{author}{P.~De~Jaeger}, \bibinfo{author}{H.~Jasak},
  \bibinfo{author}{J.~Nagy}, \bibinfo{author}{A.~Ivankovi{\'c}},
  \bibinfo{author}{{\v Z}.~Tukovi{\'c}}, \bibinfo{title}{An Open-Source Finite
  Volume Toolbox for Solid Mechanics and Fluid-Solid Interaction Simulations},
  \bibinfo{type}{Tech. Rep.}, \bibinfo{year}{2018}.

\bibitem[{foam-extend Project(2017)}]{foam-extend}
\bibinfo{author}{foam-extend Project}, \bibinfo{title}{{foam-extend Website}},
  \bibinfo{howpublished}{\url{https://sourceforge.net/projects/foam-extend/}},
  \bibinfo{note}{[Accessed 19-June-2017]}, \bibinfo{year}{2017}.

\bibitem[{Weller et~al.(1998)Weller, Tabor, Jasak, and Fureby}]{Weller1998}
\bibinfo{author}{H.~G. Weller}, \bibinfo{author}{G.~Tabor},
  \bibinfo{author}{H.~Jasak}, \bibinfo{author}{C.~Fureby}, \bibinfo{title}{A
  Tensorial Approach to Computational Continuum Mechanics Using Object-Oriented
  Techniques}, \bibinfo{journal}{Computers in Physics} \bibinfo{volume}{12}
  (\bibinfo{year}{1998}) \bibinfo{pages}{620--631}.

\bibitem[{Issa(1986)}]{Issa1986}
\bibinfo{author}{R.~I. Issa}, \bibinfo{title}{Solution of the Implicitly
  Discretised Fluid Flow Equations by Operator-Splitting},
  \bibinfo{journal}{Journal of Computational Physics}
  \bibinfo{volume}{62}~(\bibinfo{number}{1}) (\bibinfo{year}{1986})
  \bibinfo{pages}{40--65}, ISSN \bibinfo{issn}{10902716},
  \doi{\bibinfo{doi}{10.1016/0021-9991(86)90099-9}}.

\bibitem[{Jasak(1996)}]{JasakThesis1996}
\bibinfo{author}{H.~Jasak}, \bibinfo{title}{Error {{Analysis}} and
  {{Estimation}} for the {{Finite Volume Method}} with {{Applications}} to
  {{Fluid Flows}}}, Ph.D. thesis, \bibinfo{school}{Imperial College},
  \bibinfo{year}{1996}.

\bibitem[{Wheel(1999)}]{Wheel1999}
\bibinfo{author}{M.~A. Wheel}, \bibinfo{title}{A Mixed Finite Volume
  Formulation for Determining the Small Strain Deformation of Incompressible
  Materials}, \bibinfo{journal}{International Journal for Numerical Methods in
  Engineering} \bibinfo{volume}{44} (\bibinfo{year}{1999})
  \bibinfo{pages}{1843--1861}.

\bibitem[{NumPy(2017)}]{numpy}
\bibinfo{author}{NumPy}, \bibinfo{title}{{NumPy Website}},
  \bibinfo{howpublished}{\url{https://numpy.org/}}, \bibinfo{note}{[Accessed
  05-May-2022]}, \bibinfo{year}{2017}.

\bibitem[{SciPy(2017)}]{scipy}
\bibinfo{author}{SciPy}, \bibinfo{title}{{SciPy Website}},
  \bibinfo{howpublished}{\url{https://www.scipy.org}}, \bibinfo{note}{[Accessed
  05-May-2022]}, \bibinfo{year}{2017}.

\bibitem[{Castro(2013)}]{Castro2013}
\bibinfo{author}{M.~A. Castro}, \bibinfo{title}{Understanding the {{Role}} of
  {{Hemodynamics}} in the {{Initiation}}, {{Progression}}, {{Rupture}}, and
  {{Treatment Outcome}} of {{Cerebral Aneurysm}} from {{Medical Image-Based
  Computational Studies}}}, \bibinfo{journal}{ISRN Radiology}
  (\bibinfo{year}{2013})
  \bibinfo{pages}{1--17}\doi{\bibinfo{doi}{10.5402/2013/602707}}.

\bibitem[{Cebral et~al.(2015)Cebral, Vazquez, Sforza, Houzeaux, Tateshima,
  Scrivano, Bleise, Lylyk, and Putman}]{Cebral2015}
\bibinfo{author}{J.~R. Cebral}, \bibinfo{author}{M.~Vazquez},
  \bibinfo{author}{D.~M. Sforza}, \bibinfo{author}{G.~Houzeaux},
  \bibinfo{author}{S.~Tateshima}, \bibinfo{author}{E.~Scrivano},
  \bibinfo{author}{C.~Bleise}, \bibinfo{author}{P.~Lylyk},
  \bibinfo{author}{C.~M. Putman}, \bibinfo{title}{Analysis of Hemodynamics and
  Wall Mechanics at Sites of Cerebral Aneurysm Rupture},
  \bibinfo{journal}{Journal of NeuroInterventional Surgery}
  \bibinfo{volume}{7}~(\bibinfo{number}{7}), ISSN \bibinfo{issn}{1759-8478},
  \doi{\bibinfo{doi}{10.1136/neurintsurg-2014-011247}}.

\bibitem[{Holzapfel and Ogden(2003)}]{Holzapfel2003}
\bibinfo{editor}{G.~A. Holzapfel}, \bibinfo{editor}{R.~W. Ogden} (Eds.),
  \bibinfo{title}{Biomechanics of {{Soft Tissue}} in {{Cardiovascular
  Systems}}}, \bibinfo{publisher}{{Springer Vienna}},
  \bibinfo{address}{{Vienna}}, ISBN \bibinfo{isbn}{978-3-211-00455-5
  978-3-7091-2736-0}, \doi{\bibinfo{doi}{10.1007/978-3-7091-2736-0}},
  \bibinfo{year}{2003}.

\end{thebibliography}
